\documentclass[a4paper,fleqn]{cas-sc}

\usepackage[authoryear,longnamesfirst]{natbib}

\usepackage{rotating}
\usepackage{float}
\usepackage[linesnumbered,ruled]{algorithm2e}
\usepackage{caption}
\usepackage[section]{placeins}
\usepackage{times}
\usepackage{soul}
\usepackage{url}
\usepackage{algorithmic}
\usepackage{subfig}
\usepackage{diagbox}
\usepackage{makecell}

\usepackage[normalem]{ulem}
\useunder{\uline}{\ul}{}

\usepackage{xcolor}

\newcommand*\red[1]{\textcolor{red}{#1}}

\def\tsc#1{\csdef{#1}{\textsc{\lowercase{#1}}\xspace}}
\tsc{WGM}
\tsc{QE}

\usepackage[english]{babel}
\usepackage{amsthm}
\theoremstyle{definition}
\newtheorem{definition}{Definition}[section]

\begin{document}
\let\WriteBookmarks\relax
\def\floatpagepagefraction{1}
\def\textpagefraction{.001}

\shorttitle{Solution Space Exploring and Descent for PESS}    

\shortauthors{Anonymous}

\title [mode = title]{
An Efficient Solution Space Exploring and Descent Method for Packing Equal Spheres in a Sphere
}


\author[1]{Jianrong Zhou}[]
\author[1]{Shuo Ren}[]
\author[1]{Kun He$^*$}[]
\author[2]{Yanli Liu} 
\author[3]{Chu-Min Li}[]

\affiliation[1]{organization={School of Computer Science and Technology, Huazhong University of Science and Technology},
            city={Wuhan},
            postcode={430074}, 
            country={China}}
            
\affiliation[2]{organization={School of Science, Wuhan University of Science and Technology},
            city={Wuhan},
            postcode={430081}, 
            country={China}}
            
\affiliation[3]{organization={MIS, University of Picardie Jules Verne},
            city={Amiens},
            postcode={80039}, 
            country={France}}

%
















\begin{abstract}
The problem of packing equal spheres in a spherical container is a classic global optimization problem, which has attracted enormous studies in academia and found various applications in industry. This problem is computationally challenging, and many efforts focus on small-scale instances with the number of spherical items less than 200 in the literature. In this work, we propose an efficient local search heuristic algorithm named solution space exploring and descent for solving this problem, which can quantify the solution's quality to determine the number of exploring actions and quickly discover a high-quality solution. Besides, we propose an adaptive neighbor object maintenance method to speed up the convergence of the continuous optimization process and reduce the time consumption. Computational experiments on a large number of benchmark instances with $5 \leq n \leq 400$ spherical items show that our algorithm significantly outperforms the state-of-the-art algorithm. In particular, it improves the 274 best-known results and matches the 84 best-known results out of the 396 well-known benchmark instances. 
\end{abstract}


\maketitle

\section{Introduction} \label{sec:01intro}
The Sphere Packing Problem (SPP) is a classic optimization problem that seeks an arrangement of non-overlapping spherical objects within a container or a bounded or unbounded space where the arrangement is usually required to be as dense as possible. SPP is
an important 
topic in many research fields and has a variety of real-world applications.
In chemistry, a dense configuration of SPP can assist researchers in realizing the structures of chemical compounds~\citep{o2011new}.
In material science, the random sphere packing and particle packing models are adopted to analyze the structure of powders, liquids, proteins, colloidal suspensions, and porous materials~\citep{clarke1993structural, silbert2002geometry, aste2005geometrical, klumov2011structural, klumov2014structural}, where these models are used in the study of microscopic particle arrangement and the phenomena of fluid flow, electrical conductivity, stress distribution and other physical characteristics. 
In digital communication, SPP is a model in the study of telecommunication systems, optical communications and classical-quantum channels~\citep{valembois2004sphere, fazeli2014generalized, chaaban2016free, cheng2019quantum}.
SPP also has an application on radiosurgical treatment planning~\citep{wang1999packing}.

Especially, in mathematics, SPP is a classic and famous research topic. The well-known highest density $\rho^* = \pi / \sqrt{18} \approx 0.74048$ is proven to be obtained by the Face-Centred Cubic (FCC) or Hexagonal Close Packing (HCP) arrangements of the congruent sphere packing in unbounded space. And many efforts have been 
devoted to finding a tighter bound, contact number and other mathematical features in various instances and spaces, such as $n$-dimensional ($n \geq 3$) Euclidean space~\citep{cohn2003new, bezdek2012contact, viazovska2017sphere, cohn2017sphere, cohn2017conceptual}, 3-dimensional non-Euclidean space~\citep{kazakov2018sphere}, sphere packing on error-correcting codes~\citep{leech1971sphere, fazeli2015generalized} or spherical codes~\citep{cohn2014sphere}.

SPP has a lot of variants, some of which require packing equal or unequal sphere objects in a container with a specified geometric shape, such as rectangular container~\citep{akeb2016two, hifi2019local, hifi2023threshold}, spherical container~\citep{liu2009effective, zeng2012algorithm, hifi2018hybrid}, cylindrical container~\citep{han2005sphere, mueller2005numerically}, and other shapes of containers~\citep{birgin2008minimizing, labra2009high, stoyan2013packing, stoyan2020optimized}. 
Besides, the hypersphere packing (i.e., high-dimensional sphere packing)~\citep{stoyan2012packing} can be considered as an SPP variant. 
Furthermore, the 2-dimensional variant of SPP degenerates into circle packing problems~\citep{he2018efficient, he2021adaptive, lai2022iterated}, which are well-known and hot topics in academia and industry.

One important variant of SPP 
is Packing Equal Spheres in a Sphere (PESS)~\citep{huang2011quasi, m2013packing, hifi2017solving}. Given $n$ unit sphere objects, PESS aims to pack the $n$ sphere objects into a spherical container with the least radius. Due to its simplicity in form, PESS is a classic and representative problem in the SPP family.

However, even the restricted models of geometric packing problems (e.g., circle packing problems) have been proven to be NP-hard~\citep{fowler1981optimal, demaine2016circle}. Thus, solving SPP, which can be regarded as an extension of circle packing problems, is computationally very challenging. Most works aim to design an algorithm for solving SPP on small-scale instances where the number of packing objects is less than 200 ($n \leq 200$). 
Only a few studies have attempted to solve SPP on a large scale, and the algorithms proposed in these studies require a significant amount of time to obtain a dense configuration of SPP.

In this work, we address the PESS variant of SPP on small and moderate scales with the number of packing objects up to 400. We employ the elastic model (also known as the Quasi-Physical Quasi-Human model, QPQH)~\citep{huang1999two, huang2011quasi} of PESS.
This model allows for the intersection of spheres and containers, and an elastic metric function is defined to calculate the degree of intersection.
Then, the algorithm can adapt the continuous optimization method to minimize the intersection area, such as gradient descent~\citep{huang2011quasi} and quasi-Newton method~\citep{he2018efficient, lai2022iterated}. 

Based on the PESS elastic system, we propose an advanced local search heuristic, termed Solution-space Exploring and Descent (SED), to solve the PESS problem. SED iteratively performs two processes, ``Exploring'' and ``Descent'', to improve the best found solution during the heuristic search. 
In the ``Exploring'' process, SED perturbs the operated solution to obtain several candidate solutions, and the number of perturbations depends on a new metric function that uses a new expression to quantify the solution's quality. 
In the ``Descent'' process, SED employs a strategy to select a high-quality solution from the several candidate solutions as the offspring solution in the next iteration. 
Experiments show that SED could efficiently find a dense configuration by executing the iteration process on small and moderate instances. 

In addition, we propose an Adaptive Neighbor object Maintenance (ANM) method to maintain the neighbor structure~\citep{he2018efficient} for the PESS elastic system. The neighbor structure stores adjacent object messages for each packing object to reduce the time consumption of the elastic function and gradient calculation. 
ANM uses two variables ``counter'' and ``deferring length'' to achieve the adaptive feature, which maintains the neighbor structure in the continuous optimization process. 
When the layout changes significantly, ANM reconstructs the neighbor structure at each iteration. Otherwise, ANM defers the neighbor structure maintenance. 
In this way, ANM can reduce the time consumption of the neighbor structure construction during the continuous optimization process.

The experimental results show that our proposed algorithm significantly outperforms the state-of-the-art algorithm. Specifically, our algorithm yields 14 better, 30 equal and 2 worse results than the state-of-the-art algorithm out of the 46 comparing instances ($5 \leq n \leq 50$). 
In addition, our algorithm improves the best-known results for 274 instances, matches the best-known results for 84 instances, and obtains the worse results for 38 instances out of the 396 benchmark instances.
Besides, the extensive experiments show our proposed ANM module can defer over 50\%  unnecessary maintenance and reduce over 30\% time consumption in the continuous optimization process. 

The main contributions of this work are summarized as follows:
\begin{itemize}
    \item We propose an efficient Solution space Exploring and Descent (SED) heuristic for solving the PESS problem.
    \item We propose an Adaptive Neighbor object Maintenance (ANM) method of maintaining the neighbor structure for solving the PESS problem, which is a general method and can be easily adapted for other packing problems.
    \item Extensive experiments on a large number of the benchmark instances with up to $n = 400$ demonstrate the excellent performance and efficiency of our proposed algorithm, gaining new best solutions on many instances.
\end{itemize}

The rest of this paper is organized as follows. 
Section~\ref{sec:02RW} presents the related works of the PSS and PESS problems, including some construction methods, some typical models, heuristics and metaheuristics, and the recent work for solving the PESS problem.
Section~\ref{sec:03pre} introduces the mathematical formula of the PESS problem and the classic elastic model (QPQH) for solving PESS.  
Section~\ref{sec:04sed} presents the main framework of our algorithm and other components, including the initialization, the SED heuristic, the container adjustment method, the neighbor structure, and the ANM module.
Section~\ref{sec:05exp} presents the experimental results of our proposed algorithm compared with the state-of-the-art algorithm and the best-known results, parameter study, and the analysis of the ANM module.
The conclusion is drawn in the end. 

\section{Related Work} \label{sec:02RW}
There are two main categories of algorithms for solving the SPP, namely random sphere packing and dense sphere packing. 

Algorithms of random sphere packing are designed for specified purposes, such as various container filling, particle microstructure analysis, physical characteristic analysis, etc. The main idea of these algorithms belongs to the simulation method~\citep{silbert2002geometry, wouterse2006geometrical, shi2008simulation, liu2017equation} or construction method~\citep{han2005sphere, soontrapa2013mono, chen2022quasi}. Due to the characteristic of these algorithms, they can easily obtain a configuration with the number of packing spheres up to hundreds or thousands, but the packing density is not very high. 

The algorithms for dense sphere packing aim to solve SPP as an optimization problem, with the goal of finding a configuration that is as dense as possible, packing as many objects as possible into a specified container, and using as few bins as possible (in the case of bin packing problems), among other objectives. Our literature review focuses on the optimization version of dense SPP.

One of the most popular SPP variants is packing spheres in a rectangular container, and many efforts are devoted to solving this problem. 
\citet{hifi2014width} propose a width-beam search heuristic to pack the spheres one by one into the container to find a feasible solution, and a hill-climbing strategy is proposed for improving the width-beam heuristic. Meanwhile, A dichotomous search, which can be regarded as a binary search, is adapted to find a dense configuration
and is followed by several 
researches~\citep{hifi2015dichotomous, hifi2016handling, hifi2018global} based on similar ideas that improve the algorithm performance, which can be regarded as the extension of the previous work. \citet{hifi2019local} propose an efficient local search-based method with multi-strategies, including using a basic greedy local strategy to ensure a feasible solution, employing the elastic model (QPQH) to solve the decision problem of SPP, and a drop and rebuild method for perturbation. \citet{akeb2016two} proposes a multi-level look-ahead strategy and some population-based algorithms~\citep{hifi2022population, hifi2023threshold} are proposed for solving this problem.

Packing spheres in a cubic or spherical container is also a popular and representative problem in the SPP family. \citet{huang2011quasi} present the classic and powerful elastic model (QPQH) for solving the Packing Equal Spheres in a Sphere (PESS) problem and packing equal spheres in a cube problem, and a serial symmetrical relocation strategy is proposed for perturbation. \citet{zeng2012algorithm} extend this model to solve the unequal spheres packing problem. \citet{liu2009effective} propose an energy landscape paving method and combine it with the gradient descent method based on the elastic model to solve the circle and sphere packing problem. \citet{hifi2017solving} propose an adaptive particle swarm optimization algorithm based on the elastic model to solve the PESS problem, and an extension work~\citep{hifi2018hybrid} is presented for other equal sphere packing problems. \citet{m2012packing} and \citet{m2013packing} present using the non-linear program and variable neighborhood search method to solve the equal sphere packing problem. There are also some works based on the non-linear program method to solve the sphere packing problem in various containers~\citep{birgin2008minimizing} and hyperspaces~\citep{stoyan2012packing, stoyan2020optimized}.

The well-known Packomania website~\citep{Spechtweb} maintained by Specht presents many circle and sphere packing problems and records their best-known solutions. From the recently updated history on the PESS problem at Packomania, the best-known solutions for $1 \leq n \leq 25$ are provided by mathematical analysis  except for $n = 24$, \citet{huang2011quasi} hold several best-known solutions for $26 \leq n \leq 160$, and the remaining best-known solutions are held by Specht with unpublished methods. 

In summary, most efforts are based on the local search, heuristic, metaheuristic and non-linear programming methods to solve the optimization version of SPP. And the elastic model (QPQH) based methods~\citep{huang2011quasi, hifi2018hybrid} can be regarded as the state-of-the-art algorithms for solving the PESS problem.

\section{Preliminaries} \label{sec:03pre}
\subsection{Problem Formulation} 

The PESS problem aims to pack $n$ unit spheres $\{ s_1, s_2, ..., s_n\}$ into a spherical container and minimizes the radius of the container while subjected to two constraints: 
(I) No pair of unit spheres overlap with each other; (II) No unit sphere exceeds the spherical container. 
The PESS problem can be formulated in three-dimensions Cartesian coordinate system as a non-linear constrained optimization problem:

\begin{align}
    \mathrm{Minimize} \quad & R & \nonumber \\
    \textit{s.t.} \quad & \sqrt{(x_i - x_j)^2 + (y_i - y_j)^2 + (z_i - z_j)^2} \geq 2, \quad 1 \leq i, j \leq n, \ i \neq j  \label{eq-1}, \\
    \quad &  \sqrt{x_i^2 + y_i^2 + z_i^2} + 1 \leq R, \quad 1 \leq i \leq n \label{eq-2},
\end{align}
where $R$ is the radius of the spherical container centered at the origin $(0, 0)$, and the center of the unit sphere $s_i$ is located at $(x_i, y_i, z_i)$. Eqs.~\eqref{eq-1} and \eqref{eq-2} correspond to the two constraints (I) and (II), respectively. 

\subsection{The Elastic Model for PESS} 

Given a fixed spherical container of radius $R$, it is difficult to find a feasible solution or determine whether there exists a feasible solution. To cope with this difficulty, the elastic model~\citep{huang1999two, huang2011quasi}, which can be regarded as a relaxation of PESS, is proposed for solving the PESS problem. The overlapping of sphere-sphere and sphere-container is allowed in this model, and a metric function called elastic energy $E$ is designed to quantify the overlapping degree of a candidate solution. 
The goal of the algorithm based on this model is to minimize the metric function, it is equivalent to minimizing the overlapping area. 
In this way, the elastic model converts the PESS problem to an unconstrained non-convex continuous optimization problem.

\begin{figure}[tb]
    \centering
    \begin{minipage}[b]{0.49\linewidth}
        \centering
        \subfloat[Three-dimensional view]{\includegraphics[width=1\linewidth]{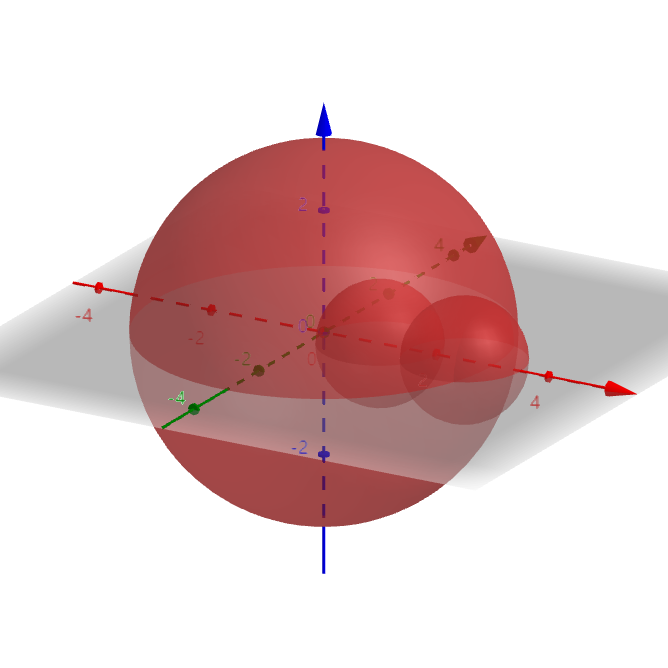}}
    \end{minipage}
    \begin{minipage}[b]{0.49\linewidth}
        \centering
        \subfloat[Sectional view for $z=0$]{\includegraphics[width=1\linewidth]{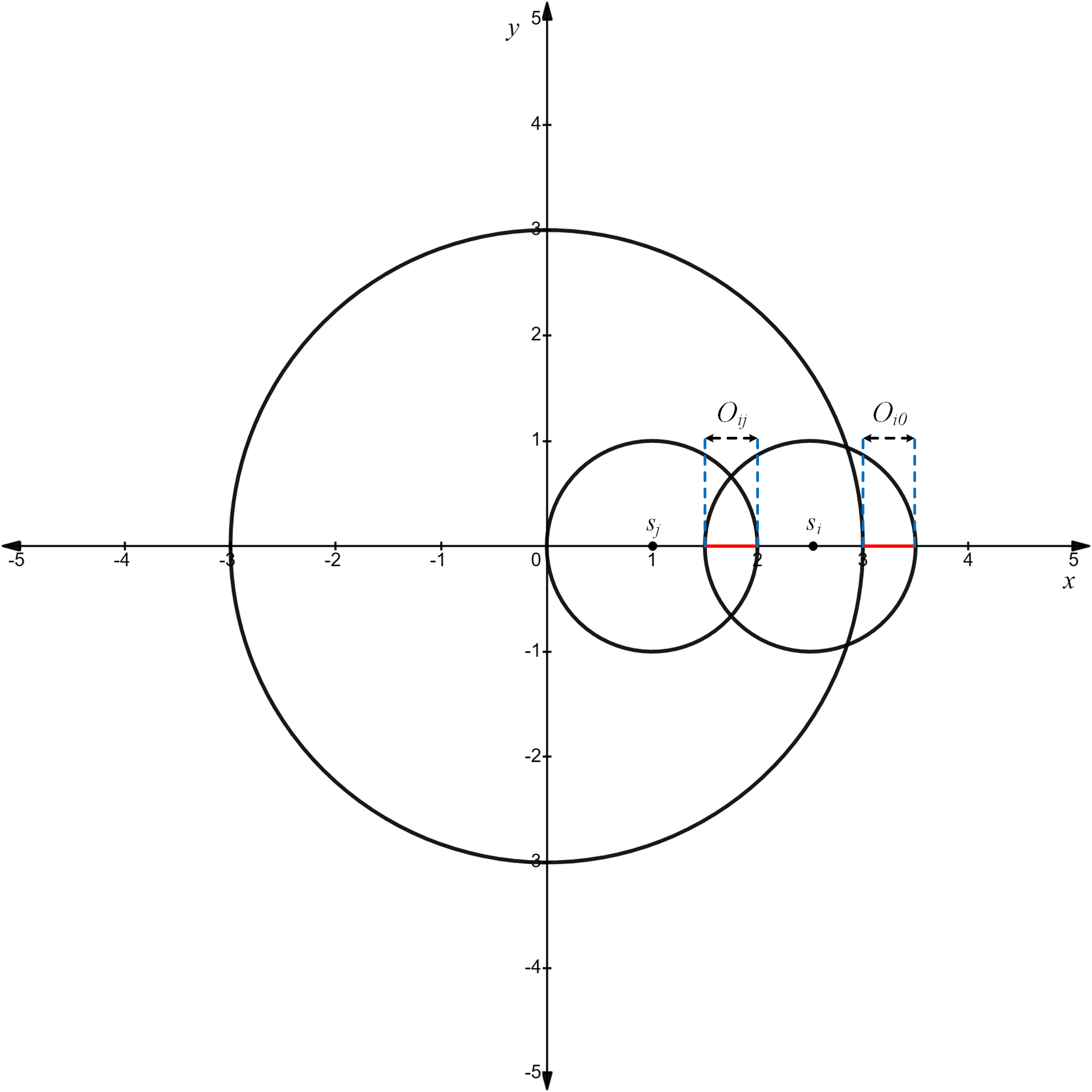}}
    \end{minipage}    

    \caption{Illustration of an example with two types of overlaps where the center of the container is located at the origin with $R = 3$ and two spherical items $s_i$ and $s_j$ are located at $(2.5, 0, 0)$ and $(1, 0, 0)$, respectively, with $r=1$. (a) shows the 3-dimensional illustration and (b) the sectional view for $z=0$ where $O_{ij}$ indicates the sphere-sphere overlapping and $O_{i0}$ indicates the sphere-container overlapping.}
    \label{fig_pre}
\end{figure}

\begin{definition}[Overlapping distance]
The overlapping distance of two unit spheres $s_i$ and $s_j$, denoted as $O_{ij}$, is defined as follows:
\begin{equation} \label{eq-3}
    O_{ij} = \max \left( 0, \ 2 - \sqrt{ (x_i - x_j)^2 + (y_i - y_j)^2 + (z_i - z_j)^2 } \right), 
\end{equation}
And the overlapping distance of a unit sphere $s_i$ to the spherical hollow container, denoted as $O_{i0}$, is defined as follows:
\begin{equation} \label{eq-4}
    O_{i0} = \max \left( 0, \ \sqrt{x_i^2 + y_i^2 + z_i^2} + 1 - R \right).
\end{equation}
\end{definition}

As is shown in the Figure~\ref{fig_pre}, the square of the embedded distance can be proportionally reflected by the elastic potential energy, according to the following elastic energy definition.

\begin{definition}[Elastic energy]
The total elastic energy $E$ of the PESS system is defined as follows:
\begin{equation} \label{eq-5}
    E_R(\boldsymbol{x}) = E(\boldsymbol{x}, R) = \sum_{i=1}^{n} \sum_{j=i+1}^{n} O_{ij}^2 + \sum_{i=1}^{n} O_{i0}^2,
\end{equation}
where $R$ is the container radius, 
and $\boldsymbol{x} = [x_1, y_1, z_1, x_2, y_2, z_2, ... , x_n, y_n, z_n]^{\rm T}$ is a vector, $\boldsymbol{x} \in \mathbb{R}^{3n}$, representing a candidate solution.
\end{definition}

Note that the elastic energy $E_R(\boldsymbol{x})$ quantifies the overlapping degree of candidate solution $\boldsymbol{x}$. If the condition $E_R(\boldsymbol{x}) = 0$ is met, 
then there is no overlap in the corresponding solution $\boldsymbol{x}$ 
(i.e., Eqs.~\ref{eq-1} and \ref{eq-2} are satisfied), indicating that $\boldsymbol{x}$ is a feasible solution for the PESS problem. 
Since we solve the problem with the fixed container of radius $R$ and use a sequential unconstrained optimization approach to 
do adjustment, 
we omit the subscript $R$ in $E_R(\boldsymbol{x})$ throughout the rest of the paper and simply denote it as $E(\boldsymbol{x})$ for the sake of readability.

\section{The Proposed Algorithm for PESS} \label{sec:04sed}
In this section, we introduce the main framework of our algorithms, including the Solution-space Exploring and Descent (SED) heuristic, the Adaptive Neighbor object Maintenance (ANM) method, and other minor components.

\begin{algorithm}[b]
\caption{The framework for solving PESS} \label{alg_frame}
\textbf{Input}: A number of unit spheres $n$; A cut-off time $T_{cut}$ \\
\textbf{Output}: A feasible solution with the minimal container radius $(\boldsymbol{x}^{*}, R^{*})$
\begin{algorithmic}[1] 
\STATE $(\boldsymbol{x}, R) \leftarrow \mathrm{initialize}(n)$
\STATE $\boldsymbol{x}^* \leftarrow \boldsymbol{x}, \ \  R^* \leftarrow R$
\WHILE {$\mathrm{time}() \leq T_{cut}$}
\STATE $R \leftarrow R^*$
\STATE $\boldsymbol{x} \leftarrow \mathrm{SED}(n, R)$
\STATE $(\boldsymbol{x}, R) \leftarrow \mathrm{adjust\_container}(\boldsymbol{x}, R)$
\IF {$R < R^{*}$}
\STATE $\boldsymbol{x}^{*} \leftarrow \boldsymbol{x}, \ \  R^{*} \leftarrow R$
\ENDIF 
\ENDWHILE
\STATE \textbf{return} $(\boldsymbol{x}^{*}, R^{*})$
\end{algorithmic}
\end{algorithm}

\subsection{Main Framework} \label{ssec:04-01frame}

We first introduce our algorithm framework, of which the pseudocode is depicted in Algorithm~\ref{alg_frame}. 

The algorithm first obtains an initial solution (line 1) by employing the initialization procedure that is described in Section~\ref{ssec:04-02init}, and the initial solution is recorded as the current best solution $(\boldsymbol{x}^*, R^*)$ (line 2). Subsequently, the algorithm performs an iterative search process to improve the current best solution until the cut-off time $T_{cut}$ is reached (lines 3-10). 

In the iterative search process, the current radius is set as the best radius found so far (line 4). The algorithm employs the SED heuristic described in Section~\ref{ssec:04-03sed} to obtain a feasible solution or an infeasible solution with the smallest energy (i.e., the smallest overlapping area) (line 5). Then the radius adjustment approach is employed to adjust the container radius of the current operated solution until the overlaps are eliminated and 
a feasible solution with a minimal container radius (line 6) is obtained. 
And the best solution is updated when a better solution is found (lines 7-9). Finally, the algorithm returns the solution $(\boldsymbol{x}^*, R^*)$ as the final feasible solution (line 11).

\subsection{Initialization} \label{ssec:04-02init}

The initialization module of the algorithm aims to rapidly provide a high-quality feasible solution to serve as a good starting point for the iterative search process.

The search algorithm based on the elastic model for solving the PESS problem has a limitation 
in that it needs a fixed container radius. 
Thus, it is necessary to estimate a lower bound radius to serve as the fixed search radius. Based on the density equation, one method to obtain a lower bound radius is as follows:
\begin{align}
    & \rho = \frac{n V_{sph.}}{V_{con.}}, \quad V_{sph.} = \frac{4}{3} \pi r^3, \quad V_{con.} = \frac{4}{3} \pi R^3, \label{eq_density} \\ 
    & \Longrightarrow \rho = \frac{n r^3}{R^3}, \nonumber \\ 
    & \Longrightarrow R = \sqrt[3]{\frac{n}{\rho}}, \quad (r = 1). \nonumber
\end{align} 
Here $n$ is the number of packing spheres, $r$ and $R$ represent the radius of packing spheres and the spherical container where $r=1$ corresponds to the PESS problem. $V_{sph.}$ and $V_{con.}$ indicate the volume of packing spheres and the spherical container, respectively, and $\rho$ denotes the density that is the ratio of the total volume of packing spheres to the volume of the spherical container. 

The highest density $\rho \simeq 0.74048$ is proved amongst all possible spherical lattice packings.
However, the lower bound based on this density is too tight for the PESS problem on small and moderate scales. 
Thus, in this work, we reference the density of the best-known results of PESS on the Packomania website~\citep{Spechtweb} for $n \leq 400$ and empirically set $\rho = 0.6$ so as to estimate the lower bound of the radius $R$, so that, the pseudocode of the initialization module depicted in Algorithm~\ref{alg_init} uses the density of $\rho = 0.6$ to estimate a lower bound radius that is $R = \sqrt[3]{\frac{n}{0.6}}$.


Then, SED is employed with the fixed radius $R$ to find a candidate solution with the smallest energy. 
Following that, the container adjustment module is applied to obtain 
a feasible solution with the smallest container radius $(\boldsymbol{x}, R)$, and it is returned as a good start for the iterative search process. 

\begin{algorithm}[tb]
\caption{Initialize ($n$)} \label{alg_init}
\textbf{Input}: A number of unit spheres $n$ \\
\textbf{Output}: A feasible solution with the minimal container radius $(\boldsymbol{x}, R)$
\begin{algorithmic}[1] 
\STATE $R \leftarrow \sqrt[3]{\frac{n}{0.6}}$
\STATE $\boldsymbol{x} \leftarrow \mathrm{SED}(n, R)$
\STATE $(\boldsymbol{x}, R) \leftarrow \mathrm{adjust\_container}(\boldsymbol{x}, R)$
\STATE \textbf{return} $(\boldsymbol{x}, R)$
\end{algorithmic}
\end{algorithm}

\subsection{SED Heuristic} \label{ssec:04-03sed}

\begin{algorithm}[tb]
\caption{SED($n, R$)} \label{alg_sed}
\textbf{Input}: A number of unit spheres $n$; A container radius $R$ \\
\textbf{Output}: A smallest energy solution found so far $\boldsymbol{x}^*$
\begin{algorithmic}[1] 
\STATE $\boldsymbol{x} \leftarrow \mathrm{random\_layout}(n, R)$
\STATE $\boldsymbol{x} \leftarrow \mathrm{optimizer}(E, \boldsymbol{x})$
\STATE $\boldsymbol{x}^* \leftarrow \boldsymbol{x}$
\FOR {$i$ from $1$ to $S_{iter}$} 
\IF {$E(\boldsymbol{x}^*) \leq 10^{-25}$}
\STATE \textbf{break}
\ENDIF
\STATE $m \leftarrow \mathrm{max}(1, J(\boldsymbol{x}))$
\STATE $C \leftarrow \emptyset$
\FOR {$j$ from $1$ to $m$} 
\STATE $\boldsymbol{x}' \leftarrow \mathrm{perturbing}(\boldsymbol{x})$ 
\STATE $\boldsymbol{x}' \leftarrow \mathrm{optimizer}(E, \boldsymbol{x}')$
\STATE $C \leftarrow C \cup \{ \boldsymbol{x}' \}$
\ENDFOR
\STATE $\boldsymbol{x} \leftarrow \mathrm{select}(C)$
\IF {$E(\boldsymbol{x}) < E(\boldsymbol{x}^*)$}
\STATE $\boldsymbol{x}^* \leftarrow \boldsymbol{x}$
\ENDIF
\ENDFOR
\STATE \textbf{return} $\boldsymbol{x}^*$
\end{algorithmic}
\end{algorithm}

Given that we adopt the elastic model, our proposed SED heuristic aims to solve the following problem: given a fixed container radius, the objective is to find a feasible solution for the PESS problem. If a feasible solution is found, the algorithm returns it immediately. Otherwise, the algorithm returns an infeasible solution with the smallest energy (i.e., the minimum overlapping area) during the search process, which can be regarded as a decision problem for the PESS problem. 

To design an efficient local search heuristic, the key is to design an efficient solution exploration strategy. Starting from an initial low-quality solution, the algorithm is expected to find a high-quality solution by performing explorations and iterations as relatively few as possible. Given that a feasible solution is more potentially located in the neighborhood of the high-quality solution in the solution space than the low-quality one, it is worth making more exploration actions based on high-quality solutions to discover a feasible solution.  
According to this idea, we define a new metric function $J(\boldsymbol{x})$, formulated as follows:
\begin{equation} \label{eq-6}
    J(\boldsymbol{x}) = \lceil - c \log_{2} E(\boldsymbol{x}) \rceil,
\end{equation}
where function $J$ is the ceiling of the negative logarithm of the energy $E$, and $c$ is a coefficient that controls the value of function $J$. The function $J$ maps the energy $E$ to an integer which is applied to control the exploration number in the heuristic search process where a low-quality solution with larger energy is assigned to a small exploration number and a high-quality solution with smaller energy is assigned to a large exploration number. 
The coefficient is set to $c = 7$ as default. 

Now, we introduce our proposed SED heuristic depicted in Algorithm~\ref{alg_sed}. The call of SED ($n, R$) returns a feasible solution or an infeasible solution with the smallest energy if the feasible solution can not be found. Initially, SED generates a random layout as the initial solution $\boldsymbol{x}$ where the unit spheres are randomly packed into the container with possible overlaps (line 1). Then, SED employs the classic Limited-memory Broyden–Fletcher–Goldfarb–Shanno algorithm~\citep{Liu1989OnTL}, denoted as L-BFGS, as the basic optimizer to minimize the energy of the initial solution $\boldsymbol{x}$ (line 2), and it is recorded as the best solution $\boldsymbol{x}^*$ (line 3). Subsequently, SED performs several iterative search processes to improve the best solution $\boldsymbol{x}^*$. 

At each iterative search process, SED uses the current operated solution $\boldsymbol{x}$ to calculate the exploration number $m$ by Eq.~\ref{eq-6} (line 8). Then SED perturbs the current operated solution $\boldsymbol{x}$ to obtain a perturbed solution $\boldsymbol{x}'$ (line 11) and employs the L-BFGS algorithm to minimize the energy of the perturbed solution $\boldsymbol{x}'$ (line 12). SED repeats this perturbing operation for $m$ times until a perturbed solution set $C$, $|C| = m$, is obtained (lines 9-14). After that, SED adopts a ``select'' operator to choose a candidate solution from the perturbed solution set $C$ 
as the offspring solution $\boldsymbol{x}$ (line 15). 
The best solution $\boldsymbol{x}^*$ is updated if the energy of the operated solution $\boldsymbol{x}$ is smaller than that of the current best solution $\boldsymbol{x}^*$ (lines 16-18).
If a feasible solution is found, SED returns the solution immediately (lines 5-7), where the solution $\boldsymbol{x}^*$ is regarded as a feasible solution when $E(\boldsymbol{x}^*)$ is tiny enough ($E(\boldsymbol{x}^*) \leq 10^{-25}$ in this work). Otherwise, SED returns the best found solution $\boldsymbol{x}^*$ as the result when reaching the maximum number of iteration steps $S_{iter}$ ($S_{iter} = 700$ is set as default). 

To obtain a perturbed solution $\boldsymbol{x}'$ (line 11), we randomly shift the coordinate of the unit spheres in the operated solution $\boldsymbol{x}$ which can be described as follows, $x_i' \leftarrow x_i + r_x$, $y_i' \leftarrow y_i + r_y$ and $z_i' \leftarrow z_i + r_y$ ($1 \leq i \leq n$), where $r_x$ $r_y$ and $r_z$ are the random numbers, $r_x, r_y, r_z \in U(-\theta, \theta)$, $U$ stands for uniform distribution and $\theta = 0.8$ is set as default. 

The strategy of the ``select'' operator is described as follows:
\begin{align}
    \mathrm{select}(C) = 
    \begin{cases}
        \underset{\boldsymbol{x}' \in C}{\arg\min} \ E(\boldsymbol{x}'), &
        \mathrm{if} \ \ \underset{\boldsymbol{x}' \in C}{\min} \ E(\boldsymbol{x}') < E(\boldsymbol{x}) \\
        P\left( X = \boldsymbol{x}' \mid p_{\boldsymbol{x}'} = \mathrm{softmax}(J(\boldsymbol{x}'))  \right), &
        \mathrm{otherwise} 
    \end{cases} \nonumber
\end{align}
where the ``softmax'' function is described as follows:
\begin{equation}
    \mathrm{softmax}(J(\boldsymbol{x}')) = \frac{\exp(J(\boldsymbol{x}'))}{\sum_{\boldsymbol{y} \in C} \exp(J(\boldsymbol{y}))}.  \nonumber
\end{equation}
The ``select'' operator first compares the candidate solution $\boldsymbol{x}'$ with the smallest energy in the perturbed solution set $C$ with the current operated solution $\boldsymbol{x}$. 
The solution $\boldsymbol{x}$ is replaced by $\boldsymbol{x}'$ if solution $\boldsymbol{x}'$ has smaller energy (i.e., the overlaps) than solution $\boldsymbol{x}$, which means a better solution is found. 
Otherwise, the operator employs a ``softmax'' function to choose a candidate solution in $C$. Finally, the selected solution $\boldsymbol{x}$ becomes the operated solution in the next iteration.

\subsection{Container Adjustment Method} \label{ssec:04-04adju}

Assuming we obtain a solution by employing the SED heuristic, the solution may be feasible but probably contain overlaps, thus being infeasible. Now, we aim to solve another problem described as follows. (I) If the solution is feasible, the problem is adjusting the packing spheres' position and shrinking the container to obtain a better solution. (II) If the solution is infeasible, the problem is adjusting the packing spheres' position and expanding the container until the overlap is eliminated to obtain a feasible solution with the smallest container radius.

The most intuitive and popular method to deal with this problem is the binary search approach~\citep{huang2011quasi}. Inspired by the packing equal circles in a circle problem, we adopt a smart and significantly faster method~\citep{lai2022iterated} to solve the PESS problem, which is presented as follows. 

Let vector $\boldsymbol{z} = [x_1, y_1, z_1, x_2, y_2, z_2, ..., x_n, y_n, z_n, R]^{\mathrm{T}}$, $\boldsymbol{z} \in \mathbb{R}^{3n+1}$, be a candidate solution with the container radius $R$ being a variable. And a new elastic energy $U$ can be reformulated as follows:
\begin{equation}
    U_{\lambda}(\boldsymbol{z}) = U(\boldsymbol{z}, \lambda) = \sum_{i=1}^{n} \sum_{j=i+1}^{n} O_{ij} + \sum_{i=1}^{n} O_{i0} + \lambda R^2, \nonumber
\end{equation}
where $O_{ij}$ and $O_{i0}$ is defined in Eqs.~\eqref{eq-3} and \eqref{eq-4}, $\lambda R^2$ is a penalty term and $\lambda$ is a penalty coefficient. Starting from an empirical $\lambda$ value setting, we employ the L-BFGS algorithm to minimize the energy $U$ of a candidate solution, and the solution tries to shrink the container and contains overlaps because of the penalty term $\lambda R^2$. Then, we consecutively decrease the $\lambda$ value and  minimize the energy $U$ by employing the L-BFGS algorithm, the solution tries to expand the container and eliminates overlaps. When the penalty term $\lambda R^2$ is tiny enough, the energy $U$ degenerates to the energy $E$ (Eq.~\eqref{eq-5}) and the algorithm forces to minimize the energy $E$ (i.e., eliminate overlaps) without fixed radius constraint to obtain a feasible solution with a minimal container radius. 

The pseudocode of the container adjustment approach is presented in Algorithm~\ref{alg_adj}. Given a candidate solution $\boldsymbol{x}$ and a container radius $R$, the algorithm combines $(\boldsymbol{x}, R)$ to obtain a new solution $\boldsymbol{z}$ and initializes the coefficient $\lambda$ to an empirical value $10^{-4}$. Then, the algorithm performs several iterations to obtain a feasible solution with a minimal container radius. 
At each iteration, the algorithm employs the L-BFGS algorithm to minimize the energy $U_{\lambda}(\boldsymbol{z})$ and updates the solution $\boldsymbol{z}$. Then, the coefficient $\lambda$ is halved and the algorithm continually adjusts the solution $\boldsymbol{z}$ in the next iteration. After several iterations, the energy $U_{\lambda}(\boldsymbol{z})$ converges to 0 so that the overlaps are tiny enough in solution $\boldsymbol{z}$, which is split into $(\boldsymbol{x}^*, R^*)$ returned as the result. 

\begin{algorithm}[tb]
\caption{adjust\_container($\boldsymbol{x}, R$)} \label{alg_adj}
\textbf{Input}: A candidate solution $\boldsymbol{x}$; A container radius $R$ \\
\textbf{Output}: A feasible solution with a minimal container radius $(\boldsymbol{x}^*, R^*)$
\begin{algorithmic}[1] 
\STATE $\boldsymbol{z} \leftarrow \mathrm{combine}(\boldsymbol{x}, R), \ \ \lambda \leftarrow 10^{-4}$
\FOR {$i$ from $1$ to $35$} 
\STATE $\boldsymbol{z} \leftarrow \mathrm{optimizer}(U_{\lambda}, \boldsymbol{z})$
\STATE $\lambda \leftarrow 0.5 \times \lambda$
\ENDFOR
\STATE $(\boldsymbol{x}^*, R^*) \leftarrow \mathrm{split}(\boldsymbol{z})$
\STATE \textbf{return} $(\boldsymbol{x}^*, R^*)$
\end{algorithmic}
\end{algorithm}

Note that the coefficient $\lambda$ is less than $3 \times 10^{-15}$ after 35 iterations, where the influence of the penalty term $\lambda R^2$ is considered tiny enough in this work.

\subsection{Adaptive Neighbor Object Maintenance} \label{ssec:04-05anm}

\citet{he2018efficient} propose an efficient neighbor structure to solve the problem of packing equal circles in a circle. 
Their method can significantly reduce the time complexity of calculating the energy $E$ and its gradient (from $O(n^2)$ to $O(n)$) and accelerate the convergence of the continuous optimization process. In this work, we adopt it to solve the PESS problem and propose an adaptive method, called ANM, to maintain this structure. First, we introduce the efficient neighbor structure as follows.

Let $l_{ij}$ indicates the Euclidean distance between the centers of two unit spheres $s_i$ and $s_j$, given by the following equation:
\begin{equation}
    l_{ij} = \sqrt{(x_i - x_j)^2 + (y_i - y_j)^2 + (z_i - z_j)^2}.  \nonumber
\end{equation}
Recall that $O_{ij}$ denotes the overlapping distance between two unit spheres $s_i$ and $s_j$ defined by Eq.~\ref{eq-3}. It is clear that $O_{ij} > 0$ when $l_{ij} < 2$, and $O_{ij} = 0$ otherwise. 

Now, we define the neighbor $\Gamma(i)$ of unit sphere $s_i$ to be a subset of $n$ unit spheres $\{s_1, s_2, ..., s_n\}$ as follows:
\begin{equation}
    \Gamma(i) = \left\{ s_j \mid \forall j: 1 \leq j \leq n, \ i \neq j, \ l_{ij} < l_{cut} \right\}, \nonumber
\end{equation}
where $l_{cut}$ is a distance controlling hyperparameter. If $l_{cut}$ is set to 2, then all the unit spheres $\{s_j\}$ overlapping with unit sphere $s_i$ are contained in neighbor $\Gamma(i)$. Therefore, the energy concerning sphere $s_i$ can be calculated by enumerating the spheres $s_j$ in $\Gamma(i)$ instead of enumerating all the unit spheres, and the energy $E(\boldsymbol{x})$ (Eq.~\ref{eq-5}) can be reformulated as follows:
\begin{equation}
    E(\boldsymbol{x}) = \sum_{i=1}^{n} \sum_{s_{j} \in \Gamma(i)} O_{ij}^2 [i < j] + \sum_{i=1}^{n} O_{i0}^2, \nonumber
\end{equation}
where ``$[]$'' is the Iverson bracket that $[P] = 1$ if statement $P$ is true, otherwise $[P] = 0$. The statement ``$i < j$'' guarantees the overlaps of pairwise spheres are only calculated once. 

\begin{figure}[t]
    \centering
    \includegraphics[width=0.7\columnwidth]{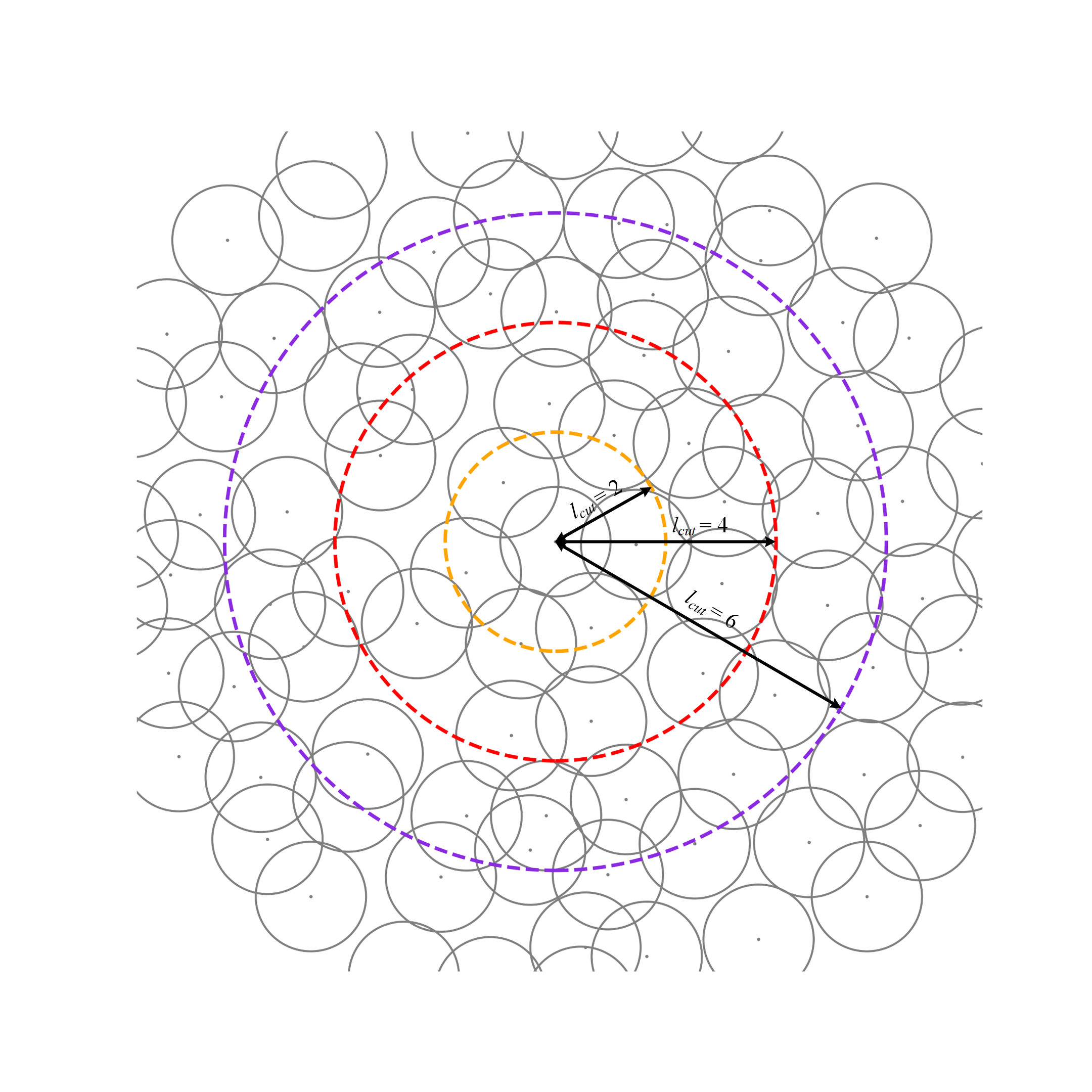}
    \caption{Illustration of the neighbors on the equal circle packing problem with different $l_{cut}$ settings. This illustrative example gives three settings for $l_{cut} = $ 2, 4 and 6 on a conflicting layout. We empirically set $l_{cut} = 4$ as a trade-off setting for the PESS problem in this work.}
    \label{fig_neighbor}
\end{figure}


We empirically set the default value of $l_{cut}$ to 4 in this work, as setting it too large can result in the neighbor set containing many unnecessary packing spheres, increasing the time cost of energy and gradient computation. Conversely, if $l_{cut}$ is set too small, the correctness of energy and gradient calculation cannot be guaranteed without maintaining the neighbor set when the packing spheres have minor shifts.



By adopting the neighbor structure, the time complexity of the energy and gradient calculation can be reduced to $O(n)$~\citep{he2018efficient} instead of enumerating pairwise spheres $O(n^2)$. Figure~\ref{fig_neighbor} gives an example to show the neighbor of a packing item on the equal circle packing problems with three different $l_{cut}$ settings. 

\begin{algorithm}[tb]
\caption{optimizer($f, \boldsymbol{x}$)} \label{alg_opti}
\textbf{Input}: An objective function $f$; A variable $\boldsymbol{x}$ \\
\textbf{Output}: A variable $\boldsymbol{x}^*$ with the value $f(\boldsymbol{x}^*)$ is a local minimum
\begin{algorithmic}[1] 
\STATE $cnt \leftarrow 0, \ \ len \leftarrow 1$
\STATE construct a current neighbor $\Gamma$
\FOR {$k$ from 0 to $MaxIter$} 
\STATE $\boldsymbol{d}_k \leftarrow \mathrm{two\_loop\_recursion}()$
\STATE $\alpha_k \leftarrow \underset{\boldsymbol{\alpha} \in \mathbb{R}}{\arg\min} \ f(\boldsymbol{x} + \alpha \boldsymbol{d}_k)$
\STATE $\boldsymbol{x} \leftarrow \boldsymbol{x} + \alpha_{k} \boldsymbol{d}_k$
\STATE $cnt \leftarrow cnt + 1$
\IF {$cnt \geq len$} 
\STATE construct a new neighbor $\Gamma'$
\IF {$\Gamma \neq \Gamma'$}
\STATE $cnt \leftarrow 0, \ \ len \leftarrow 1, \ \ \Gamma \leftarrow \Gamma'$
\ELSE
\STATE $cnt \leftarrow 0, \ \ len \leftarrow 2 \times len$
\ENDIF
\ENDIF
\IF {${\|g(\boldsymbol{x})\|_2} \leq 10^{-12}$}
\STATE \textbf{break}
\ENDIF
\ENDFOR
\STATE $\boldsymbol{x}^* \leftarrow \boldsymbol{x}$
\STATE \textbf{return} $\boldsymbol{x}^*$
\end{algorithmic}
\end{algorithm}

Now, we introduce our proposed ANM module, which is combined in the optimizer and presented in Algorithm~\ref{alg_opti}. The call of $\mathrm{optimizer}(f, \boldsymbol{x})$ returns a variable $\boldsymbol{x}^*$ with a local minimum of the objective function $f(\boldsymbol{x}^*)$. To accomplish the adaptive maintenance feature, ANM maintains two variables, the deferring counter $cnt$ and the deferring length $len$. 

At the beginning of the optimizer procedure, the counter $cnt$ and length $len$ are initialized to 0 and 1, respectively, and a neighbor set $\Gamma$ is constructed (lines 1-2). Then, the optimizer uses the neighbors to calculate the energy and gradient to iteratively update the variable $\boldsymbol{x}$ based on the L-BFGS algorithm (lines 3-19).

At each iteration, the optimizer employs the classic two-loop recursion approach, which is the core of L-BFGS, to obtain the descent direction $d_k$ (line 4). Then, a line search approach is adopted for obtaining a step length $\alpha_k$ (line 5) and the variable $\boldsymbol{x}$ is updated (line 6). Followed by the main part of the ANM module (lines 7-15), the counter $cnt$ is incremented by 1 and compared with the deferring length $len$. The neighbor maintenance process is triggered if the counter $cnt$ reaches the deferring length $len$. 

In the maintenance process, a new neighbor structure $\Gamma'$ is constructed and compared with the historical neighbor structure $\Gamma$. If two neighbor structures $\Gamma$ and $\Gamma'$ are different, the current neighbor structure is updated by $\Gamma'$, and the counter $cnt$ and the deferring length are reset to 0 and 1, respectively. Otherwise, the counter $cnt$ is reset to 0 and the deferring length $len$ is multiplied by 2. 

At the end of the iteration, the algorithm checks the norm of the gradient, and the iterative process is terminated if the norm is tiny enough (lines 16-18) or the maximum iteration step is reached. Finally, the objective function $f(\boldsymbol{x})$ can be regarded as reaching a local minimum, and the corresponding variable $\boldsymbol{x}$ is returned as the final result.

The mechanism of the ANM module can be described as follows. If the layout is unstable, the neighbors obtained at each iteration are different, which causes the counter $cnt$ and the deferring length $len$ to be constantly reset to 0 and 1, respectively, and the maintenance process is triggered when the ANM module is called. Otherwise, the length $len$ grows exponentially to defer the maintenance process. Note that ANM is a general adaptive method, which can be adapted to other problems like dynamic packing problems and online packing problems~\citep{ye2009note, ye2011online, brubach2015improved, hokama2016bounded, fekete2019online, lintzmayer2019online, epstein2019lower}. 

Some works also employ the neighbor structure to solve packing problems and they also propose some maintenance strategies. \citet{he2018efficient} use a simple method to maintain the neighbor, consisting in reconstructing the neighbor every 10 iterations. Reconstructing the neighbor is unnecessary when the layout is stable. In this case, this method will waste computational resources. \citet{lai2022iterated} propose a two-phase strategy to maintain the neighbor. In the first phase, it calculates the energy and gradient by enumerating all the pairwise circles without using the neighbor structure. In the second phase, it constructs a neighbor at the beginning and uses the neighbor to calculate the energy and gradient without updating until the end. This strategy has several disadvantages: 1) The enumeration method in the first phase is computationally expensive; 2) The changing condition from the first phase to the second phase requires particular expert experience and needs to be fine-tuned in different problems; 3) The neighbor  is not updated in the second phase so that if the solution falls into a saddle point, the correctness of this strategy can not be guaranteed. 
In contrast, our ANM method can effectively address these issues.

\section{Experiments and Discussions} \label{sec:05exp}
In this section, we first present our experimental setup. Then, we compare our algorithm with the state-of-the-art PSO-BA method~\citep{hifi2018hybrid}, evaluate our algorithm's performance on a large number of benchmark instances, and make a comparison with the best-known results from Packomania~\citep{Spechtweb} (downloaded date: 2023/03/01). Finally, we present the parameter study and the performance analysis of the ANM module. 

\subsection{Experimental Setup}
Our algorithm was implemented in the C++ language and compiled using g++ 5.4.0.
Experiments were performed on a server with Intel® Xeon® E5-2650 v3 CPU and 256 GBytes RAM, running on the Linux OS. 
Due to the randomness, we ran our algorithm 10 times independently with different random seeds (CPU timestamps) for each instance. 

We set the different cut-off times $T_{cut}$ (refer to Algorithm~\ref{alg_frame} line 3) for instances of different scales, as described below: $T_{cut}$ is set to 2 hours for the small scale instances ($n \leq 100$); $T_{cut}$ is set to 6 hours for the moderate I scale instances ($101 \leq n \leq 200$); and $T_{cut}$ is set to 12 hours for the moderate II scale instances ($201 \leq n \leq 400$). These cut-off time settings are comparable with the state-of-the-art PSO-BA method and other similar works in the literature (e.g., algorithms for circle packing problems). For example, the PSO-BA method set the cut-off time to 2 hours for the instances with $n \leq 50$; \citet{huang2011quasi} obtain the results for $n \leq 100$ and $101 \leq n \leq 200$ within the running time of 3,378 seconds (0.94 hours) and 50,982 seconds (14.16 hours), respectively, while solving the PESS problem; \citet{lai2022iterated} set 2, 8 and 12 hours for the instances with $n \leq 100$, $101 \leq n \leq 200$ and $201 \leq n \leq 320$, respectively, while solving the packing equal circles in a circle problem.

The default settings for the remaining parameters are described as follows. 
The maximum iteration step of the heuristic SED is set to $S_{iter} = 700$ (as used in Algorithm~\ref{alg_sed}); the controlling coefficient of function $J$ is set to $c = 7$ (as used in Section~\ref{ssec:04-03sed}); and the perturbing parameter of the uniform distribution $U$ is set to $\theta = 0.8$ (as used in Section~\ref{ssec:04-03sed}). The tuning analysis and parameter study of these parameters are presented in Section~\ref{ssec:05-03PS}.

\begin{table}[tb]
\centering
\caption{
Below are the computational results and a comparison of the best-known records $R^*$ from Packomania (accessed on 2023/03/01), the state-of-the-art PSO-BA method, and our algorithm SED for small-scale instances ($3 \leq n \leq 50$). 
The best result values are presented in bold among the compared results.}
\label{tb_cmp_PSO}

\scalebox{0.69}{
\begin{tabular}{lllllllrlllllrlrlr}
\toprule
         &                       &  & \multicolumn{5}{l}{PSO-BA}                                    &  & \multicolumn{5}{l}{SED (this work)}                                       &  &           &  &           \\ \cline{4-8} \cline{10-14}
$n$        & $R^*$                    &  & $R_{best}$           &  & $R_{avg}$            &  & $Time~(s)$ &  & $R_{best}$           &  & $R_{avg}$            &  & $Time~(s)$ &  & $\Delta_{best}$      &  & $\Delta_{avg}$      \\ \midrule
5        & \textbf{2.4142135624} &  & \textbf{2.4142135624} &  & \textbf{2.4142135624} &  & 473.91  &  & \textbf{2.4142135624} &  & \textbf{2.4142135624} &  & 0.00    &  & 0         &  & 0         \\
6        & \textbf{2.4142135624} &  & \textbf{2.4142135624} &  & \textbf{2.4142135624} &  & 66.88   &  & \textbf{2.4142135624} &  & \textbf{2.4142135624} &  & 0.00    &  & 0         &  & 0         \\
7        & \textbf{2.5912538723} &  & \textbf{2.5912538723} &  & \textbf{2.5912538723} &  & 827.55  &  & \textbf{2.5912538723} &  & \textbf{2.5912538723} &  & 0.00    &  & 0         &  & 0         \\
8        & \textbf{2.6453287760} &  & \textbf{2.6453287760} &  & \textbf{2.6453287760} &  & 310.62  &  & \textbf{2.6453287760} &  & \textbf{2.6453287760} &  & 1.00    &  & 0         &  & 0         \\
9        & \textbf{2.7320508076} &  & \textbf{2.7320508076} &  & \textbf{2.7320508076} &  & 206.78  &  & \textbf{2.7320508076} &  & \textbf{2.7320508076} &  & 1.00    &  & 0         &  & 0         \\
10       & \textbf{2.8324645611} &  & \textbf{2.8324645611} &  & \textbf{2.8324645611} &  & 1290.20 &  & \textbf{2.8324645611} &  & \textbf{2.8324645611} &  & 23.30   &  & 0         &  & 0         \\
11       & \textbf{2.9021130326} &  & \textbf{2.9021130326} &  & \textbf{2.9021130326} &  & 153.82  &  & \textbf{2.9021130326} &  & \textbf{2.9021130326} &  & 1.00    &  & 0         &  & 0         \\
12       & \textbf{2.9021130326} &  & \textbf{2.9021130326} &  & 2.9314798912          &  & 1179.29 &  & \textbf{2.9021130326} &  & \textbf{2.9021130326} &  & 2.00    &  & 0         &  & -2.94E-02 \\
13       & \textbf{3.0000000000} &  & \textbf{3.0000000000} &  & \textbf{3.0000000000} &  & 488.90  &  & \textbf{3.0000000000} &  & \textbf{3.0000000000} &  & 4.00    &  & 0         &  & 0         \\
14       & \textbf{3.0911454449} &  & \textbf{3.0911454449} &  & \textbf{3.0911454449} &  & 458.23  &  & \textbf{3.0911454449} &  & \textbf{3.0911454449} &  & 8.80    &  & 0         &  & 0         \\
15       & \textbf{3.1416426249} &  & \textbf{3.1416426249} &  & \textbf{3.1416426249} &  & 6282.91 &  & \textbf{3.1416426249} &  & \textbf{3.1416426249} &  & 10.00   &  & 0         &  & 0         \\
16       & \textbf{3.2156830320} &  & \textbf{3.2156830320} &  & 3.2157562243          &  & 1124.03 &  & \textbf{3.2156830320} &  & \textbf{3.2156830320} &  & 5.40    &  & 0         &  & -7.32E-05 \\
17       & \textbf{3.2712455117} &  & \textbf{3.2712455117} &  & \textbf{3.2712455117} &  & 92.19   &  & \textbf{3.2712455117} &  & \textbf{3.2712455117} &  & 7.00    &  & 0         &  & 0         \\
18       & \textbf{3.3189887817} &  & \textbf{3.3189887817} &  & \textbf{3.3189887817} &  & 707.66  &  & \textbf{3.3189887817} &  & \textbf{3.3189887817} &  & 5.00    &  & 0         &  & 0         \\
19       & \textbf{3.3860159733} &  & \textbf{3.3860159733} &  & 3.3860161307          &  & 6738.12 &  & \textbf{3.3860159733} &  & \textbf{3.3860159733} &  & 5.00    &  & 0         &  & -1.57E-07 \\
20       & \textbf{3.4735389622} &  & \textbf{3.4735389622} &  & 3.4748202071          &  & 7000.69 &  & \textbf{3.4735389622} &  & \textbf{3.4735389622} &  & 164.40  &  & 0         &  & -1.28E-03 \\
21       & \textbf{3.4863514104} &  & \textbf{3.4863514104} &  & \textbf{3.4863514104} &  & 2732.54 &  & \textbf{3.4863514104} &  & \textbf{3.4863514104} &  & 6.00    &  & 0         &  & 0         \\
22       & \textbf{3.5798331912} &  & \textbf{3.5798331912} &  & 3.5798511534          &  & 2500.96 &  & \textbf{3.5798331912} &  & \textbf{3.5798331912} &  & 640.40  &  & 0         &  & -1.80E-05 \\
23       & \textbf{3.6275164365} &  & \textbf{3.6275164365} &  & 3.6284425467          &  & 5236.08 &  & \textbf{3.6275164365} &  & \textbf{3.6275164365} &  & 29.10   &  & 0         &  & -9.26E-04 \\
24       & \textbf{3.6853949355} &  & \textbf{3.6853949355} &  & 3.6861597918          &  & 6538.98 &  & \textbf{3.6853949355} &  & \textbf{3.6853949355} &  & 205.50  &  & 0         &  & -7.65E-04 \\
25       & \textbf{3.6874267475} &  & \textbf{3.6874267475} &  & \textbf{3.6874267475} &  & 2661.64 &  & \textbf{3.6874267475} &  & \textbf{3.6874267475} &  & 185.80  &  & 0         &  & 0         \\
26       & \textbf{3.7474057765} &  & \textbf{3.7474057765} &  & 3.7474079315          &  & 1805.77 &  & \textbf{3.7474057765} &  & \textbf{3.7474057765} &  & 9.00    &  & 0         &  & -2.16E-06 \\
27       & \textbf{3.8134159569} &  & \textbf{3.8134159569} &  & 3.8154473167          &  & 4003.44 &  & \textbf{3.8134159569} &  & \textbf{3.8134159569} &  & 266.80  &  & 0         &  & -2.03E-03 \\
28       & \textbf{3.8416402781} &  & \textbf{3.8416402781} &  & 3.8427270479          &  & 6877.24 &  & \textbf{3.8416402781} &  & \textbf{3.8416402781} &  & 580.20  &  & 0         &  & -1.09E-03 \\
29       & \textbf{3.8770891032} &  & \textbf{3.8770891032} &  & 3.8773011833          &  & 6802.59 &  & \textbf{3.8770891032} &  & \textbf{3.8770891032} &  & 561.70  &  & 0         &  & -2.12E-04 \\
30       & \textbf{3.9164916616} &  & 3.9164916887          &  & 3.9182908180          &  & 6661.05 &  & \textbf{3.9164916616} &  & \textbf{3.9164916616} &  & 349.70  &  & -2.71E-08 &  & -1.80E-03 \\
31       & \textbf{3.9507544849} &  & \textbf{3.9507544849} &  & 3.9508604111          &  & 6794.01 &  & \textbf{3.9507544849} &  & \textbf{3.9507544849} &  & 1639.20 &  & 0         &  & -1.06E-04 \\
32       & \textbf{3.9874403893} &  & \textbf{3.9874403893} &  & 3.9885139480          &  & 6783.05 &  & \textbf{3.9874403893} &  & \textbf{3.9874403893} &  & 309.90  &  & 0         &  & -1.07E-03 \\
33       & \textbf{4.0199009160} &  & 4.0199009254          &  & 4.0200538386          &  & 6897.21 &  & \textbf{4.0199009160} &  & \textbf{4.0199009160} &  & 633.20  &  & -9.40E-09 &  & -1.53E-04 \\
34       & \textbf{4.0477199712} &  & \textbf{4.0477199712} &  & 4.0492757675          &  & 5846.87 &  & \textbf{4.0477199712} &  & \textbf{4.0477199712} &  & 436.90  &  & 0         &  & -1.56E-03 \\
35       & \textbf{4.0844057408} &  & 4.0844077658          &  & 4.0854592377          &  & 6852.70 &  & \textbf{4.0844057408} &  & \textbf{4.0844057408} &  & 2410.10 &  & -2.02E-06 &  & -1.05E-03 \\
36       & \textbf{4.1129893297} &  & \textbf{4.1129893297} &  & 4.1284894708          &  & 5000.74 &  & \textbf{4.1129893297} &  & \textbf{4.1129893297} &  & 322.30  &  & 0         &  & -1.55E-02 \\
37       & \textbf{4.1547812520} &  & 4.1548030417          &  & 4.1564599544          &  & 7045.52 &  & \textbf{4.1547812520} &  & \textbf{4.1547812520} &  & 4287.00 &  & -2.18E-05 &  & -1.68E-03 \\
38       & \textbf{4.1576692600} &  & 4.1576724213          &  & 4.1578290898          &  & 6661.83 &  & \textbf{4.1576692600} &  & \textbf{4.1576692600} &  & 7200.00 &  & -3.16E-06 &  & -1.60E-04 \\
39       & \textbf{4.2239497563} &  & \textbf{4.2239497563} &  & 4.2271089635          &  & 6833.07 &  & \textbf{4.2239497563} &  & \textbf{4.2239497563} &  & 631.70  &  & 0         &  & -3.16E-03 \\
40       & \textbf{4.2553329537} &  & 4.2553331430          &  & 4.2568592733          &  & 6588.66 &  & \textbf{4.2553329537} &  & \textbf{4.2553329537} &  & 569.40  &  & -1.89E-07 &  & -1.53E-03 \\
41       & 4.2963450048          &  & \textbf{4.2553329537} &  & 4.2844863118          &  & 7198.97 &  & 4.2963450048          &  & 4.2963450048          &  & 1197.90 &  & 4.10E-02  &  & 1.19E-02  \\
42       & 4.3081420430          &  & \textbf{4.2553329537} &  & 4.2893547713          &  & 5984.61 &  & 4.3081420430          &  & 4.3081420430          &  & 33.00   &  & 5.28E-02  &  & 1.88E-02  \\
43       & \textbf{4.3528798324} &  & 4.3530264887          &  & 4.3576909767          &  & 5153.70 &  & \textbf{4.3528798324} &  & \textbf{4.3528798324} &  & 2055.90 &  & -1.47E-04 &  & -4.81E-03 \\
44       & \textbf{4.3828308379} &  & 4.3828454090          &  & 4.3845814567          &  & 6339.39 &  & \textbf{4.3828308379} &  & \textbf{4.3828308379} &  & 6145.30 &  & -1.46E-05 &  & -1.75E-03 \\
45       & \textbf{4.4070031477} &  & 4.4070031605          &  & 4.4118673635          &  & 6762.92 &  & \textbf{4.4070031477} &  & \textbf{4.4070031477} &  & 1397.60 &  & -1.28E-08 &  & -4.86E-03 \\
46       & \textbf{4.4411244747} &  & 4.4417854293          &  & 4.4449129285          &  & 6146.29 &  & \textbf{4.4411244747} &  & \textbf{4.4411244747} &  & 1232.30 &  & -6.61E-04 &  & -3.79E-03 \\
47       & \textbf{4.4741318035} &  & 4.4744654770          &  & 4.4769235843          &  & 5523.97 &  & \textbf{4.4741318035} &  & \textbf{4.4741318035} &  & 1531.70 &  & -3.34E-04 &  & -2.79E-03 \\
48       & \textbf{4.4962827447} &  & 4.4963283111          &  & 4.4975725203          &  & 5194.01 &  & \textbf{4.4962827447} &  & \textbf{4.4962827447} &  & 6323.20 &  & -4.56E-05 &  & -1.29E-03 \\
49       & \textbf{4.5191984746} &  & 4.5192891866          &  & 4.5217695981          &  & 7102.21 &  & \textbf{4.5191984746} &  & \textbf{4.5191984746} &  & 2341.90 &  & -9.07E-05 &  & -2.57E-03 \\
50       & \textbf{4.5504543407} &  & 4.5509544053          &  & 4.5517945731          &  & 6346.64 &  & \textbf{4.5504543407} &  & \textbf{4.5504543407} &  & 974.70  &  & -5.00E-04 &  & -1.34E-03 \\ \midrule
\#Better &                       &  &                       &  &                       &  &         &  & 14                    &  & 30                    &  &         &  &           &  &           \\
\#Equal  &                       &  &                       &  &                       &  &         &  & 30                    &  & 14                    &  &         &  &           &  &           \\
\#Worse  &                       &  &                       &  &                       &  &         &  & ~~2                     &  & ~~2                     &  &         &  &           &  &           \\ \bottomrule
\end{tabular}
}
\end{table}

\subsection{Comparison of State-of-the-Art Method}

The optimal solutions are trivial for $1 \leq n \leq 4$ and the optimality can be easily proven by mathematical analysis where the optimal radii are $R_1^* = 1$, $R_2^* = 2$, $R_3^* = 2 / \sqrt{3} + 1 \approx 2.1547$ and $R_4^* = \sqrt{6} / 2 + 1 \approx 2.2247$. Therefore, we performed our algorithm beginning from $n = 5$. The comparison results of the best-known records reported on Packomania, the state-of-the-art PSO-BA method~\citep{hifi2018hybrid} and our algorithm SED are shown in Table~\ref{tb_cmp_PSO}.

In Table~\ref{tb_cmp_PSO}, the column of $n$ corresponds to the number of packing items in the instance, followed by the best-known records $R^*$ reported at Packomania. 
PSO-BA corresponds to the results reported in the work of~\citep{hifi2018hybrid}: $R_{best}$ and $R_{avg}$ indicate the best and average results obtained by PSO-BA over 10 independent runs, and $Time~(s)$ indicates the corresponding runtime of $R_{best}$. 
SED corresponds to the result of our algorithm: $R_{best}$ and $R_{avg}$ indicate the best results obtained by our algorithm over 10 independent runs, and $Time~(s)$ indicates the corresponding runtime of $R_{best}$. 
The last two columns of $\Delta_{best}$ and $\Delta_{avg}$ correspond to the difference of the best and average result of SED and PSO-BA (i.e., $\Delta_{best} = R_{best}^{\mathrm{SED}} - R_{best}^{\mathrm{PSO-BA}}$ and $\Delta_{avg} = R_{avg}^{\mathrm{SED}} - R_{avg}^{\mathrm{PSO-BA}}$), the negative value indicates that SED yields a better result than PSO-BA. 
At the bottom of the table, the rows of ``\#Better'', ``\#Equal'' and ``\#Worse'' show the number of instances for which SED obtains the better, equal and worse best and average results compared to PSO-BA.

Note that PSO-BA was performed on an Intel Core 2 Duo (2.53 GHz and with 4Gb of RAM) environment. Since PSO-BA and SED are performed on different computing platforms, the runtime information is only provided for indicative purposes.

From Table~\ref{tb_cmp_PSO}, we have the following observations:
\begin{itemize}
    \item SED yields 14 better, 30 equal and 2 worse results in terms of $R_{best}$ than PSO-BA with $5 \leq n \leq 50$. And the 46 best results $R_{best}$ obtained by SED reach the best-known records at Packomania with much shorter computation time. It demonstrates that SED has a strong solving ability on the small-scale instances of PESS and significantly outperforms the state-of-the-art PSO-BA method. 
    \item SED yields 30 better, 14 equal and 2 worse results in terms of $R_{avg}$ than PSO-BA with $5 \leq n \leq 50$. And the 46 average results obtained by SED also reached the best-known records at Packomania, which means SED has a 100\% success rate to obtain the best result on the small-scale instances. It also demonstrates that SED significantly outperforms PSO-BA, and SED is an efficient and powerful heuristic algorithm for solving the PESS problem that can stably obtain most best solutions on the small-scale PESS instances.
\end{itemize}

\subsection{Comparison of Best-Known Records} 

We further evaluate our proposed SED algorithm's performance on the small and moderate scale instances with $5 \leq n \leq 400$ by making a comparison with the best-known records $R^*$ reported at Packomania~\citep{Spechtweb}. The computational results and comparison of the SED algorithm are summarized in Tables~\ref{tb_cmp_best1}-\ref{tb_cmp_best4}, respectively for $5 \leq n \leq 100$, $101 \leq n \leq 200$, $201 \leq n \leq 300$ and $301 \leq n \leq 400$, for which we provide more details to analyze the SED algorithm's performance. 

In each table, the column labeled $n$ corresponds to the number of items being packed in each instance, followed by the column labeled $R^*$, which reports the best-known record for that instance as reported on Packomania. 
SED corresponds to the results of our algorithm: $R_{best}$ and $R_{avg}$ show the best and average results over the 10 independent runs, $\Delta_{best}$ and $\Delta_{avg}$ show the difference of the best and average result of SED and the best-known record $R^*$ (i.e., $\Delta_{best} = R_{best} - R^*$ and $\Delta_{avg} = R_{avg} - R^*$) for which the negative value indicates SED yields an improved result compared with the best-known result, $HR$ shows the ratio of hitting the best result $R_{best}$ and $RR$ shows the ratio of obtaining 
results not inferior to the best-known result $R^*$, and $Time~(s)$ shows the average runtime of obtaining the best result. The last three rows of ``\#Improved'', ``\#Equal'' and ``\#Worse'' show the number of instances for which SED obtained an improved, equal and worse result compared to the best-known result $R^*$.

\textbf{On small scale instances.}  
We observe from Table~\ref{tb_cmp_best1} that SED obtains 4 improved and 71 equal results compared to $R^*$ for $5 \leq n \leq 79$ with 100\% success ratio, and SED obtains 15 improved, 5 equal and 1 worse results compared to $R^*$ for $80 \leq n \leq 100$ where there are 14 results with $RR$ values over 5/10 among the 21 instances, which means over half of the runs SED can obtain an improved or equal result. In summary, SED has 19 improved, 76 equal and 1 worse results on the 96 small-scale instances, demonstrating that SED has excellent performance for solving the small-scale PESS problem.

\clearpage
\begin{sidewaystable}[!h]
\centering
\caption{Computational results and comparison of our SED algorithm with the best-known results recorded at Packomania on the small scale instances ($5 \leq n \leq 100$). The improved and equal results appear in bold and underlined, respectively, compared with the best-known results $R^*$ in terms of $R_{best}$ and $R_{avg}$. The symbol ``-'' means the data is unavailable.}
\label{tb_cmp_best1}

\resizebox{1.0\textwidth}{!}{
\begin{tabular}{lllllllrlrlllllrlllllllllrlrlllllr}
\toprule
           &              &  & \multicolumn{13}{l}{SED (this work)}                                                                                &  &  &     &              &  & \multicolumn{13}{l}{SED (this work)}                                                                                                \\ \cline{4-16} \cline{22-34} 
$n$          & $R^*$           &  & $R_{best}$        &           & $R_{avg}$         &           & $\Delta_{best}$ &  & $\Delta_{avg}$ &  & $HR$    &  & $RR$    &  & $Time~(s)$ &  &  & $n$   & $R^*$           &  & $R_{best}$           &           & $R_{avg}$            &           &  $\Delta_{best}$     &  & $\Delta_{avg}$      &  & $HR$    &  & $RR$    &  & $Time~(s)$ \\ \midrule
1          & 1.0000000000 &  & -                  &           & -                  &           & -    &  & -    &  & -     &  & -     &  & -       &  &  & 51  & 4.5756057950 &  & {\ul 4.5756057950}    & \textbf{} & {\ul 4.5756057950}    & \textbf{} & 0         &  & 0         &  & 10/10 &  & 10/10 &  & 1518.60 \\
2          & 2.0000000000 &  & -                  &           & -                  &           & -    &  & -    &  & -     &  & -     &  & -       &  &  & 52  & 4.6097732930 &  & {\ul 4.6097732930}    & \textbf{} & {\ul 4.6097732930}    & \textbf{} & 0         &  & 0         &  & 10/10 &  & 10/10 &  & 6043.10 \\
3          & 2.1547005384 &  & -                  &           & -                  &           & -    &  & -    &  & -     &  & -     &  & -       &  &  & 53  & 4.6234637833 &  & {\ul 4.6234637833}    & \textbf{} & {\ul 4.6234637833}    & \textbf{} & 0         &  & 0         &  & 10/10 &  & 10/10 &  & 1291.10 \\
4          & 2.2247448714 &  & -                  &           & -                  &           & -    &  & -    &  & -     &  & -     &  & -       &  &  & 54  & 4.6528795754 &  & {\ul 4.6528795754}    & \textbf{} & {\ul 4.6528795754}    & \textbf{} & 0         &  & 0         &  & 10/10 &  & 10/10 &  & 2799.80 \\
5          & 2.4142135624 &  & {\ul 2.4142135624} & \textbf{} & {\ul 2.4142135624} & \textbf{} & 0    &  & 0    &  & 10/10 &  & 10/10 &  & 0.00    &  &  & 55  & 4.6851203515 &  & {\ul 4.6851203515}    & \textbf{} & {\ul 4.6851203515}    & \textbf{} & 0         &  & 0         &  & 10/10 &  & 10/10 &  & 2504.70 \\
6          & 2.4142135624 &  & {\ul 2.4142135624} & \textbf{} & {\ul 2.4142135624} & \textbf{} & 0    &  & 0    &  & 10/10 &  & 10/10 &  & 0.00    &  &  & 56  & 4.6933402171 &  & {\ul 4.6933402171}    & \textbf{} & {\ul 4.6933402171}    & \textbf{} & 0         &  & 0         &  & 10/10 &  & 10/10 &  & 7202.40 \\
7          & 2.5912538723 &  & {\ul 2.5912538723} & \textbf{} & {\ul 2.5912538723} & \textbf{} & 0    &  & 0    &  & 10/10 &  & 10/10 &  & 0.00    &  &  & 57  & 4.7322376049 &  & \textbf{4.7319976099} & \textbf{} & \textbf{4.7319976099} & \textbf{} & -2.40E-04 &  & -2.40E-04 &  & 10/10 &  & 10/10 &  & 1996.00 \\
8          & 2.6453287760 &  & {\ul 2.6453287760} & \textbf{} & {\ul 2.6453287760} & \textbf{} & 0    &  & 0    &  & 10/10 &  & 10/10 &  & 1.00    &  &  & 58  & 4.7510429431 &  & {\ul 4.7510429431}    & \textbf{} & {\ul 4.7510429431}    & \textbf{} & 0         &  & 0         &  & 10/10 &  & 10/10 &  & 1779.10 \\
9          & 2.7320508076 &  & {\ul 2.7320508076} & \textbf{} & {\ul 2.7320508076} & \textbf{} & 0    &  & 0    &  & 10/10 &  & 10/10 &  & 1.00    &  &  & 59  & 4.7673656459 &  & {\ul 4.7673656459}    & \textbf{} & {\ul 4.7673656459}    & \textbf{} & 0         &  & 0         &  & 10/10 &  & 10/10 &  & 7202.00 \\
10         & 2.8324645611 &  & {\ul 2.8324645611} & \textbf{} & {\ul 2.8324645611} & \textbf{} & 0    &  & 0    &  & 10/10 &  & 10/10 &  & 23.30   &  &  & 60  & 4.7749335903 &  & {\ul 4.7749335903}    & \textbf{} & {\ul 4.7749335903}    & \textbf{} & 0         &  & 0         &  & 10/10 &  & 10/10 &  & 922.90  \\
11         & 2.9021130326 &  & {\ul 2.9021130326} & \textbf{} & {\ul 2.9021130326} & \textbf{} & 0    &  & 0    &  & 10/10 &  & 10/10 &  & 1.00    &  &  & 61  & 4.7822027257 &  & {\ul 4.7822027257}    & \textbf{} & {\ul 4.7822027257}    & \textbf{} & 0         &  & 0         &  & 10/10 &  & 10/10 &  & 1144.80 \\
12         & 2.9021130326 &  & {\ul 2.9021130326} & \textbf{} & {\ul 2.9021130326} & \textbf{} & 0    &  & 0    &  & 10/10 &  & 10/10 &  & 2.00    &  &  & 62  & 4.8370131185 &  & {\ul 4.8370131185}    & \textbf{} & {\ul 4.8370131185}    & \textbf{} & 0         &  & 0         &  & 10/10 &  & 10/10 &  & 1414.60 \\
13         & 3.0000000000 &  & {\ul 3.0000000000} & \textbf{} & {\ul 3.0000000000} & \textbf{} & 0    &  & 0    &  & 10/10 &  & 10/10 &  & 4.00    &  &  & 63  & 4.8554314461 &  & {\ul 4.8554314461}    & \textbf{} & {\ul 4.8554314461}    & \textbf{} & 0         &  & 0         &  & 10/10 &  & 10/10 &  & 1432.70 \\
14         & 3.0911454449 &  & {\ul 3.0911454449} & \textbf{} & {\ul 3.0911454449} & \textbf{} & 0    &  & 0    &  & 10/10 &  & 10/10 &  & 8.80    &  &  & 64  & 4.8993857428 &  & {\ul 4.8993857428}    & \textbf{} & {\ul 4.8993857428}    & \textbf{} & 0         &  & 0         &  & 10/10 &  & 10/10 &  & 1685.60 \\
15         & 3.1416426249 &  & {\ul 3.1416426249} & \textbf{} & {\ul 3.1416426249} & \textbf{} & 0    &  & 0    &  & 10/10 &  & 10/10 &  & 10.00   &  &  & 65  & 4.9242635534 &  & {\ul 4.9242635534}    & \textbf{} & {\ul 4.9242635534}    & \textbf{} & 0         &  & 0         &  & 10/10 &  & 10/10 &  & 7201.80 \\
16         & 3.2156830320 &  & {\ul 3.2156830320} & \textbf{} & {\ul 3.2156830320} & \textbf{} & 0    &  & 0    &  & 10/10 &  & 10/10 &  & 5.40    &  &  & 66  & 4.9478309581 &  & {\ul 4.9478309581}    & \textbf{} & {\ul 4.9478309581}    & \textbf{} & 0         &  & 0         &  & 10/10 &  & 10/10 &  & 2818.70 \\
17         & 3.2712455117 &  & {\ul 3.2712455117} & \textbf{} & {\ul 3.2712455117} & \textbf{} & 0    &  & 0    &  & 10/10 &  & 10/10 &  & 7.00    &  &  & 67  & 4.9697680835 &  & {\ul 4.9697680835}    & \textbf{} & {\ul 4.9697680835}    & \textbf{} & 0         &  & 0         &  & 10/10 &  & 10/10 &  & 4560.10 \\
18         & 3.3189887817 &  & {\ul 3.3189887817} & \textbf{} & {\ul 3.3189887817} & \textbf{} & 0    &  & 0    &  & 10/10 &  & 10/10 &  & 5.00    &  &  & 68  & 4.9999442473 &  & {\ul 4.9999442473}    & \textbf{} & {\ul 4.9999442473}    & \textbf{} & 0         &  & 0         &  & 10/10 &  & 10/10 &  & 6503.30 \\
19         & 3.3860159733 &  & {\ul 3.3860159733} & \textbf{} & {\ul 3.3860159733} & \textbf{} & 0    &  & 0    &  & 10/10 &  & 10/10 &  & 5.00    &  &  & 69  & 5.0180264442 &  & {\ul 5.0180264442}    & \textbf{} & {\ul 5.0180264442}    & \textbf{} & 0         &  & 0         &  & 10/10 &  & 10/10 &  & 2559.50 \\
20         & 3.4735389622 &  & {\ul 3.4735389622} & \textbf{} & {\ul 3.4735389622} & \textbf{} & 0    &  & 0    &  & 10/10 &  & 10/10 &  & 164.40  &  &  & 70  & 5.0329981188 &  & {\ul 5.0329981188}    & \textbf{} & {\ul 5.0329981188}    & \textbf{} & 0         &  & 0         &  & 10/10 &  & 10/10 &  & 6742.30 \\
21         & 3.4863514104 &  & {\ul 3.4863514104} & \textbf{} & {\ul 3.4863514104} & \textbf{} & 0    &  & 0    &  & 10/10 &  & 10/10 &  & 6.00    &  &  & 71  & 5.0704246177 &  & {\ul 5.0704246177}    & \textbf{} & {\ul 5.0704246177}    & \textbf{} & 0         &  & 0         &  & 10/10 &  & 10/10 &  & 4901.90 \\
22         & 3.5798331912 &  & {\ul 3.5798331912} & \textbf{} & {\ul 3.5798331912} & \textbf{} & 0    &  & 0    &  & 10/10 &  & 10/10 &  & 640.40  &  &  & 72  & 5.0946090432 &  & {\ul 5.0946090432}    & \textbf{} & {\ul 5.0946090432}    & \textbf{} & 0         &  & 0         &  & 10/10 &  & 10/10 &  & 7202.30 \\
23         & 3.6275164365 &  & {\ul 3.6275164365} & \textbf{} & {\ul 3.6275164365} & \textbf{} & 0    &  & 0    &  & 10/10 &  & 10/10 &  & 29.10   &  &  & 73  & 5.1117410063 &  & \textbf{5.1117310540} & \textbf{} & \textbf{5.1117310540} & \textbf{} & -9.95E-06 &  & -9.95E-06 &  & 10/10 &  & 10/10 &  & 7204.00 \\
24         & 3.6853949355 &  & {\ul 3.6853949355} & \textbf{} & {\ul 3.6853949355} & \textbf{} & 0    &  & 0    &  & 10/10 &  & 10/10 &  & 205.50  &  &  & 74  & 5.1243475815 &  & {\ul 5.1243475815}    & \textbf{} & {\ul 5.1243475815}    & \textbf{} & 0         &  & 0         &  & 10/10 &  & 10/10 &  & 186.90  \\
25         & 3.6874267475 &  & {\ul 3.6874267475} & \textbf{} & {\ul 3.6874267475} & \textbf{} & 0    &  & 0    &  & 10/10 &  & 10/10 &  & 185.80  &  &  & 75  & 5.1555845175 &  & {\ul 5.1555845175}    & \textbf{} & {\ul 5.1555845175}    & \textbf{} & 0         &  & 0         &  & 10/10 &  & 10/10 &  & 165.40  \\
26         & 3.7474057765 &  & {\ul 3.7474057765} & \textbf{} & {\ul 3.7474057765} & \textbf{} & 0    &  & 0    &  & 10/10 &  & 10/10 &  & 9.00    &  &  & 76  & 5.1827544049 &  & \textbf{5.1827522689} & \textbf{} & \textbf{5.1827522689} & \textbf{} & -2.14E-06 &  & -2.14E-06 &  & 10/10 &  & 10/10 &  & 7203.50 \\
27         & 3.8134159569 &  & {\ul 3.8134159569} & \textbf{} & {\ul 3.8134159569} & \textbf{} & 0    &  & 0    &  & 10/10 &  & 10/10 &  & 266.80  &  &  & 77  & 5.2014502549 &  & \textbf{5.2014494794} & \textbf{} & \textbf{5.2014494794} & \textbf{} & -7.76E-07 &  & -7.76E-07 &  & 10/10 &  & 10/10 &  & 7203.00 \\
28         & 3.8416402781 &  & {\ul 3.8416402781} & \textbf{} & {\ul 3.8416402781} & \textbf{} & 0    &  & 0    &  & 10/10 &  & 10/10 &  & 580.20  &  &  & 78  & 5.2230523740 &  & {\ul 5.2230523740}    & \textbf{} & {\ul 5.2230523740}    & \textbf{} & 0         &  & 0         &  & 10/10 &  & 10/10 &  & 5712.10 \\
29         & 3.8770891032 &  & {\ul 3.8770891032} & \textbf{} & {\ul 3.8770891032} & \textbf{} & 0    &  & 0    &  & 10/10 &  & 10/10 &  & 561.70  &  &  & 79  & 5.2449457291 &  & {\ul 5.2449457291}    & \textbf{} & {\ul 5.2449457291}    & \textbf{} & 0         &  & 0         &  & 10/10 &  & 10/10 &  & 2759.30 \\
30         & 3.9164916616 &  & {\ul 3.9164916616} & \textbf{} & {\ul 3.9164916616} & \textbf{} & 0    &  & 0    &  & 10/10 &  & 10/10 &  & 349.70  &  &  & 80  & 5.2707282625 &  & \textbf{5.2707228597} & \textbf{} & \textbf{5.2707230129} & \textbf{} & -5.40E-06 &  & -5.25E-06 &  & 6/10  &  & 10/10 &  & 5329.10 \\
31         & 3.9507544849 &  & {\ul 3.9507544849} & \textbf{} & {\ul 3.9507544849} & \textbf{} & 0    &  & 0    &  & 10/10 &  & 10/10 &  & 1639.20 &  &  & 81  & 5.2918921098 &  & \textbf{5.2918269759} & \textbf{} & \textbf{5.2918490181} & \textbf{} & -6.51E-05 &  & -4.31E-05 &  & 1/10  &  & 9/10  &  & 5812.40 \\
32         & 3.9874403893 &  & {\ul 3.9874403893} & \textbf{} & {\ul 3.9874403893} & \textbf{} & 0    &  & 0    &  & 10/10 &  & 10/10 &  & 309.90  &  &  & 82  & 5.3108551826 &  & \textbf{5.3107500414} & \textbf{} & \textbf{5.3107710697} & \textbf{} & -1.05E-04 &  & -8.41E-05 &  & 8/10  &  & 10/10 &  & 4509.00 \\
33         & 4.0199009160 &  & {\ul 4.0199009160} & \textbf{} & {\ul 4.0199009160} & \textbf{} & 0    &  & 0    &  & 10/10 &  & 10/10 &  & 633.20  &  &  & 83  & 5.3293460126 &  & {\ul 5.3293460126}    & \textbf{} & 5.3295343781          & \textbf{} & 0         &  & 1.88E-04  &  & 1/10  &  & 1/10  &  & 4292.80 \\
34         & 4.0477199712 &  & {\ul 4.0477199712} & \textbf{} & {\ul 4.0477199712} & \textbf{} & 0    &  & 0    &  & 10/10 &  & 10/10 &  & 436.90  &  &  & 84  & 5.3488647209 &  & \textbf{5.3488512301} & \textbf{} & \textbf{5.3488635868} & \textbf{} & -1.35E-05 &  & -1.13E-06 &  & 4/10  &  & 5/10  &  & 5430.70 \\
35         & 4.0844057408 &  & {\ul 4.0844057408} & \textbf{} & {\ul 4.0844057408} & \textbf{} & 0    &  & 0    &  & 10/10 &  & 10/10 &  & 2410.10 &  &  & 85  & 5.3644833789 &  & {\ul 5.3644833789}    & \textbf{} & 5.3670410532          & \textbf{} & 0         &  & 2.56E-03  &  & 3/10  &  & 3/10  &  & 5159.50 \\
36         & 4.1129893297 &  & {\ul 4.1129893297} & \textbf{} & {\ul 4.1129893297} & \textbf{} & 0    &  & 0    &  & 10/10 &  & 10/10 &  & 322.30  &  &  & 86  & 5.3876129492 &  & {\ul 5.3876129492}    & \textbf{} & 5.3876242092          & \textbf{} & 0         &  & 1.13E-05  &  & 7/10  &  & 7/10  &  & 7003.80 \\
37         & 4.1547812520 &  & {\ul 4.1547812520} & \textbf{} & {\ul 4.1547812520} & \textbf{} & 0    &  & 0    &  & 10/10 &  & 10/10 &  & 4287.00 &  &  & 87  & 5.4073256708 &  & \textbf{5.4067811877} & \textbf{} & \textbf{5.4067811877} & \textbf{} & -5.44E-04 &  & -5.44E-04 &  & 10/10 &  & 10/10 &  & 4652.30 \\
38         & 4.1576692600 &  & {\ul 4.1576692600} & \textbf{} & {\ul 4.1576692600} & \textbf{} & 0    &  & 0    &  & 10/10 &  & 10/10 &  & 7200.00 &  &  & 88  & 5.4266705368 &  & \textbf{5.4266460069} & \textbf{} & 5.4267151869          & \textbf{} & -2.45E-05 &  & 4.47E-05  &  & 2/10  &  & 5/10  &  & 4285.80 \\
39         & 4.2239497563 &  & {\ul 4.2239497563} & \textbf{} & {\ul 4.2239497563} & \textbf{} & 0    &  & 0    &  & 10/10 &  & 10/10 &  & 631.70  &  &  & 89  & 5.4466807175 &  & {\ul 5.4466807175}    & \textbf{} & {\ul 5.4466807175}    & \textbf{} & 0         &  & 0         &  & 10/10 &  & 10/10 &  & 3603.90 \\
40         & 4.2553329537 &  & {\ul 4.2553329537} & \textbf{} & {\ul 4.2553329537} & \textbf{} & 0    &  & 0    &  & 10/10 &  & 10/10 &  & 569.40  &  &  & 90  & 5.4663957267 &  & \textbf{5.4663528574} & \textbf{} & 5.4666204162          & \textbf{} & -4.29E-05 &  & 2.25E-04  &  & 2/10  &  & 2/10  &  & 4929.10 \\
41         & 4.2963450048 &  & {\ul 4.2963450048} & \textbf{} & {\ul 4.2963450048} & \textbf{} & 0    &  & 0    &  & 10/10 &  & 10/10 &  & 1197.90 &  &  & 91  & 5.4840116687 &  & \textbf{5.4838608716} & \textbf{} & 5.4840178565          & \textbf{} & -1.51E-04 &  & 6.19E-06  &  & 2/10  &  & 4/10  &  & 3718.90 \\
42         & 4.3081420430 &  & {\ul 4.3081420430} & \textbf{} & {\ul 4.3081420430} & \textbf{} & 0    &  & 0    &  & 10/10 &  & 10/10 &  & 33.00   &  &  & 92  & 5.5010819544 &  & 5.5012326826          & \textbf{} & 5.5016577070          & \textbf{} & 1.51E-04  &  & 5.76E-04  &  & 1/10  &  & 0/10  &  & 5714.70 \\
43         & 4.3528798324 &  & {\ul 4.3528798324} & \textbf{} & {\ul 4.3528798324} & \textbf{} & 0    &  & 0    &  & 10/10 &  & 10/10 &  & 2055.90 &  &  & 93  & 5.5167123871 &  & \textbf{5.5161208996} & \textbf{} & \textbf{5.5166205614} & \textbf{} & -5.91E-04 &  & -9.18E-05 &  & 1/10  &  & 8/10  &  & 3782.20 \\
44         & 4.3828308379 &  & {\ul 4.3828308379} & \textbf{} & {\ul 4.3828308379} & \textbf{} & 0    &  & 0    &  & 10/10 &  & 10/10 &  & 6145.30 &  &  & 94  & 5.5307641333 &  & \textbf{5.5304774705} & \textbf{} & 5.5313809950          & \textbf{} & -2.87E-04 &  & 6.17E-04  &  & 2/10  &  & 2/10  &  & 5457.00 \\
45         & 4.4070031477 &  & {\ul 4.4070031477} & \textbf{} & {\ul 4.4070031477} & \textbf{} & 0    &  & 0    &  & 10/10 &  & 10/10 &  & 1397.60 &  &  & 95  & 5.5516834671 &  & \textbf{5.5391547512} & \textbf{} & \textbf{5.5508888816} & \textbf{} & -1.25E-02 &  & -7.95E-04 &  & 1/10  &  & 8/10  &  & 5089.50 \\
46         & 4.4411244747 &  & {\ul 4.4411244747} & \textbf{} & {\ul 4.4411244747} & \textbf{} & 0    &  & 0    &  & 10/10 &  & 10/10 &  & 1232.30 &  &  & 96  & 5.5689451565 &  & \textbf{5.5689435455} & \textbf{} & \textbf{5.5689443510} & \textbf{} & -1.61E-06 &  & -8.06E-07 &  & 5/10  &  & 10/10 &  & 3326.90 \\
47         & 4.4741318035 &  & {\ul 4.4741318035} & \textbf{} & {\ul 4.4741318035} & \textbf{} & 0    &  & 0    &  & 10/10 &  & 10/10 &  & 1531.70 &  &  & 97  & 5.5859794729 &  & {\ul 5.5859794729}    & \textbf{} & 5.5861566571          & \textbf{} & 0         &  & 1.77E-04  &  & 8/10  &  & 8/10  &  & 5522.60 \\
48         & 4.4962827447 &  & {\ul 4.4962827447} & \textbf{} & {\ul 4.4962827447} & \textbf{} & 0    &  & 0    &  & 10/10 &  & 10/10 &  & 6323.20 &  &  & 98  & 5.6022822809 &  & \textbf{5.6022558001} & \textbf{} & 5.6029733282          & \textbf{} & -2.65E-05 &  & 6.91E-04  &  & 3/10  &  & 9/10  &  & 5686.90 \\
49         & 4.5191984746 &  & {\ul 4.5191984746} & \textbf{} & {\ul 4.5191984746} & \textbf{} & 0    &  & 0    &  & 10/10 &  & 10/10 &  & 2341.90 &  &  & 99  & 5.6218126153 &  & \textbf{5.6217725678} & \textbf{} & 5.6223277124          & \textbf{} & -4.00E-05 &  & 5.15E-04  &  & 2/10  &  & 4/10  &  & 5889.00 \\
50         & 4.5504543407 &  & {\ul 4.5504543407} & \textbf{} & {\ul 4.5504543407} & \textbf{} & 0    &  & 0    &  & 10/10 &  & 10/10 &  & 974.70  &  &  & 100 & 5.6359808164 &  & \textbf{5.6357325219} & \textbf{} & \textbf{5.6359770664} & \textbf{} & -2.48E-04 &  & -3.75E-06 &  & 1/10  &  & 5/10  &  & 4714.50 \\ \midrule
\#Improved &              &  & ~~0                  &           & ~~0                  &           &      &  &      &  &       &  &       &  &         &  &  &     &              &  & 19                    &           & 13                    &           &           &  &           &  &       &  &       &  &         \\
\#Equal    &              &  & 46                 &           & 46                 &           &      &  &      &  &       &  &       &  &         &  &  &     &              &  & 30                    &           & 26                    &           &           &  &           &  &       &  &       &  &         \\
\#Worse    &              &  & ~~0                  &           & ~~0                  &           &      &  &      &  &       &  &       &  &         &  &  &     &              &  & ~~1                     &           & 11                    &           &           &  &           &  &       &  &       &  &         \\ \bottomrule
\end{tabular}
}

\end{sidewaystable}

\clearpage
\begin{sidewaystable}[!h]
\centering
\caption{Computational results and comparison of our SED algorithm with the best-known results recorded at Packomania on the moderate I scale instances ($101 \leq n \leq 200$). The improved and equal results appear in bold and underlined, respectively, compared with the best-known results $R^*$ in terms of $R_{best}$ and $R_{avg}$.}
\label{tb_cmp_best2}

\resizebox{1.0\textwidth}{!}{
\begin{tabular}{lllllllrlrlllllrlllllllllrlrlllllr}
\toprule
           &              &  & \multicolumn{13}{l}{SED (this work)}                                                                                &  &  &     &              &  & \multicolumn{13}{l}{SED (this work)}                                                                                                \\ \cline{4-16} \cline{22-34} 
$n$          & $R^*$           &  & $R_{best}$        &           & $R_{avg}$         &           & $\Delta_{best}$ &  & $\Delta_{avg}$ &  & $HR$    &  & $RR$    &  & $Time~(s)$ &  &  & $n$   & $R^*$           &  & $R_{best}$           &           & $R_{avg}$            &           &  $\Delta_{best}$     &  & $\Delta_{avg}$      &  & $HR$    &  & $RR$    &  & $Time~(s)$ \\ \midrule
101        & 5.6599579629 &  & \textbf{5.6597794928} & \textbf{} & \textbf{5.6598457572} &  & -1.78E-04 &  & -1.12E-04 &  & 2/10  &  & 10/10 &  & 9941.10  &  &  & 151 & 6.3985999459 &  & \textbf{6.3840816057} & \textbf{} & \textbf{6.3896769602} &  & -1.45E-02 &  & -8.92E-03 &  & 1/10 &  & 10/10 &  & 16216.80 \\
102        & 5.6748822207 &  & {\ul 5.6748822207}    & \textbf{} & 5.6761187574          &  & 0         &  & 1.24E-03  &  & 3/10  &  & 3/10  &  & 8971.30  &  &  & 152 & 6.4096321832 &  & \textbf{6.4084860226} & \textbf{} & 6.4097374079          &  & -1.15E-03 &  & 1.05E-04  &  & 2/10 &  & 5/10  &  & 10276.00 \\
103        & 5.6923913336 &  & \textbf{5.6921248921} & \textbf{} & 5.6928626608          &  & -2.66E-04 &  & 4.71E-04  &  & 2/10  &  & 5/10  &  & 11333.10 &  &  & 153 & 6.4275041177 &  & \textbf{6.4229011359} & \textbf{} & \textbf{6.4236943663} &  & -4.60E-03 &  & -3.81E-03 &  & 1/10 &  & 10/10 &  & 13533.70 \\
104        & 5.7089561454 &  & \textbf{5.7082798932} & \textbf{} & \textbf{5.7086249682} &  & -6.76E-04 &  & -3.31E-04 &  & 1/10  &  & 8/10  &  & 11103.20 &  &  & 154 & 6.4419628019 &  & \textbf{6.4363632879} & \textbf{} & \textbf{6.4392664417} &  & -5.60E-03 &  & -2.70E-03 &  & 1/10 &  & 10/10 &  & 8635.50  \\
105        & 5.7263005035 &  & \textbf{5.7250734932} & \textbf{} & \textbf{5.7251699857} &  & -1.23E-03 &  & -1.13E-03 &  & 7/10  &  & 10/10 &  & 17688.80 &  &  & 155 & 6.4564928744 &  & \textbf{6.4537179487} & \textbf{} & \textbf{6.4546576387} &  & -2.77E-03 &  & -1.84E-03 &  & 2/10 &  & 9/10  &  & 11761.80 \\
106        & 5.7420854884 &  & \textbf{5.7416014823} & \textbf{} & \textbf{5.7417349458} &  & -4.84E-04 &  & -3.51E-04 &  & 5/10  &  & 10/10 &  & 14459.00 &  &  & 156 & 6.4687030989 &  & \textbf{6.4664665398} & \textbf{} & \textbf{6.4666929851} &  & -2.24E-03 &  & -2.01E-03 &  & 9/10 &  & 9/10  &  & 5981.10  \\
107        & 5.7600955649 &  & 5.7600995060          & \textbf{} & 5.7608367948          &  & 3.94E-06  &  & 7.41E-04  &  & 2/10  &  & 0/10  &  & 14137.50 &  &  & 157 & 6.4806879723 &  & \textbf{6.4793187351} & \textbf{} & \textbf{6.4801269581} &  & -1.37E-03 &  & -5.61E-04 &  & 1/10 &  & 8/10  &  & 7711.40  \\
108        & 5.7747329463 &  & \textbf{5.7747315860} & \textbf{} & 5.7753922938          &  & -1.36E-06 &  & 6.59E-04  &  & 1/10  &  & 4/10  &  & 11369.90 &  &  & 158 & 6.4901644711 &  & \textbf{6.4899861268} & \textbf{} & 6.4906221569          &  & -1.78E-04 &  & 4.58E-04  &  & 1/10 &  & 5/10  &  & 9042.90  \\
109        & 5.7924398458 &  & {\ul 5.7924398458}    & \textbf{} & 5.7927828765          &  & 0         &  & 3.43E-04  &  & 1/10  &  & 1/10  &  & 19340.40 &  &  & 159 & 6.5063519438 &  & \textbf{6.5050243930} & \textbf{} & \textbf{6.5054759138} &  & -1.33E-03 &  & -8.76E-04 &  & 1/10 &  & 10/10 &  & 7050.70  \\
110        & 5.8043281002 &  & 5.8043790103          & \textbf{} & 5.8070774340          &  & 5.09E-05  &  & 2.75E-03  &  & 1/10  &  & 0/10  &  & 13714.30 &  &  & 160 & 6.5216146623 &  & \textbf{6.5187621450} & \textbf{} & \textbf{6.5203667820} &  & -2.85E-03 &  & -1.25E-03 &  & 1/10 &  & 9/10  &  & 9556.50  \\
111        & 5.8226237949 &  & \textbf{5.8225938205} & \textbf{} & 5.8231336777          &  & -3.00E-05 &  & 5.10E-04  &  & 1/10  &  & 3/10  &  & 12636.80 &  &  & 161 & 6.5318675905 &  & \textbf{6.5315929690} & \textbf{} & 6.5330487499          &  & -2.75E-04 &  & 1.18E-03  &  & 1/10 &  & 4/10  &  & 11714.50 \\
112        & 5.8367934883 &  & 5.8381793014          & \textbf{} & 5.8409569163          &  & 1.39E-03  &  & 4.16E-03  &  & 2/10  &  & 0/10  &  & 9841.40  &  &  & 162 & 6.5459496371 &  & \textbf{6.5450672093} & \textbf{} & 6.5460087564          &  & -8.82E-04 &  & 5.91E-05  &  & 3/10 &  & 3/10  &  & 5743.30  \\
113        & 5.8529505194 &  & {\ul 5.8529505194}    & \textbf{} & 5.8537121832          &  & 0         &  & 7.62E-04  &  & 1/10  &  & 1/10  &  & 10367.80 &  &  & 163 & 6.5645506055 &  & \textbf{6.5577283543} & \textbf{} & \textbf{6.5584221542} &  & -6.82E-03 &  & -6.13E-03 &  & 1/10 &  & 10/10 &  & 7448.80  \\
114        & 5.8668408971 &  & \textbf{5.8668377535} & \textbf{} & 5.8711734848          &  & -3.14E-06 &  & 4.33E-03  &  & 1/10  &  & 1/10  &  & 7843.10  &  &  & 164 & 6.5747462297 &  & \textbf{6.5695683516} & \textbf{} & \textbf{6.5697443165} &  & -5.18E-03 &  & -5.00E-03 &  & 2/10 &  & 10/10 &  & 15182.80 \\
115        & 5.8911773387 &  & \textbf{5.8877955184} & \textbf{} & \textbf{5.8890441963} &  & -3.38E-03 &  & -2.13E-03 &  & 1/10  &  & 8/10  &  & 8656.50  &  &  & 165 & 6.5868295847 &  & \textbf{6.5794341232} & \textbf{} & \textbf{6.5795512291} &  & -7.40E-03 &  & -7.28E-03 &  & 1/10 &  & 10/10 &  & 14182.70 \\
116        & 5.9043232575 &  & \textbf{5.9026298032} & \textbf{} & 5.9045535851          &  & -1.69E-03 &  & 2.30E-04  &  & 1/10  &  & 4/10  &  & 9062.60  &  &  & 166 & 6.6002228950 &  & \textbf{6.5951677305} & \textbf{} & \textbf{6.5958877885} &  & -5.06E-03 &  & -4.34E-03 &  & 1/10 &  & 10/10 &  & 12318.40 \\
117        & 5.9151425496 &  & \textbf{5.9151356434} & \textbf{} & 5.9155961822          &  & -6.91E-06 &  & 4.54E-04  &  & 1/10  &  & 1/10  &  & 9359.70  &  &  & 167 & 6.6104638217 &  & \textbf{6.6098597108} & \textbf{} & \textbf{6.6102879842} &  & -6.04E-04 &  & -1.76E-04 &  & 1/10 &  & 5/10  &  & 7921.20  \\
118        & 5.9268531543 &  & {\ul 5.9268531543}    & \textbf{} & {\ul 5.9268531543}    &  & 0         &  & 0         &  & 10/10 &  & 10/10 &  & 7391.40  &  &  & 168 & 6.6242823896 &  & \textbf{6.6230917029} & \textbf{} & \textbf{6.6231154345} &  & -1.19E-03 &  & -1.17E-03 &  & 2/10 &  & 10/10 &  & 4024.90  \\
119        & 5.9491701550 &  & \textbf{5.9490619877} & \textbf{} & \textbf{5.9491643600} &  & -1.08E-04 &  & -5.79E-06 &  & 4/10  &  & 8/10  &  & 11968.90 &  &  & 169 & 6.6386257645 &  & \textbf{6.6363389670} & \textbf{} & \textbf{6.6372393674} &  & -2.29E-03 &  & -1.39E-03 &  & 1/10 &  & 10/10 &  & 12286.80 \\
120        & 5.9668074819 &  & \textbf{5.9636279898} & \textbf{} & \textbf{5.9645070091} &  & -3.18E-03 &  & -2.30E-03 &  & 1/10  &  & 9/10  &  & 12676.80 &  &  & 170 & 6.6482796492 &  & \textbf{6.6466042999} & \textbf{} & 6.6489536764          &  & -1.68E-03 &  & 6.74E-04  &  & 1/10 &  & 2/10  &  & 9279.90  \\
121        & 5.9827857339 &  & \textbf{5.9802333759} & \textbf{} & \textbf{5.9811283874} &  & -2.55E-03 &  & -1.66E-03 &  & 1/10  &  & 9/10  &  & 11757.50 &  &  & 171 & 6.6610392083 &  & \textbf{6.6606281796} & \textbf{} & 6.6619873814          &  & -4.11E-04 &  & 9.48E-04  &  & 2/10 &  & 3/10  &  & 10573.10 \\
122        & 5.9960642820 &  & \textbf{5.9936598394} & \textbf{} & \textbf{5.9947179557} &  & -2.40E-03 &  & -1.35E-03 &  & 1/10  &  & 10/10 &  & 7227.90  &  &  & 172 & 6.6745006321 &  & \textbf{6.6714601988} & \textbf{} & \textbf{6.6732211904} &  & -3.04E-03 &  & -1.28E-03 &  & 1/10 &  & 8/10  &  & 10023.20 \\
123        & 6.0090785558 &  & \textbf{6.0059664994} & \textbf{} & \textbf{6.0063563277} &  & -3.11E-03 &  & -2.72E-03 &  & 4/10  &  & 10/10 &  & 5469.00  &  &  & 173 & 6.6866891682 &  & \textbf{6.6818155183} & \textbf{} & \textbf{6.6837432870} &  & -4.87E-03 &  & -2.95E-03 &  & 1/10 &  & 10/10 &  & 10865.60 \\
124        & 6.0225351062 &  & \textbf{6.0224402824} & \textbf{} & 6.0231508009          &  & -9.48E-05 &  & 6.16E-04  &  & 1/10  &  & 3/10  &  & 9234.90  &  &  & 174 & 6.6981105869 &  & \textbf{6.6973820953} & \textbf{} & 6.6983253058          &  & -7.28E-04 &  & 2.15E-04  &  & 1/10 &  & 4/10  &  & 7223.20  \\
125        & 6.0380168643 &  & \textbf{6.0347337708} & \textbf{} & \textbf{6.0356903597} &  & -3.28E-03 &  & -2.33E-03 &  & 1/10  &  & 10/10 &  & 8789.40  &  &  & 175 & 6.7110188626 &  & \textbf{6.7101346800} & \textbf{} & 6.7111654896          &  & -8.84E-04 &  & 1.47E-04  &  & 1/10 &  & 4/10  &  & 12185.90 \\
126        & 6.0521543613 &  & \textbf{6.0519428829} & \textbf{} & 6.0523238658          &  & -2.11E-04 &  & 1.70E-04  &  & 1/10  &  & 2/10  &  & 14472.60 &  &  & 176 & 6.7228800450 &  & \textbf{6.7228629814} & \textbf{} & 6.7245784136          &  & -1.71E-05 &  & 1.70E-03  &  & 2/10 &  & 2/10  &  & 8947.00  \\
127        & 6.0668536575 &  & \textbf{6.0665232674} & \textbf{} & 6.0669545910          &  & -3.30E-04 &  & 1.01E-04  &  & 3/10  &  & 6/10  &  & 14369.10 &  &  & 177 & 6.7412137384 &  & \textbf{6.7374499882} & \textbf{} & \textbf{6.7390210504} &  & -3.76E-03 &  & -2.19E-03 &  & 1/10 &  & 10/10 &  & 9173.10  \\
128        & 6.0806366813 &  & \textbf{6.0797285101} & \textbf{} & 6.0807445392          &  & -9.08E-04 &  & 1.08E-04  &  & 1/10  &  & 6/10  &  & 10502.80 &  &  & 178 & 6.7537260778 &  & \textbf{6.7500785679} & \textbf{} & \textbf{6.7519579397} &  & -3.65E-03 &  & -1.77E-03 &  & 1/10 &  & 10/10 &  & 10426.40 \\
129        & 6.0932666869 &  & \textbf{6.0922757381} & \textbf{} & \textbf{6.0930011686} &  & -9.91E-04 &  & -2.66E-04 &  & 2/10  &  & 7/10  &  & 10997.50 &  &  & 179 & 6.7668261653 &  & \textbf{6.7629223995} & \textbf{} & \textbf{6.7637985814} &  & -3.90E-03 &  & -3.03E-03 &  & 1/10 &  & 10/10 &  & 8457.30  \\
130        & 6.1076531690 &  & \textbf{6.1072029065} & \textbf{} & 6.1076651361          &  & -4.50E-04 &  & 1.20E-05  &  & 1/10  &  & 4/10  &  & 14599.50 &  &  & 180 & 6.7728585654 &  & 6.7729312241          & \textbf{} & 6.7753088647          &  & 7.27E-05  &  & 2.45E-03  &  & 1/10 &  & 0/10  &  & 8947.10  \\
131        & 6.1178856994 &  & \textbf{6.1146523017} & \textbf{} & \textbf{6.1157575742} &  & -3.23E-03 &  & -2.13E-03 &  & 1/10  &  & 8/10  &  & 4180.80  &  &  & 181 & 6.7924484211 &  & \textbf{6.7866497775} & \textbf{} & \textbf{6.7889594255} &  & -5.80E-03 &  & -3.49E-03 &  & 1/10 &  & 10/10 &  & 7381.40  \\
132        & 6.1269393558 &  & \textbf{6.1262395723} & \textbf{} & \textbf{6.1267578890} &  & -7.00E-04 &  & -1.81E-04 &  & 1/10  &  & 8/10  &  & 16765.70 &  &  & 182 & 6.8047492229 &  & \textbf{6.7999756316} & \textbf{} & \textbf{6.8017976377} &  & -4.77E-03 &  & -2.95E-03 &  & 1/10 &  & 10/10 &  & 11125.60 \\
133        & 6.1363940823 &  & \textbf{6.1363253673} & \textbf{} & \textbf{6.1363643585} &  & -6.87E-05 &  & -2.97E-05 &  & 6/10  &  & 7/10  &  & 19921.60 &  &  & 183 & 6.8160062463 &  & \textbf{6.8129617420} & \textbf{} & \textbf{6.8141746128} &  & -3.04E-03 &  & -1.83E-03 &  & 1/10 &  & 10/10 &  & 5952.90  \\
134        & 6.1412839494 &  & \textbf{6.1412827373} & \textbf{} & 6.1413091994          &  & -1.21E-06 &  & 2.53E-05  &  & 5/10  &  & 9/10  &  & 19320.70 &  &  & 184 & 6.8283189424 &  & \textbf{6.8249182583} & \textbf{} & \textbf{6.8276016309} &  & -3.40E-03 &  & -7.17E-04 &  & 1/10 &  & 9/10  &  & 4433.20  \\
135        & 6.1508293815 &  & {\ul 6.1508293815}    & \textbf{} & 6.1508295653          &  & 0         &  & 1.84E-07  &  & 3/10  &  & 3/10  &  & 21618.80 &  &  & 185 & 6.8424112657 &  & \textbf{6.8374332956} & \textbf{} & \textbf{6.8396361609} &  & -4.98E-03 &  & -2.78E-03 &  & 1/10 &  & 10/10 &  & 5675.60  \\
136        & 6.1584969194 &  & \textbf{6.1584884950} & \textbf{} & \textbf{6.1584894041} &  & -8.42E-06 &  & -7.52E-06 &  & 2/10  &  & 10/10 &  & 17476.90 &  &  & 186 & 6.8552427473 &  & \textbf{6.8504870996} & \textbf{} & \textbf{6.8532734673} &  & -4.76E-03 &  & -1.97E-03 &  & 1/10 &  & 9/10  &  & 10833.40 \\
137        & 6.1815082562 &  & {\ul 6.1815082562}    & \textbf{} & {\ul 6.1815082562}    &  & 0         &  & 0         &  & 10/10 &  & 10/10 &  & 16313.80 &  &  & 187 & 6.8661850359 &  & \textbf{6.8631014993} & \textbf{} & \textbf{6.8657471912} &  & -3.08E-03 &  & -4.38E-04 &  & 1/10 &  & 5/10  &  & 11367.00 \\
138        & 6.2059663809 &  & \textbf{6.2059641354} & \textbf{} & 6.2060674407          &  & -2.25E-06 &  & 1.01E-04  &  & 6/10  &  & 7/10  &  & 19133.70 &  &  & 188 & 6.8783760115 &  & \textbf{6.8753773078} & \textbf{} & \textbf{6.8776170850} &  & -3.00E-03 &  & -7.59E-04 &  & 1/10 &  & 7/10  &  & 9281.30  \\
139        & 6.2210255874 &  & \textbf{6.2209771473} & \textbf{} & 6.2210550401          &  & -4.84E-05 &  & 2.95E-05  &  & 1/10  &  & 4/10  &  & 21004.10 &  &  & 189 & 6.8869865649 &  & \textbf{6.8853029301} & \textbf{} & 6.8898744153          &  & -1.68E-03 &  & 2.89E-03  &  & 1/10 &  & 3/10  &  & 11578.30 \\
140        & 6.2341312659 &  & \textbf{6.2341210439} & \textbf{} & 6.2342623742          &  & -1.02E-05 &  & 1.31E-04  &  & 6/10  &  & 7/10  &  & 21423.80 &  &  & 190 & 6.8992268282 &  & 6.9009579733          & \textbf{} & 6.9027526865          &  & 1.73E-03  &  & 3.53E-03  &  & 1/10 &  & 0/10  &  & 15317.90 \\
141        & 6.2490415957 &  & \textbf{6.2483731721} & \textbf{} & \textbf{6.2483929733} &  & -6.68E-04 &  & -6.49E-04 &  & 3/10  &  & 10/10 &  & 21345.70 &  &  & 191 & 6.9119625946 &  & \textbf{6.9104619032} & \textbf{} & 6.9136540117          &  & -1.50E-03 &  & 1.69E-03  &  & 1/10 &  & 2/10  &  & 12324.30 \\
142        & 6.2617416866 &  & {\ul 6.2617416866}    & \textbf{} & 6.2617448548          &  & 0         &  & 3.17E-06  &  & 4/10  &  & 4/10  &  & 20381.20 &  &  & 192 & 6.9229251057 &  & \textbf{6.9203258423} & \textbf{} & 6.9236192514          &  & -2.60E-03 &  & 6.94E-04  &  & 1/10 &  & 4/10  &  & 14234.80 \\
143        & 6.2716966903 &  & \textbf{6.2715956530} & \textbf{} & \textbf{6.2715957051} &  & -1.01E-04 &  & -1.01E-04 &  & 3/10  &  & 10/10 &  & 21629.90 &  &  & 193 & 6.9364631911 &  & \textbf{6.9329148694} & \textbf{} & 6.9370827791          &  & -3.55E-03 &  & 6.20E-04  &  & 1/10 &  & 4/10  &  & 7599.70  \\
144        & 6.2832371987 &  & \textbf{6.2832364663} & \textbf{} & \textbf{6.2832370480} &  & -7.32E-07 &  & -1.51E-07 &  & 2/10  &  & 10/10 &  & 21109.70 &  &  & 194 & 6.9506991452 &  & \textbf{6.9405887493} & \textbf{} & \textbf{6.9481145753} &  & -1.01E-02 &  & -2.58E-03 &  & 1/10 &  & 8/10  &  & 11619.90 \\
145        & 6.3051401291 &  & \textbf{6.3048057891} & \textbf{} & \textbf{6.3048623098} &  & -3.34E-04 &  & -2.78E-04 &  & 4/10  &  & 10/10 &  & 19513.10 &  &  & 195 & 6.9631085477 &  & \textbf{6.9551013467} & \textbf{} & \textbf{6.9603931174} &  & -8.01E-03 &  & -2.72E-03 &  & 1/10 &  & 7/10  &  & 12510.10 \\
146        & 6.3192938478 &  & \textbf{6.3191616112} & \textbf{} & \textbf{6.3191867233} &  & -1.32E-04 &  & -1.07E-04 &  & 7/10  &  & 9/10  &  & 19888.90 &  &  & 196 & 6.9772480051 &  & \textbf{6.9651197544} & \textbf{} & \textbf{6.9704765550} &  & -1.21E-02 &  & -6.77E-03 &  & 1/10 &  & 10/10 &  & 8656.90  \\
147        & 6.3377734986 &  & {\ul 6.3377734986}    & \textbf{} & 6.3377957984          &  & 0         &  & 2.23E-05  &  & 6/10  &  & 6/10  &  & 20428.90 &  &  & 197 & 6.9878958414 &  & \textbf{6.9790672127} & \textbf{} & \textbf{6.9832654056} &  & -8.83E-03 &  & -4.63E-03 &  & 1/10 &  & 10/10 &  & 12163.60 \\
148        & 6.3535817257 &  & \textbf{6.3465377175} & \textbf{} & \textbf{6.3515717482} &  & -7.04E-03 &  & -2.01E-03 &  & 1/10  &  & 10/10 &  & 15163.10 &  &  & 198 & 6.9979276863 &  & \textbf{6.9892007374} & \textbf{} & \textbf{6.9949396268} &  & -8.73E-03 &  & -2.99E-03 &  & 1/10 &  & 9/10  &  & 12900.40 \\
149        & 6.3663869913 &  & \textbf{6.3654709825} & \textbf{} & \textbf{6.3663756849} &  & -9.16E-04 &  & -1.13E-05 &  & 1/10  &  & 4/10  &  & 15256.50 &  &  & 199 & 7.0101428170 &  & \textbf{7.0033775577} & \textbf{} & \textbf{7.0065604316} &  & -6.77E-03 &  & -3.58E-03 &  & 1/10 &  & 9/10  &  & 17047.00 \\
150        & 6.3814711453 &  & \textbf{6.3800749408} & \textbf{} & \textbf{6.3809350553} &  & -1.40E-03 &  & -5.36E-04 &  & 1/10  &  & 10/10 &  & 14089.80 &  &  & 200 & 7.0226524557 &  & \textbf{7.0160475056} & \textbf{} & \textbf{7.0199497597} &  & -6.60E-03 &  & -2.70E-03 &  & 1/10 &  & 10/10 &  & 20248.00 \\ \midrule
\#Improved &              &  & 39                    &           & 24                    &  &           &  &           &  &       &  &       &  &          &  &  &     &              &  & 48                    &           & 35                    &  &           &  &           &  &      &  &       &  &          \\
\#Equal    &              &  & ~~8                     &           & ~~2                     &  &           &  &           &  &       &  &       &  &          &  &  &     &              &  & ~~0                     &           & ~~0                     &  &           &  &           &  &      &  &       &  &          \\
\#Worse    &              &  & ~~3                     &           & 24                    &  &           &  &           &  &       &  &       &  &          &  &  &     &              &  & ~~2                     &           & 15                    &  &           &  &           &  &      &  &       &  &          \\ \bottomrule
\end{tabular}
}

\end{sidewaystable}
\clearpage
\begin{sidewaystable}[!h]
\centering
\caption{Computational results and comparison of our SED algorithm with the best-known results recorded at Packomania on the first part of the moderate II scale instances ($201 \leq n \leq 300$). The improved and equal results appear in bold and underlined, respectively, compared with the best-known results $R^*$ in terms of $R_{best}$ and $R_{avg}$.}
\label{tb_cmp_best3}

\resizebox{1.0\textwidth}{!}{
\begin{tabular}{lllllllrlrlllllrlllllllllrlrlllllr}
\toprule
           &              &  & \multicolumn{13}{l}{SED (this work)}                                                                                &  &  &     &              &  & \multicolumn{13}{l}{SED (this work)}                                                                                                \\ \cline{4-16} \cline{22-34} 
$n$          & $R^*$           &  & $R_{best}$        &           & $R_{avg}$         &           & $\Delta_{best}$ &  & $\Delta_{avg}$ &  & $HR$    &  & $RR$    &  & $Time~(s)$ &  &  & $n$   & $R^*$           &  & $R_{best}$           &           & $R_{avg}$            &           &  $\Delta_{best}$     &  & $\Delta_{avg}$      &  & $HR$    &  & $RR$    &  & $Time~(s)$ \\ \midrule
201        & 7.0311579735 &  & \textbf{7.0255544330} & \textbf{} & \textbf{7.0291874305} &  & -5.60E-03 &  & -1.97E-03 &  & 1/10 &  & 8/10  &  & 21949.40 &  &  & 251 & 7.5340290155 &  & \textbf{7.5263662601} & \textbf{} & \textbf{7.5272731866} &  & -7.66E-03 &  & -6.76E-03 &  & 1/10 &  & 10/10 &  & 12987.90 \\
202        & 7.0452775617 &  & \textbf{7.0358154477} & \textbf{} & \textbf{7.0400322271} &  & -9.46E-03 &  & -5.25E-03 &  & 1/10 &  & 10/10 &  & 27886.50 &  &  & 252 & 7.5424648986 &  & \textbf{7.5340412095} & \textbf{} & \textbf{7.5343693954} &  & -8.42E-03 &  & -8.10E-03 &  & 1/10 &  & 10/10 &  & 24414.60 \\
203        & 7.0547724371 &  & \textbf{7.0484216078} & \textbf{} & \textbf{7.0511581917} &  & -6.35E-03 &  & -3.61E-03 &  & 1/10 &  & 10/10 &  & 25251.60 &  &  & 253 & 7.5521308100 &  & \textbf{7.5424247553} & \textbf{} & \textbf{7.5426571038} &  & -9.71E-03 &  & -9.47E-03 &  & 6/10 &  & 10/10 &  & 26944.60 \\
204        & 7.0654812476 &  & \textbf{7.0590273102} & \textbf{} & \textbf{7.0617579008} &  & -6.45E-03 &  & -3.72E-03 &  & 1/10 &  & 9/10  &  & 24885.50 &  &  & 254 & 7.5621077128 &  & \textbf{7.5482890171} & \textbf{} & \textbf{7.5495184489} &  & -1.38E-02 &  & -1.26E-02 &  & 2/10 &  & 10/10 &  & 19186.80 \\
205        & 7.0793921434 &  & \textbf{7.0632644486} & \textbf{} & \textbf{7.0702898675} &  & -1.61E-02 &  & -9.10E-03 &  & 1/10 &  & 10/10 &  & 18411.00 &  &  & 255 & 7.5718630254 &  & \textbf{7.5633379207} & \textbf{} & \textbf{7.5647885091} &  & -8.53E-03 &  & -7.07E-03 &  & 3/10 &  & 10/10 &  & 21495.70 \\
206        & 7.0896011259 &  & \textbf{7.0812834485} & \textbf{} & \textbf{7.0855049261} &  & -8.32E-03 &  & -4.10E-03 &  & 1/10 &  & 10/10 &  & 20135.40 &  &  & 256 & 7.5793149724 &  & \textbf{7.5721483921} & \textbf{} & \textbf{7.5751185507} &  & -7.17E-03 &  & -4.20E-03 &  & 1/10 &  & 8/10  &  & 23250.50 \\
207        & 7.1003436731 &  & \textbf{7.0909103971} & \textbf{} & \textbf{7.0945016181} &  & -9.43E-03 &  & -5.84E-03 &  & 1/10 &  & 10/10 &  & 29044.60 &  &  & 257 & 7.5923599786 &  & \textbf{7.5806235741} & \textbf{} & \textbf{7.5823156930} &  & -1.17E-02 &  & -1.00E-02 &  & 1/10 &  & 10/10 &  & 26719.70 \\
208        & 7.1114371819 &  & \textbf{7.1043854402} & \textbf{} & \textbf{7.1072611271} &  & -7.05E-03 &  & -4.18E-03 &  & 1/10 &  & 10/10 &  & 29433.20 &  &  & 258 & 7.6081822216 &  & \textbf{7.5937072918} & \textbf{} & \textbf{7.5963308481} &  & -1.45E-02 &  & -1.19E-02 &  & 1/10 &  & 10/10 &  & 22982.20 \\
209        & 7.1225774947 &  & \textbf{7.1158587077} & \textbf{} & \textbf{7.1192283319} &  & -6.72E-03 &  & -3.35E-03 &  & 1/10 &  & 10/10 &  & 29498.50 &  &  & 259 & 7.6232794657 &  & \textbf{7.6054822855} & \textbf{} & \textbf{7.6071154633} &  & -1.78E-02 &  & -1.62E-02 &  & 1/10 &  & 10/10 &  & 25211.30 \\
210        & 7.1340172788 &  & \textbf{7.1249463415} & \textbf{} & \textbf{7.1287760840} &  & -9.07E-03 &  & -5.24E-03 &  & 1/10 &  & 10/10 &  & 28973.70 &  &  & 260 & 7.6303474460 &  & \textbf{7.6147896386} & \textbf{} & \textbf{7.6204840351} &  & -1.56E-02 &  & -9.86E-03 &  & 1/10 &  & 10/10 &  & 29490.10 \\
211        & 7.1453393266 &  & \textbf{7.1353927996} & \textbf{} & \textbf{7.1402314896} &  & -9.95E-03 &  & -5.11E-03 &  & 1/10 &  & 10/10 &  & 24398.70 &  &  & 261 & 7.6389568630 &  & \textbf{7.6249382747} & \textbf{} & \textbf{7.6309926787} &  & -1.40E-02 &  & -7.96E-03 &  & 1/10 &  & 9/10  &  & 28613.60 \\
212        & 7.1571882237 &  & \textbf{7.1474588978} & \textbf{} & \textbf{7.1490794335} &  & -9.73E-03 &  & -8.11E-03 &  & 1/10 &  & 10/10 &  & 24013.30 &  &  & 262 & 7.6485119939 &  & \textbf{7.6350999584} & \textbf{} & \textbf{7.6405314425} &  & -1.34E-02 &  & -7.98E-03 &  & 1/10 &  & 9/10  &  & 26678.30 \\
213        & 7.1672317136 &  & \textbf{7.1573496074} & \textbf{} & \textbf{7.1604642324} &  & -9.88E-03 &  & -6.77E-03 &  & 1/10 &  & 10/10 &  & 24727.20 &  &  & 263 & 7.6627443277 &  & \textbf{7.6430467452} & \textbf{} & \textbf{7.6509048689} &  & -1.97E-02 &  & -1.18E-02 &  & 1/10 &  & 10/10 &  & 35999.60 \\
214        & 7.1767228307 &  & \textbf{7.1668129415} & \textbf{} & \textbf{7.1696479254} &  & -9.91E-03 &  & -7.07E-03 &  & 1/10 &  & 10/10 &  & 31065.30 &  &  & 264 & 7.6713523051 &  & \textbf{7.6609426956} & \textbf{} & \textbf{7.6651209911} &  & -1.04E-02 &  & -6.23E-03 &  & 1/10 &  & 10/10 &  & 32310.60 \\
215        & 7.1858107754 &  & \textbf{7.1779075067} & \textbf{} & \textbf{7.1808898146} &  & -7.90E-03 &  & -4.92E-03 &  & 1/10 &  & 10/10 &  & 30391.20 &  &  & 265 & 7.6835789969 &  & \textbf{7.6668328972} & \textbf{} & \textbf{7.6727525059} &  & -1.67E-02 &  & -1.08E-02 &  & 1/10 &  & 10/10 &  & 34413.50 \\
216        & 7.1962064445 &  & \textbf{7.1876566450} & \textbf{} & \textbf{7.1930848655} &  & -8.55E-03 &  & -3.12E-03 &  & 1/10 &  & 10/10 &  & 29824.10 &  &  & 266 & 7.6903773179 &  & \textbf{7.6776988774} & \textbf{} & \textbf{7.6825349053} &  & -1.27E-02 &  & -7.84E-03 &  & 1/10 &  & 10/10 &  & 37073.50 \\
217        & 7.2037038757 &  & \textbf{7.1977035204} & \textbf{} & \textbf{7.2018444199} &  & -6.00E-03 &  & -1.86E-03 &  & 1/10 &  & 8/10  &  & 29494.90 &  &  & 267 & 7.6944927612 &  & \textbf{7.6860055622} & \textbf{} & \textbf{7.6922616494} &  & -8.49E-03 &  & -2.23E-03 &  & 1/10 &  & 8/10  &  & 37946.30 \\
218        & 7.2119969745 &  & \textbf{7.2089282580} & \textbf{} & 7.2138131185          &  & -3.07E-03 &  & 1.82E-03  &  & 1/10 &  & 2/10  &  & 28066.90 &  &  & 268 & 7.6988832962 &  & \textbf{7.6964397347} & \textbf{} & 7.7030194411          &  & -2.44E-03 &  & 4.14E-03  &  & 1/10 &  & 1/10  &  & 33625.30 \\
219        & 7.2248960523 &  & \textbf{7.2203773197} & \textbf{} & \textbf{7.2230900357} &  & -4.52E-03 &  & -1.81E-03 &  & 1/10 &  & 8/10  &  & 33260.30 &  &  & 269 & 7.7044148510 &  & 7.7048369985          & \textbf{} & 7.7101553679          &  & 4.22E-04  &  & 5.74E-03  &  & 1/10 &  & 0/10  &  & 38941.10 \\
220        & 7.2365800516 &  & \textbf{7.2296491612} & \textbf{} & \textbf{7.2347172798} &  & -6.93E-03 &  & -1.86E-03 &  & 1/10 &  & 8/10  &  & 21758.60 &  &  & 270 & 7.7082910864 &  & 7.7112854992          & \textbf{} & 7.7191886111          &  & 2.99E-03  &  & 1.09E-02  &  & 1/10 &  & 0/10  &  & 34779.00 \\
221        & 7.2462659092 &  & \textbf{7.2412540521} & \textbf{} & \textbf{7.2435271826} &  & -5.01E-03 &  & -2.74E-03 &  & 1/10 &  & 9/10  &  & 27656.20 &  &  & 271 & 7.7123271806 &  & 7.7202832347          & \textbf{} & 7.7291912698          &  & 7.96E-03  &  & 1.69E-02  &  & 1/10 &  & 0/10  &  & 39066.90 \\
222        & 7.2551034048 &  & \textbf{7.2500633764} & \textbf{} & \textbf{7.2527363680} &  & -5.04E-03 &  & -2.37E-03 &  & 1/10 &  & 10/10 &  & 34211.40 &  &  & 272 & 7.7217513975 &  & 7.7333575768          & \textbf{} & 7.7384232551          &  & 1.16E-02  &  & 1.67E-02  &  & 1/10 &  & 0/10  &  & 37883.30 \\
223        & 7.2656309049 &  & \textbf{7.2580609579} & \textbf{} & \textbf{7.2627742904} &  & -7.57E-03 &  & -2.86E-03 &  & 1/10 &  & 9/10  &  & 29684.30 &  &  & 273 & 7.7283368304 &  & 7.7411887098          & \textbf{} & 7.7465249044          &  & 1.29E-02  &  & 1.82E-02  &  & 1/10 &  & 0/10  &  & 38376.30 \\
224        & 7.2773063486 &  & \textbf{7.2696697784} & \textbf{} & \textbf{7.2763985458} &  & -7.64E-03 &  & -9.08E-04 &  & 1/10 &  & 5/10  &  & 17439.50 &  &  & 274 & 7.7386433178 &  & 7.7493245537          & \textbf{} & 7.7550211192          &  & 1.07E-02  &  & 1.64E-02  &  & 1/10 &  & 0/10  &  & 29294.40 \\
225        & 7.2841046441 &  & \textbf{7.2813658072} & \textbf{} & 7.2850444400          &  & -2.74E-03 &  & 9.40E-04  &  & 1/10 &  & 4/10  &  & 25900.90 &  &  & 275 & 7.7472814344 &  & 7.7574168452          & \textbf{} & 7.7646436754          &  & 1.01E-02  &  & 1.74E-02  &  & 1/10 &  & 0/10  &  & 35441.30 \\
226        & 7.2923021211 &  & \textbf{7.2868642085} & \textbf{} & 7.2929506330          &  & -5.44E-03 &  & 6.49E-04  &  & 1/10 &  & 3/10  &  & 34965.30 &  &  & 276 & 7.7633757754 &  & \textbf{7.7537790480} & \textbf{} & 7.7736485773          &  & -9.60E-03 &  & 1.03E-02  &  & 1/10 &  & 1/10  &  & 36941.20 \\
227        & 7.3067387277 &  & \textbf{7.3013688606} & \textbf{} & \textbf{7.3035699495} &  & -5.37E-03 &  & -3.17E-03 &  & 1/10 &  & 10/10 &  & 26158.50 &  &  & 277 & 7.7685794251 &  & 7.7711489967          & \textbf{} & 7.7806041894          &  & 2.57E-03  &  & 1.20E-02  &  & 1/10 &  & 0/10  &  & 36317.00 \\
228        & 7.3163529791 &  & \textbf{7.3086523632} & \textbf{} & \textbf{7.3128750428} &  & -7.70E-03 &  & -3.48E-03 &  & 1/10 &  & 10/10 &  & 30731.00 &  &  & 278 & 7.7753213557 &  & \textbf{7.7705949667} & \textbf{} & 7.7889657025          &  & -4.73E-03 &  & 1.36E-02  &  & 1/10 &  & 1/10  &  & 34971.70 \\
229        & 7.3304940328 &  & \textbf{7.3230948777} & \textbf{} & \textbf{7.3244515439} &  & -7.40E-03 &  & -6.04E-03 &  & 1/10 &  & 10/10 &  & 28738.70 &  &  & 279 & 7.7828242480 &  & 7.7925896802          & \textbf{} & 7.7985418673          &  & 9.77E-03  &  & 1.57E-02  &  & 1/10 &  & 0/10  &  & 32826.30 \\
230        & 7.3387673249 &  & \textbf{7.3304753817} & \textbf{} & \textbf{7.3323861104} &  & -8.29E-03 &  & -6.38E-03 &  & 1/10 &  & 10/10 &  & 31189.50 &  &  & 280 & 7.7895974859 &  & \textbf{7.7850390707} & \textbf{} & 7.8009574613          &  & -4.56E-03 &  & 1.14E-02  &  & 1/10 &  & 4/10  &  & 34835.80 \\
231        & 7.3478959531 &  & \textbf{7.3395741975} & \textbf{} & \textbf{7.3433421654} &  & -8.32E-03 &  & -4.55E-03 &  & 1/10 &  & 10/10 &  & 26296.50 &  &  & 281 & 7.7970295470 &  & \textbf{7.7935255540} & \textbf{} & 7.8046367242          &  & -3.50E-03 &  & 7.61E-03  &  & 1/10 &  & 4/10  &  & 27388.20 \\
232        & 7.3606440021 &  & \textbf{7.3514522524} & \textbf{} & \textbf{7.3538407475} &  & -9.19E-03 &  & -6.80E-03 &  & 1/10 &  & 10/10 &  & 29602.60 &  &  & 282 & 7.8094475872 &  & \textbf{7.8025455856} & \textbf{} & 7.8136332483          &  & -6.90E-03 &  & 4.19E-03  &  & 1/10 &  & 5/10  &  & 28618.10 \\
233        & 7.3660814854 &  & \textbf{7.3597164787} & \textbf{} & \textbf{7.3633262498} &  & -6.37E-03 &  & -2.76E-03 &  & 1/10 &  & 8/10  &  & 24688.20 &  &  & 283 & 7.8169572685 &  & \textbf{7.8094014911} & \textbf{} & 7.8241488493          &  & -7.56E-03 &  & 7.19E-03  &  & 1/10 &  & 4/10  &  & 23766.10 \\
234        & 7.3783260883 &  & \textbf{7.3706429647} & \textbf{} & \textbf{7.3737163256} &  & -7.68E-03 &  & -4.61E-03 &  & 1/10 &  & 10/10 &  & 22580.70 &  &  & 284 & 7.8239824233 &  & \textbf{7.8157677410} & \textbf{} & 7.8256414787          &  & -8.21E-03 &  & 1.66E-03  &  & 1/10 &  & 6/10  &  & 29747.60 \\
235        & 7.3870071716 &  & \textbf{7.3818305285} & \textbf{} & \textbf{7.3847166386} &  & -5.18E-03 &  & -2.29E-03 &  & 1/10 &  & 9/10  &  & 26662.10 &  &  & 285 & 7.8308005585 &  & \textbf{7.8251346857} & \textbf{} & 7.8311060285          &  & -5.67E-03 &  & 3.05E-04  &  & 4/10 &  & 8/10  &  & 30489.20 \\
236        & 7.3941962819 &  & \textbf{7.3915014587} & \textbf{} & 7.3961418565          &  & -2.69E-03 &  & 1.95E-03  &  & 1/10 &  & 2/10  &  & 24757.80 &  &  & 286 & 7.8409629540 &  & \textbf{7.8335735779} & \textbf{} & \textbf{7.8399883454} &  & -7.39E-03 &  & -9.75E-04 &  & 1/10 &  & 8/10  &  & 28673.30 \\
237        & 7.4033253417 &  & \textbf{7.3993046206} & \textbf{} & 7.4056526724          &  & -4.02E-03 &  & 2.33E-03  &  & 1/10 &  & 2/10  &  & 26060.40 &  &  & 287 & 7.8498687648 &  & \textbf{7.8428652701} & \textbf{} & \textbf{7.8441472545} &  & -7.00E-03 &  & -5.72E-03 &  & 1/10 &  & 10/10 &  & 32973.60 \\
238        & 7.4151037322 &  & \textbf{7.4144807701} & \textbf{} & 7.4181620386          &  & -6.23E-04 &  & 3.06E-03  &  & 1/10 &  & 2/10  &  & 32024.30 &  &  & 288 & 7.8614608232 &  & \textbf{7.8508039705} & \textbf{} & \textbf{7.8521664822} &  & -1.07E-02 &  & -9.29E-03 &  & 1/10 &  & 10/10 &  & 29031.00 \\
239        & 7.4305286960 &  & \textbf{7.4202453485} & \textbf{} & \textbf{7.4256091484} &  & -1.03E-02 &  & -4.92E-03 &  & 1/10 &  & 10/10 &  & 37016.80 &  &  & 289 & 7.8700993136 &  & \textbf{7.8605264864} & \textbf{} & \textbf{7.8626203894} &  & -9.57E-03 &  & -7.48E-03 &  & 1/10 &  & 10/10 &  & 31146.20 \\
240        & 7.4396535786 &  & \textbf{7.4295984609} & \textbf{} & \textbf{7.4369271123} &  & -1.01E-02 &  & -2.73E-03 &  & 1/10 &  & 8/10  &  & 34259.00 &  &  & 290 & 7.8790360850 &  & \textbf{7.8698725201} & \textbf{} & \textbf{7.8715351972} &  & -9.16E-03 &  & -7.50E-03 &  & 1/10 &  & 10/10 &  & 32901.30 \\
241        & 7.4493181639 &  & \textbf{7.4352855259} & \textbf{} & \textbf{7.4436434751} &  & -1.40E-02 &  & -5.67E-03 &  & 1/10 &  & 10/10 &  & 32379.40 &  &  & 291 & 7.8887911114 &  & \textbf{7.8793971860} & \textbf{} & \textbf{7.8800645151} &  & -9.39E-03 &  & -8.73E-03 &  & 1/10 &  & 10/10 &  & 31435.40 \\
242        & 7.4583263208 &  & \textbf{7.4483780897} & \textbf{} & \textbf{7.4533697575} &  & -9.95E-03 &  & -4.96E-03 &  & 1/10 &  & 9/10  &  & 33125.20 &  &  & 292 & 7.8973927600 &  & \textbf{7.8877228593} & \textbf{} & \textbf{7.8888227597} &  & -9.67E-03 &  & -8.57E-03 &  & 1/10 &  & 10/10 &  & 31993.50 \\
243        & 7.4662779719 &  & \textbf{7.4543237759} & \textbf{} & \textbf{7.4601566836} &  & -1.20E-02 &  & -6.12E-03 &  & 1/10 &  & 10/10 &  & 34035.40 &  &  & 293 & 7.9052966809 &  & \textbf{7.8963644352} & \textbf{} & \textbf{7.8982809450} &  & -8.93E-03 &  & -7.02E-03 &  & 1/10 &  & 10/10 &  & 39189.50 \\
244        & 7.4755562089 &  & \textbf{7.4638668575} & \textbf{} & \textbf{7.4693595567} &  & -1.17E-02 &  & -6.20E-03 &  & 1/10 &  & 9/10  &  & 40302.80 &  &  & 294 & 7.9140909168 &  & \textbf{7.9055450950} & \textbf{} & \textbf{7.9060972510} &  & -8.55E-03 &  & -7.99E-03 &  & 1/10 &  & 10/10 &  & 34029.60 \\
245        & 7.4855419380 &  & \textbf{7.4754936428} & \textbf{} & \textbf{7.4787240443} &  & -1.00E-02 &  & -6.82E-03 &  & 1/10 &  & 10/10 &  & 28282.10 &  &  & 295 & 7.9215067329 &  & \textbf{7.9125134312} & \textbf{} & \textbf{7.9133814366} &  & -8.99E-03 &  & -8.13E-03 &  & 1/10 &  & 10/10 &  & 36254.10 \\
246        & 7.4908312394 &  & \textbf{7.4833047976} & \textbf{} & \textbf{7.4887762093} &  & -7.53E-03 &  & -2.06E-03 &  & 1/10 &  & 6/10  &  & 26785.80 &  &  & 296 & 7.9301910270 &  & \textbf{7.9218499738} & \textbf{} & \textbf{7.9221697952} &  & -8.34E-03 &  & -8.02E-03 &  & 1/10 &  & 10/10 &  & 33714.60 \\
247        & 7.4975449341 &  & \textbf{7.4923815998} & \textbf{} & \textbf{7.4954534948} &  & -5.16E-03 &  & -2.09E-03 &  & 1/10 &  & 7/10  &  & 38049.30 &  &  & 297 & 7.9439217715 &  & \textbf{7.9296931750} & \textbf{} & \textbf{7.9306822749} &  & -1.42E-02 &  & -1.32E-02 &  & 1/10 &  & 10/10 &  & 34350.10 \\
248        & 7.5072043746 &  & \textbf{7.4995291317} & \textbf{} & \textbf{7.5059041739} &  & -7.68E-03 &  & -1.30E-03 &  & 1/10 &  & 5/10  &  & 17319.50 &  &  & 298 & 7.9562684663 &  & \textbf{7.9392349019} & \textbf{} & \textbf{7.9398510754} &  & -1.70E-02 &  & -1.64E-02 &  & 1/10 &  & 10/10 &  & 36449.80 \\
249        & 7.5203328104 &  & \textbf{7.5092769099} & \textbf{} & \textbf{7.5117939937} &  & -1.11E-02 &  & -8.54E-03 &  & 1/10 &  & 9/10  &  & 18143.60 &  &  & 299 & 7.9615077588 &  & \textbf{7.9476166544} & \textbf{} & \textbf{7.9491620619} &  & -1.39E-02 &  & -1.23E-02 &  & 1/10 &  & 10/10 &  & 35998.00 \\
250        & 7.5270535066 &  & \textbf{7.5177181629} & \textbf{} & \textbf{7.5196028680} &  & -9.34E-03 &  & -7.45E-03 &  & 1/10 &  & 9/10  &  & 25389.90 &  &  & 300 & 7.9716783431 &  & \textbf{7.9589157181} & \textbf{} & \textbf{7.9604790441} &  & -1.28E-02 &  & -1.12E-02 &  & 1/10 &  & 10/10 &  & 37021.50 \\ \midrule
\#Improved &              &  & 50                    &           & 44                    &  &           &  &           &  &      &  &       &  &          &  &  &     &              &  & 41                    &           & 32                    &  &           &  &           &  &      &  &       &  &          \\
\#Equal    &              &  & ~~0                     &           & ~~0                     &  &           &  &           &  &      &  &       &  &          &  &  &     &              &  & ~~0                     &           & ~~0                     &  &           &  &           &  &      &  &       &  &          \\
\#Worse    &              &  & ~~0                     &           & ~~6                     &  &           &  &           &  &      &  &       &  &          &  &  &     &              &  & ~~9                     &           & 18                    &  &           &  &           &  &      &  &       &  &          \\ \bottomrule
\end{tabular}
}

\end{sidewaystable}
\clearpage
\begin{sidewaystable}[!h]
\centering
\caption{Computational results and comparison of our SED algorithm with the best-known results recorded at Packomania on the second part of the moderate II scale instances ($301 \leq n \leq 400$). The improved and equal results appear in bold and underlined, respectively, compared with the best-known results $R^*$ in terms of $R_{best}$ and $R_{avg}$.}
\label{tb_cmp_best4}

\resizebox{1.0\textwidth}{!}{
\begin{tabular}{lllllllrlrlllllrlllllllllrlrlllllr}
\toprule
           &              &  & \multicolumn{13}{l}{SED (this work)}                                                                                &  &  &     &              &  & \multicolumn{13}{l}{SED (this work)}                                                                                                \\ \cline{4-16} \cline{22-34} 
$n$          & $R^*$           &  & $R_{best}$        &           & $R_{avg}$         &           & $\Delta_{best}$ &  & $\Delta_{avg}$ &  & $HR$    &  & $RR$    &  & $Time~(s)$ &  &  & $n$   & $R^*$           &  & $R_{best}$           &           & $R_{avg}$            &           &  $\Delta_{best}$     &  & $\Delta_{avg}$      &  & $HR$    &  & $RR$    &  & $Time~(s)$ \\ \midrule
301        & 7.9811692972 &  & \textbf{7.9688310111} & \textbf{} & \textbf{7.9703173229} &  & -1.23E-02 &  & -1.09E-02 &  & 1/10 &  & 10/10 &  & 31465.50 &  &  & 351 & 8.4007527076 &  & \textbf{8.3921705143} & \textbf{} & \textbf{8.3976610497} &  & -8.58E-03 &  & -3.09E-03 &  & 1/10 &  & 8/10  &  & 36231.70 \\
302        & 7.9916010741 &  & \textbf{7.9761636886} & \textbf{} & \textbf{7.9780148165} &  & -1.54E-02 &  & -1.36E-02 &  & 1/10 &  & 10/10 &  & 30879.40 &  &  & 352 & 8.4129454678 &  & \textbf{8.4033387897} & \textbf{} & \textbf{8.4070109767} &  & -9.61E-03 &  & -5.93E-03 &  & 1/10 &  & 10/10 &  & 32557.90 \\
303        & 7.9983093317 &  & \textbf{7.9871618864} & \textbf{} & \textbf{7.9891388900} &  & -1.11E-02 &  & -9.17E-03 &  & 1/10 &  & 10/10 &  & 32947.10 &  &  & 353 & 8.4188352568 &  & \textbf{8.4074380625} & \textbf{} & \textbf{8.4173578479} &  & -1.14E-02 &  & -1.48E-03 &  & 1/10 &  & 7/10  &  & 36663.40 \\
304        & 8.0074253659 &  & \textbf{7.9947232300} & \textbf{} & \textbf{7.9977950015} &  & -1.27E-02 &  & -9.63E-03 &  & 1/10 &  & 10/10 &  & 35196.30 &  &  & 354 & 8.4296585287 &  & \textbf{8.4076401001} & \textbf{} & \textbf{8.4212974949} &  & -2.20E-02 &  & -8.36E-03 &  & 1/10 &  & 9/10  &  & 31278.20 \\
305        & 8.0131815877 &  & \textbf{8.0046510979} & \textbf{} & \textbf{8.0066796544} &  & -8.53E-03 &  & -6.50E-03 &  & 1/10 &  & 10/10 &  & 31234.90 &  &  & 355 & 8.4382375759 &  & \textbf{8.4283461812} & \textbf{} & \textbf{8.4341418972} &  & -9.89E-03 &  & -4.10E-03 &  & 1/10 &  & 8/10  &  & 36232.90 \\
306        & 8.0213078802 &  & \textbf{8.0124605464} & \textbf{} & \textbf{8.0161690264} &  & -8.85E-03 &  & -5.14E-03 &  & 1/10 &  & 10/10 &  & 29711.90 &  &  & 356 & 8.4440909450 &  & \textbf{8.4267054911} & \textbf{} & \textbf{8.4383669270} &  & -1.74E-02 &  & -5.72E-03 &  & 1/10 &  & 7/10  &  & 35386.90 \\
307        & 8.0301086572 &  & \textbf{8.0226883734} & \textbf{} & \textbf{8.0247586057} &  & -7.42E-03 &  & -5.35E-03 &  & 1/10 &  & 10/10 &  & 31324.90 &  &  & 357 & 8.4518644222 &  & \textbf{8.4382769178} & \textbf{} & \textbf{8.4499586603} &  & -1.36E-02 &  & -1.91E-03 &  & 1/10 &  & 6/10  &  & 36650.80 \\
308        & 8.0380891717 &  & \textbf{8.0321834308} & \textbf{} & \textbf{8.0337171490} &  & -5.91E-03 &  & -4.37E-03 &  & 1/10 &  & 10/10 &  & 32555.50 &  &  & 358 & 8.4567949768 &  & \textbf{8.4490841743} & \textbf{} & \textbf{8.4532273856} &  & -7.71E-03 &  & -3.57E-03 &  & 1/10 &  & 8/10  &  & 33469.90 \\
309        & 8.0448618089 &  & \textbf{8.0411618615} & \textbf{} & \textbf{8.0441601989} &  & -3.70E-03 &  & -7.02E-04 &  & 1/10 &  & 6/10  &  & 29683.10 &  &  & 359 & 8.4642649260 &  & \textbf{8.4548490764} & \textbf{} & \textbf{8.4634558943} &  & -9.42E-03 &  & -8.09E-04 &  & 1/10 &  & 6/10  &  & 35939.40 \\
310        & 8.0548522945 &  & \textbf{8.0464675478} & \textbf{} & \textbf{8.0496913899} &  & -8.38E-03 &  & -5.16E-03 &  & 1/10 &  & 10/10 &  & 34729.10 &  &  & 360 & 8.4721516869 &  & \textbf{8.4651558516} & \textbf{} & 8.4726874357          &  & -7.00E-03 &  & 5.36E-04  &  & 1/10 &  & 5/10  &  & 33265.00 \\
311        & 8.0606039399 &  & \textbf{8.0560797869} & \textbf{} & \textbf{8.0581898041} &  & -4.52E-03 &  & -2.41E-03 &  & 1/10 &  & 10/10 &  & 31867.30 &  &  & 361 & 8.4805536014 &  & \textbf{8.4645834364} & \textbf{} & \textbf{8.4781535429} &  & -1.60E-02 &  & -2.40E-03 &  & 1/10 &  & 6/10  &  & 37982.70 \\
312        & 8.0669842552 &  & \textbf{8.0634879963} & \textbf{} & 8.0676502918          &  & -3.50E-03 &  & 6.66E-04  &  & 1/10 &  & 3/10  &  & 32574.10 &  &  & 362 & 8.4880679606 &  & \textbf{8.4788009361} & \textbf{} & 8.4882055053          &  & -9.27E-03 &  & 1.38E-04  &  & 1/10 &  & 3/10  &  & 36328.70 \\
313        & 8.0756527378 &  & \textbf{8.0723776565} & \textbf{} & 8.0765779098          &  & -3.28E-03 &  & 9.25E-04  &  & 1/10 &  & 5/10  &  & 32459.30 &  &  & 363 & 8.4965064991 &  & \textbf{8.4830627684} & \textbf{} & \textbf{8.4963311686} &  & -1.34E-02 &  & -1.75E-04 &  & 1/10 &  & 5/10  &  & 36888.10 \\
314        & 8.0892357495 &  & \textbf{8.0840444646} & \textbf{} & \textbf{8.0861982805} &  & -5.19E-03 &  & -3.04E-03 &  & 1/10 &  & 10/10 &  & 27667.50 &  &  & 364 & 8.5030156544 &  & \textbf{8.4936601180} & \textbf{} & \textbf{8.5015802313} &  & -9.36E-03 &  & -1.44E-03 &  & 1/10 &  & 6/10  &  & 41510.80 \\
315        & 8.1013721248 &  & \textbf{8.0833362248} & \textbf{} & \textbf{8.0917078266} &  & -1.80E-02 &  & -9.66E-03 &  & 1/10 &  & 9/10  &  & 35320.70 &  &  & 365 & 8.5120741140 &  & \textbf{8.4893661721} & \textbf{} & \textbf{8.5101067518} &  & -2.27E-02 &  & -1.97E-03 &  & 1/10 &  & 5/10  &  & 36513.90 \\
316        & 8.1113304407 &  & \textbf{8.0914221909} & \textbf{} & \textbf{8.0974417541} &  & -1.99E-02 &  & -1.39E-02 &  & 1/10 &  & 10/10 &  & 37634.80 &  &  & 366 & 8.5207116927 &  & \textbf{8.5142845330} & \textbf{} & \textbf{8.5203955423} &  & -6.43E-03 &  & -3.16E-04 &  & 1/10 &  & 6/10  &  & 38102.30 \\
317        & 8.1200535298 &  & \textbf{8.0985648551} & \textbf{} & \textbf{8.1080890701} &  & -2.15E-02 &  & -1.20E-02 &  & 1/10 &  & 10/10 &  & 34461.70 &  &  & 367 & 8.5271184405 &  & \textbf{8.5206107101} & \textbf{} & 8.5276469145          &  & -6.51E-03 &  & 5.28E-04  &  & 1/10 &  & 5/10  &  & 33376.70 \\
318        & 8.1282408602 &  & \textbf{8.1078206147} & \textbf{} & \textbf{8.1160501234} &  & -2.04E-02 &  & -1.22E-02 &  & 1/10 &  & 10/10 &  & 32015.10 &  &  & 368 & 8.5351429761 &  & \textbf{8.5308028271} & \textbf{} & 8.5388016906          &  & -4.34E-03 &  & 3.66E-03  &  & 1/10 &  & 1/10  &  & 29263.30 \\
319        & 8.1367023254 &  & \textbf{8.1191101842} & \textbf{} & \textbf{8.1287896608} &  & -1.76E-02 &  & -7.91E-03 &  & 1/10 &  & 10/10 &  & 28599.30 &  &  & 369 & 8.5433733193 &  & \textbf{8.5337041371} & \textbf{} & 8.5441638262          &  & -9.67E-03 &  & 7.91E-04  &  & 1/10 &  & 3/10  &  & 33058.30 \\
320        & 8.1466752518 &  & \textbf{8.1327700764} & \textbf{} & \textbf{8.1382270265} &  & -1.39E-02 &  & -8.45E-03 &  & 1/10 &  & 10/10 &  & 33270.80 &  &  & 370 & 8.5498955957 &  & \textbf{8.5477264184} & \textbf{} & 8.5535890629          &  & -2.17E-03 &  & 3.69E-03  &  & 1/10 &  & 1/10  &  & 38641.60 \\
321        & 8.1563823395 &  & \textbf{8.1397394926} & \textbf{} & \textbf{8.1460532738} &  & -1.66E-02 &  & -1.03E-02 &  & 1/10 &  & 10/10 &  & 35082.60 &  &  & 371 & 8.5580474842 &  & 8.5583387631          & \textbf{} & 8.5637206436          &  & 2.91E-04  &  & 5.67E-03  &  & 1/10 &  & 0/10  &  & 34721.10 \\
322        & 8.1650140248 &  & \textbf{8.1477840758} & \textbf{} & \textbf{8.1557411143} &  & -1.72E-02 &  & -9.27E-03 &  & 1/10 &  & 10/10 &  & 37939.20 &  &  & 372 & 8.5646395490 &  & \textbf{8.5618715156} & \textbf{} & 8.5688457203          &  & -2.77E-03 &  & 4.21E-03  &  & 1/10 &  & 2/10  &  & 38153.50 \\
323        & 8.1723756268 &  & \textbf{8.1587592542} & \textbf{} & \textbf{8.1635652644} &  & -1.36E-02 &  & -8.81E-03 &  & 1/10 &  & 10/10 &  & 26588.60 &  &  & 373 & 8.5718304380 &  & \textbf{8.5712024323} & \textbf{} & 8.5784071217          &  & -6.28E-04 &  & 6.58E-03  &  & 1/10 &  & 1/10  &  & 29532.80 \\
324        & 8.1796327858 &  & \textbf{8.1674085081} & \textbf{} & \textbf{8.1718142906} &  & -1.22E-02 &  & -7.82E-03 &  & 1/10 &  & 10/10 &  & 33026.40 &  &  & 374 & 8.5786934269 &  & \textbf{8.5774728308} & \textbf{} & 8.5851262942          &  & -1.22E-03 &  & 6.43E-03  &  & 1/10 &  & 1/10  &  & 34824.20 \\
325        & 8.1882217082 &  & \textbf{8.1733003234} & \textbf{} & \textbf{8.1805874525} &  & -1.49E-02 &  & -7.63E-03 &  & 1/10 &  & 9/10  &  & 33231.00 &  &  & 375 & 8.5855784311 &  & \textbf{8.5820685572} & \textbf{} & 8.5927704139          &  & -3.51E-03 &  & 7.19E-03  &  & 1/10 &  & 1/10  &  & 32022.70 \\
326        & 8.1975179838 &  & \textbf{8.1817407277} & \textbf{} & \textbf{8.1914547326} &  & -1.58E-02 &  & -6.06E-03 &  & 1/10 &  & 8/10  &  & 29824.20 &  &  & 376 & 8.5940896993 &  & 8.5962563229          & \textbf{} & 8.6019527602          &  & 2.17E-03  &  & 7.86E-03  &  & 1/10 &  & 0/10  &  & 36583.20 \\
327        & 8.2058236905 &  & \textbf{8.1880247293} & \textbf{} & \textbf{8.1979037392} &  & -1.78E-02 &  & -7.92E-03 &  & 1/10 &  & 9/10  &  & 34200.90 &  &  & 377 & 8.6005441929 &  & 8.6027978057          & \textbf{} & 8.6079290231          &  & 2.25E-03  &  & 7.38E-03  &  & 1/10 &  & 0/10  &  & 36643.40 \\
328        & 8.2136443647 &  & \textbf{8.1942670899} & \textbf{} & \textbf{8.2080642404} &  & -1.94E-02 &  & -5.58E-03 &  & 1/10 &  & 6/10  &  & 32810.00 &  &  & 378 & 8.6069492023 &  & 8.6075915129          & \textbf{} & 8.6167326917          &  & 6.42E-04  &  & 9.78E-03  &  & 1/10 &  & 0/10  &  & 36768.70 \\
329        & 8.2182864526 &  & \textbf{8.2060215573} & \textbf{} & \textbf{8.2168840449} &  & -1.23E-02 &  & -1.40E-03 &  & 1/10 &  & 6/10  &  & 29943.10 &  &  & 379 & 8.6149479886 &  & 8.6184729253          & \textbf{} & 8.6232130064          &  & 3.52E-03  &  & 8.27E-03  &  & 1/10 &  & 0/10  &  & 31997.50 \\
330        & 8.2298817657 &  & \textbf{8.2157405956} & \textbf{} & \textbf{8.2270903671} &  & -1.41E-02 &  & -2.79E-03 &  & 1/10 &  & 6/10  &  & 38672.90 &  &  & 380 & 8.6213515024 &  & \textbf{8.6167407193} & \textbf{} & 8.6297682259          &  & -4.61E-03 &  & 8.42E-03  &  & 1/10 &  & 1/10  &  & 30666.80 \\
331        & 8.2373988670 &  & \textbf{8.2204413597} & \textbf{} & \textbf{8.2322238821} &  & -1.70E-02 &  & -5.17E-03 &  & 1/10 &  & 6/10  &  & 30812.60 &  &  & 381 & 8.6290602650 &  & 8.6337617241          & \textbf{} & 8.6390157316          &  & 4.70E-03  &  & 9.96E-03  &  & 1/10 &  & 0/10  &  & 31068.60 \\
332        & 8.2447022853 &  & \textbf{8.2365077203} & \textbf{} & \textbf{8.2446694472} &  & -8.19E-03 &  & -3.28E-05 &  & 1/10 &  & 6/10  &  & 39201.80 &  &  & 382 & 8.6388782340 &  & 8.6408402938          & \textbf{} & 8.6459287777          &  & 1.96E-03  &  & 7.05E-03  &  & 1/10 &  & 0/10  &  & 37316.30 \\
333        & 8.2558752036 &  & \textbf{8.2402967266} & \textbf{} & \textbf{8.2509101559} &  & -1.56E-02 &  & -4.97E-03 &  & 1/10 &  & 10/10 &  & 36149.70 &  &  & 383 & 8.6458487389 &  & 8.6513501064          & \textbf{} & 8.6540511824          &  & 5.50E-03  &  & 8.20E-03  &  & 1/10 &  & 0/10  &  & 33845.70 \\
334        & 8.2641607509 &  & \textbf{8.2624688079} & \textbf{} & 8.2653822448          &  & -1.69E-03 &  & 1.22E-03  &  & 1/10 &  & 2/10  &  & 41347.50 &  &  & 384 & 8.6537986688 &  & \textbf{8.6534665127} & \textbf{} & 8.6610714907          &  & -3.32E-04 &  & 7.27E-03  &  & 1/10 &  & 1/10  &  & 34619.30 \\
335        & 8.2716242498 &  & \textbf{8.2550643788} & \textbf{} & \textbf{8.2687134927} &  & -1.66E-02 &  & -2.91E-03 &  & 1/10 &  & 5/10  &  & 30155.00 &  &  & 385 & 8.6597468731 &  & 8.6607058843          & \textbf{} & 8.6666686067          &  & 9.59E-04  &  & 6.92E-03  &  & 1/10 &  & 0/10  &  & 31808.30 \\
336        & 8.2787743871 &  & \textbf{8.2772630110} & \textbf{} & 8.2798230222          &  & -1.51E-03 &  & 1.05E-03  &  & 1/10 &  & 4/10  &  & 32964.40 &  &  & 386 & 8.6688960219 &  & 8.6727731360          & \textbf{} & 8.6772487889          &  & 3.88E-03  &  & 8.35E-03  &  & 1/10 &  & 0/10  &  & 33264.10 \\
337        & 8.2862241400 &  & \textbf{8.2702833068} & \textbf{} & \textbf{8.2858113782} &  & -1.59E-02 &  & -4.13E-04 &  & 1/10 &  & 3/10  &  & 32532.50 &  &  & 387 & 8.6743044156 &  & 8.6831299075          & \textbf{} & 8.6851125353          &  & 8.83E-03  &  & 1.08E-02  &  & 1/10 &  & 0/10  &  & 36257.30 \\
338        & 8.2937023722 &  & 8.2939873759          & \textbf{} & 8.2968423944          &  & 2.85E-04  &  & 3.14E-03  &  & 1/10 &  & 0/10  &  & 35175.30 &  &  & 388 & 8.6806940940 &  & 8.6825360520          & \textbf{} & 8.6900621836          &  & 1.84E-03  &  & 9.37E-03  &  & 1/10 &  & 0/10  &  & 32091.90 \\
339        & 8.3019941224 &  & \textbf{8.2976655457} & \textbf{} & 8.3041915428          &  & -4.33E-03 &  & 2.20E-03  &  & 1/10 &  & 3/10  &  & 37779.40 &  &  & 389 & 8.6873874968 &  & 8.6916892658          & \textbf{} & 8.6986627004          &  & 4.30E-03  &  & 1.13E-02  &  & 1/10 &  & 0/10  &  & 34963.00 \\
340        & 8.3104116379 &  & \textbf{8.3097262711} & \textbf{} & 8.3143372656          &  & -6.85E-04 &  & 3.93E-03  &  & 1/10 &  & 1/10  &  & 31288.50 &  &  & 390 & 8.6956660896 &  & 8.6976501640          & \textbf{} & 8.7057093833          &  & 1.98E-03  &  & 1.00E-02  &  & 1/10 &  & 0/10  &  & 40716.10 \\
341        & 8.3165790385 &  & \textbf{8.3099338147} & \textbf{} & 8.3200594131          &  & -6.65E-03 &  & 3.48E-03  &  & 1/10 &  & 1/10  &  & 41373.40 &  &  & 391 & 8.7030232271 &  & 8.7036860347          & \textbf{} & 8.7117841202          &  & 6.63E-04  &  & 8.76E-03  &  & 1/10 &  & 0/10  &  & 33504.30 \\
342        & 8.3227045402 &  & \textbf{8.3135903730} & \textbf{} & 8.3271377294          &  & -9.11E-03 &  & 4.43E-03  &  & 1/10 &  & 2/10  &  & 33025.80 &  &  & 392 & 8.7102535414 &  & \textbf{8.7075529636} & \textbf{} & 8.7194373948          &  & -2.70E-03 &  & 9.18E-03  &  & 1/10 &  & 1/10  &  & 33894.60 \\
343        & 8.3315888726 &  & \textbf{8.3277561232} & \textbf{} & 8.3366690796          &  & -3.83E-03 &  & 5.08E-03  &  & 1/10 &  & 1/10  &  & 33505.00 &  &  & 393 & 8.7212637605 &  & 8.7220259806          & \textbf{} & 8.7267001491          &  & 7.62E-04  &  & 5.44E-03  &  & 1/10 &  & 0/10  &  & 38376.70 \\
344        & 8.3411770056 &  & 8.3452642590          & \textbf{} & 8.3479053529          &  & 4.09E-03  &  & 6.73E-03  &  & 1/10 &  & 0/10  &  & 37478.70 &  &  & 394 & 8.7292656343 &  & 8.7304320881          & \textbf{} & 8.7352037357          &  & 1.17E-03  &  & 5.94E-03  &  & 1/10 &  & 0/10  &  & 32689.40 \\
345        & 8.3518472673 &  & \textbf{8.3420522869} & \textbf{} & 8.3520364567          &  & -9.79E-03 &  & 1.89E-04  &  & 1/10 &  & 5/10  &  & 34106.30 &  &  & 395 & 8.7348094475 &  & \textbf{8.7307593517} & \textbf{} & 8.7414685181          &  & -4.05E-03 &  & 6.66E-03  &  & 1/10 &  & 1/10  &  & 34992.80 \\
346        & 8.3591153674 &  & \textbf{8.3537759755} & \textbf{} & 8.3604435937          &  & -5.34E-03 &  & 1.33E-03  &  & 1/10 &  & 3/10  &  & 31978.60 &  &  & 396 & 8.7405814827 &  & \textbf{8.7383102973} & \textbf{} & 8.7473704141          &  & -2.27E-03 &  & 6.79E-03  &  & 1/10 &  & 1/10  &  & 37857.60 \\
347        & 8.3672718772 &  & \textbf{8.3627048726} & \textbf{} & 8.3674485285          &  & -4.57E-03 &  & 1.77E-04  &  & 1/10 &  & 7/10  &  & 37954.30 &  &  & 397 & 8.7466387404 &  & 8.7480616080          & \textbf{} & 8.7583995060          &  & 1.42E-03  &  & 1.18E-02  &  & 1/10 &  & 0/10  &  & 32398.40 \\
348        & 8.3745515295 &  & \textbf{8.3674634360} & \textbf{} & 8.3770975392          &  & -7.09E-03 &  & 2.55E-03  &  & 1/10 &  & 3/10  &  & 33932.60 &  &  & 398 & 8.7534203080 &  & 8.7595163352          & \textbf{} & 8.7635387623          &  & 6.10E-03  &  & 1.01E-02  &  & 1/10 &  & 0/10  &  & 36624.30 \\
349        & 8.3836313552 &  & \textbf{8.3746367225} & \textbf{} & 8.3862908966          &  & -8.99E-03 &  & 2.66E-03  &  & 1/10 &  & 2/10  &  & 37856.70 &  &  & 399 & 8.7595077675 &  & 8.7652265248          & \textbf{} & 8.7728449064          &  & 5.72E-03  &  & 1.33E-02  &  & 1/10 &  & 0/10  &  & 33664.50 \\
350        & 8.3928242707 &  & \textbf{8.3867647460} & \textbf{} & \textbf{8.3926355872} &  & -6.06E-03 &  & -1.89E-04 &  & 1/10 &  & 6/10  &  & 30606.70 &  &  & 400 & 8.7674798301 &  & 8.7683321754          & \textbf{} & 8.7798413330          &  & 8.52E-04  &  & 1.24E-02  &  & 1/10 &  & 0/10  &  & 36997.10 \\ \midrule
\#Improved &              &  & 48                    &           & 34                    &  &           &  &           &  &      &  &       &  &          &  &  &     &              &  & 29                    &           & 14                    &  &           &  &           &  &      &  &       &  &          \\
\#Equal    &              &  & ~~0                     &           & ~~0                     &  &           &  &           &  &      &  &       &  &          &  &  &     &              &  & ~~0                     &           & ~~0                     &  &           &  &           &  &      &  &       &  &          \\
\#Worse    &              &  & ~~2                     &           & 16                    &  &           &  &           &  &      &  &       &  &          &  &  &     &              &  & 21                    &           & 36                    &  &           &  &           &  &      &  &       &  &          \\ \bottomrule
\end{tabular}
}

\end{sidewaystable}
\clearpage

\textbf{On moderate I scale instances.} 
From Table~\ref{tb_cmp_best2} we obverse that SED obtains 87 improved, 8 equal and 5 worse results compared with $R^*$ for $101 \leq n \leq 200$. The $HR$ values of most results equal 1/10, which means obtaining the best result is complicated. Meanwhile, SED obtains 59 improved and 2 equal average results $R_{avg}$ out of the 100 instances, and the $RR$ values of many results are over 5/10, implying that SED can obtain an improved or equal result compared with $R^*$ more than 50\% probability in many instances. These experimental results demonstrate that SED also performs excellently on the moderate scale of PESS.

\textbf{On moderate II scale instances.} 
From Tables~\ref{tb_cmp_best3} and \ref{tb_cmp_best4}, we observe that SED obtains 168 improved and 32 worse results compared with $R^*$ for the 200 moderate scale instances with $201 \leq n \leq 400$. The $HR$ values of these results equal 1/10 except for 4 instances, which implies that obtaining the best value is extremely difficult with increasing the number of packing spheres. However, SED can always obtain an improved or equal result, it can be shown as 
the $RR$ values of a number of results equal 10/10.
Meanwhile, SED obtains 124 improved results in terms of $R_{avg}$, implying a strong searching ability of our algorithm. It is worth noting that there are 21 out of the 32 worse results located at the interval $[371, 400]$, this phenomenon probably is caused by the number of packing spheres being too large and the cut-off time of 12 hours can not support SED to obtain an improved result.

\begin{figure}[tb]
    \centering
    \begin{minipage}[b]{0.49\linewidth}
        \centering
        \subfloat[][$50 \leq n \leq 100$]{\includegraphics[width=1\linewidth]{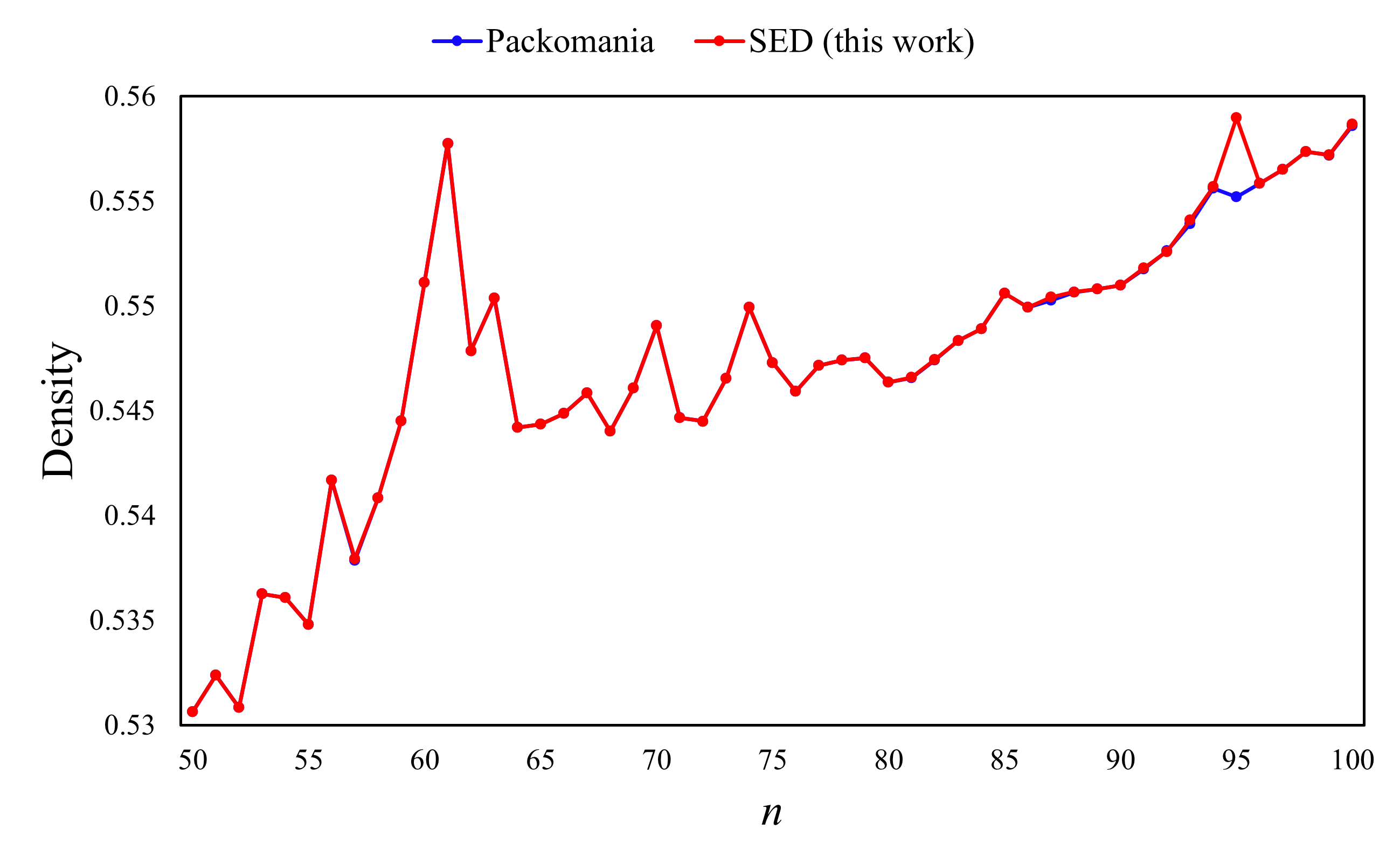}}
    \end{minipage}
    \begin{minipage}[b]{0.49\linewidth}
        \centering
        \subfloat[][$101 \leq n \leq 200$]{\includegraphics[width=1\linewidth]{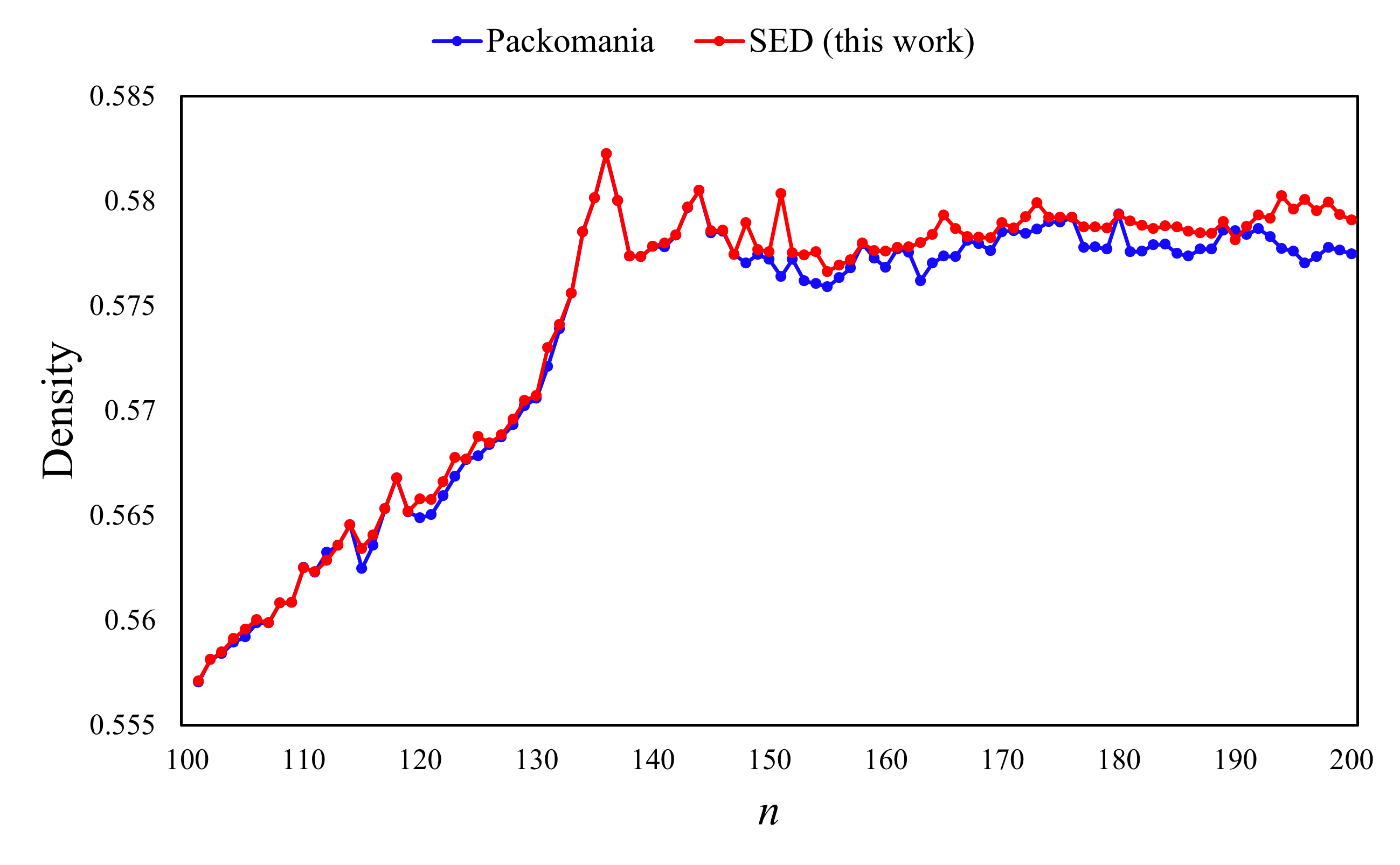}}
    \end{minipage}    
    \begin{minipage}[b]{0.49\linewidth}
        \centering
        \subfloat[][$201 \leq n \leq 300$]{\includegraphics[width=1\linewidth]{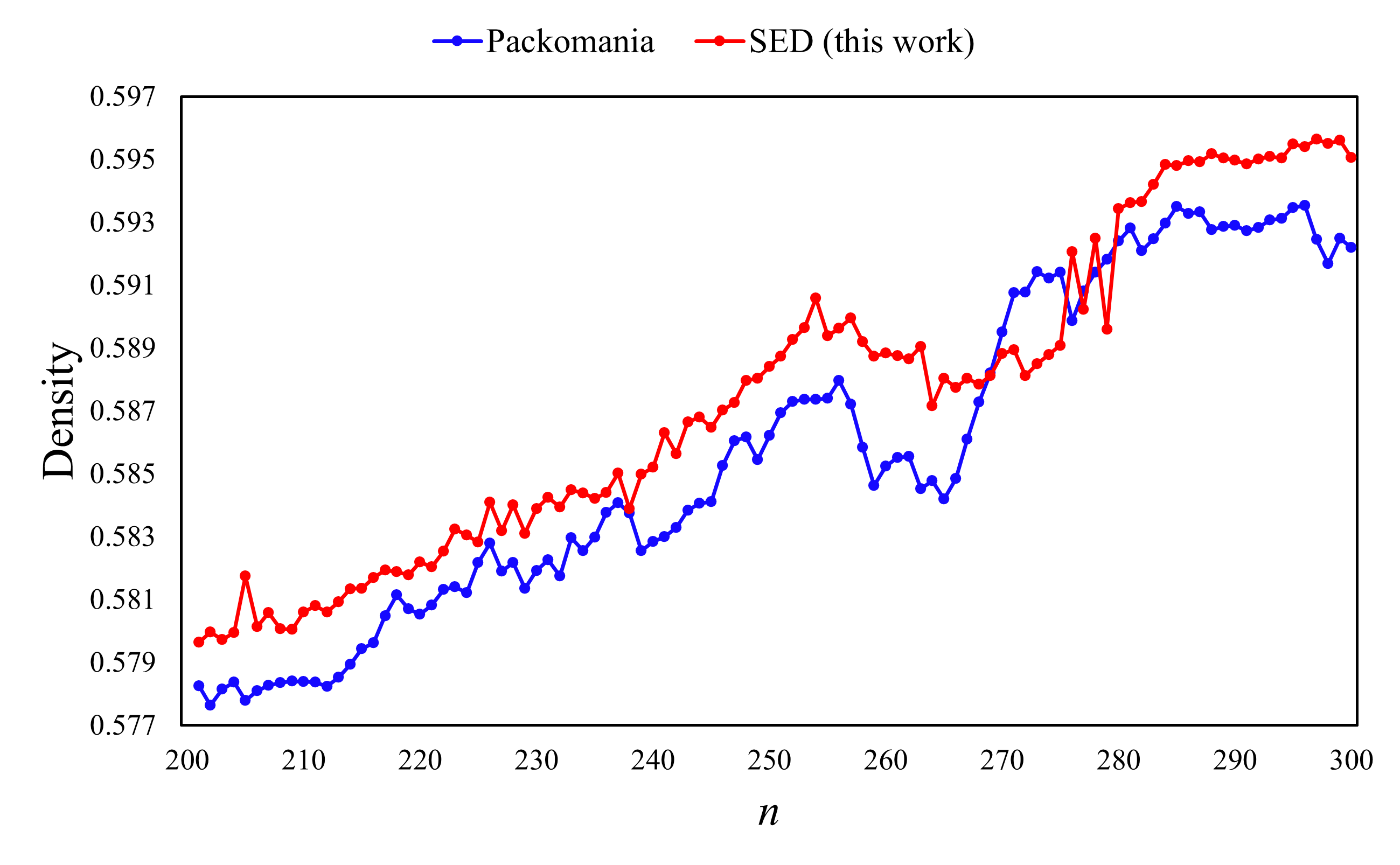}}
    \end{minipage}
    \begin{minipage}[b]{0.49\linewidth}
        \centering
        \subfloat[][$301 \leq n \leq 400$]{\includegraphics[width=1\linewidth]{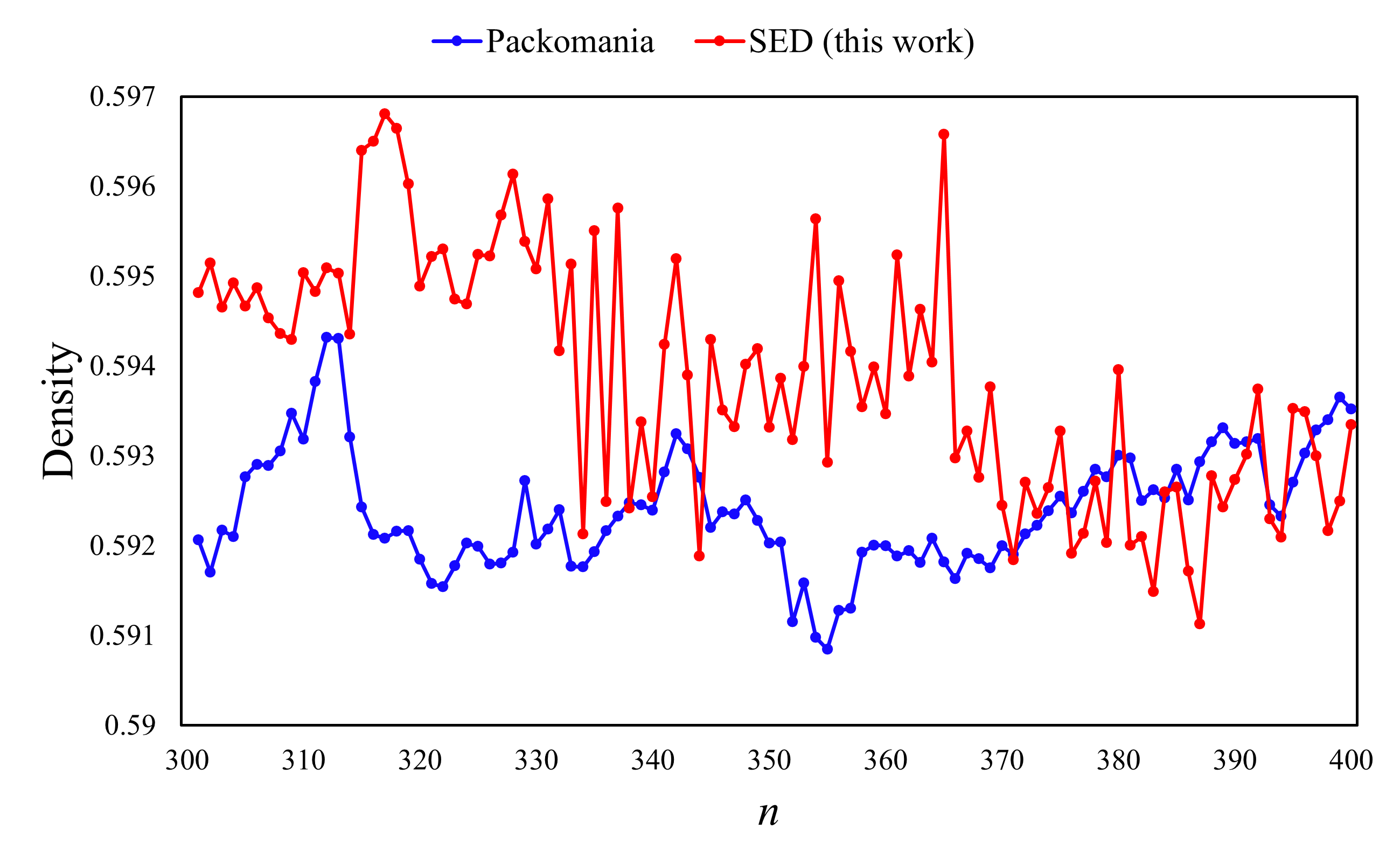}}
    \end{minipage}

    \caption{Packing density plots for the best-known results at Packomania
 and the best results of SED (this work).}
    \label{fig_cmp_density}
\end{figure}
\begin{table}[tb]
\centering
\caption{Summary of the computational results of our SED algorithm.}
\label{tb_cmp_sum}

\scalebox{0.9}{
\begin{tabular}{lllllllllll}
\toprule
           & \multicolumn{8}{l}{$n$}                                                                                                         &       &         \\ \cline{2-9}
           & $[5, 50]$ & $[51, 100]$ & $[101, 150]$ & $[151, 200]$ & $[201, 250]$ & $[251, 300]$ & $[301, 350]$ & $[351, 400]$ & Total & Ratio   \\ \midrule
\#Improved & ~~0           & 19            & 39             & 48            & 50            & 41            & 48            & 29             & 274   & 69.19\% \\
\#Equal    & 46          & 30            & ~~8              & ~~0             & ~~0             & ~~0             & ~~0             & ~~0              & ~~84    & 21.21\% \\
\#Worse    & ~~0           & ~~1             & ~~3              & ~~2             & ~~0             & ~~9             & ~~2             & 21             & ~~38    & ~~9.60\%  \\ \bottomrule
\end{tabular}
}
\end{table}

In addition, the packing density plots of the results of Packomania and SED with $50 \leq n \leq 400$ for the intuitive comparing purpose are shown in Figure~\ref{fig_cmp_density}. And the packing density $\rho$ of a feasible configuration is formulated in Eq.~\ref{eq_density}. Figure~\ref{fig_cmp_density} shows the four plots corresponding to the four different scales. The interval $[5, 49]$ is omitted because the results of Packomania and SED are the same in this interval. In each plot, the X-axis indicates the number of packing spheres, and the Y-axis indicates the packing density of a feasible configuration. Note that the higher density indicates a better configuration. From the four plots, we obverse that SED gains the obviously higher density in many instances for $n \geq 140$.

Furthermore, we summarize the results achieved by our SED algorithm on the eight intervals, presented in Table~\ref{tb_cmp_sum}. In summary, SED yields 274 improved, 84 equal and 38 worse results out of the 396 instances compared with the best-known results recorded at Packomania, and the corresponding ratios are 69.19\%, 21.21\% and 9.60\%, respectively. 
The computational results and comparisons we presented and discussed above demonstrate the excellent performance of our SED algorithm which is a powerful and efficient local search heuristic algorithm for solving the PESS problem. 

\begin{figure}[tb]
    \centering
    \begin{minipage}[b]{0.33\linewidth}
        \centering
        \subfloat[][$n = 57$]{\includegraphics[width=1\linewidth]{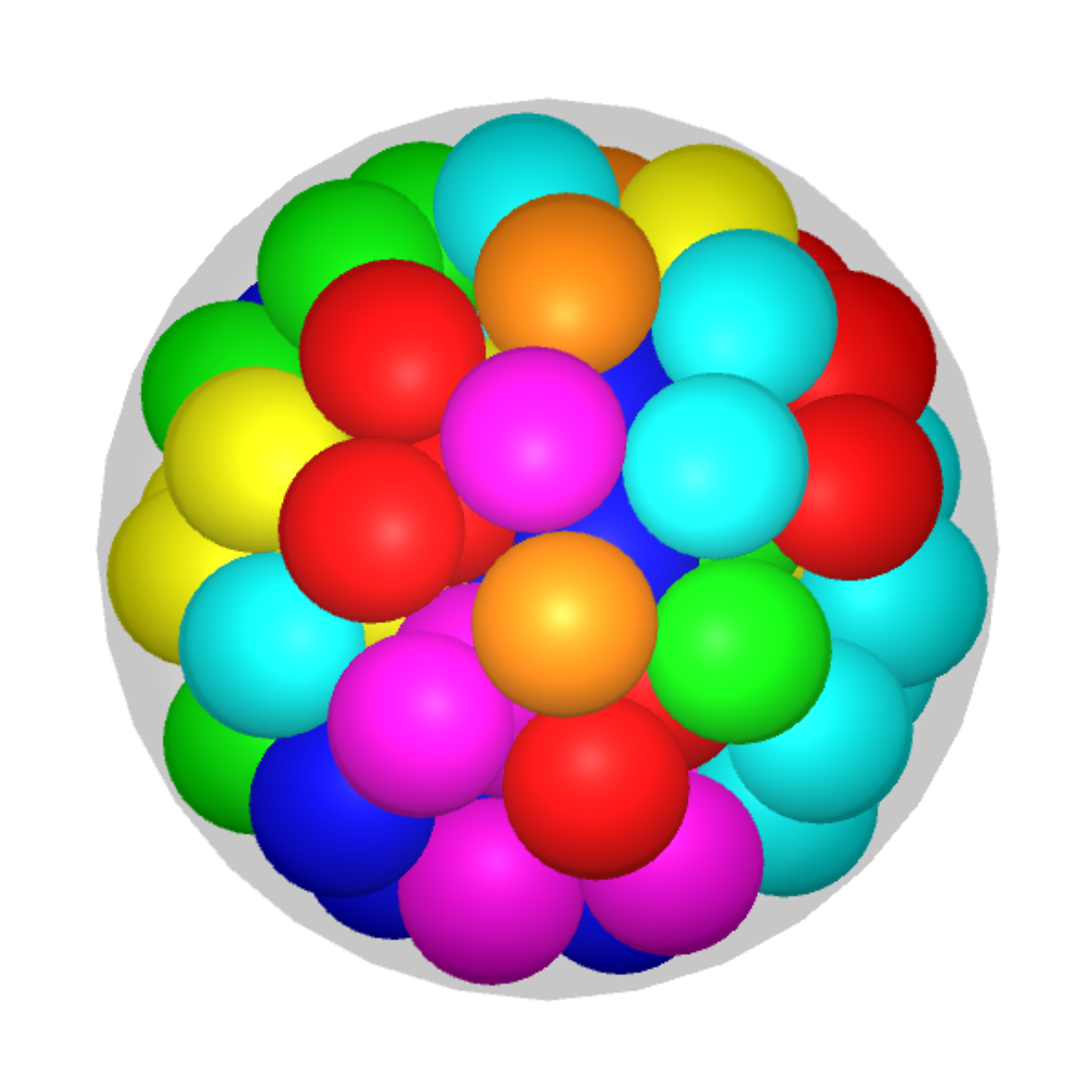}}
    \end{minipage}
    \begin{minipage}[b]{0.33\linewidth}
        \centering
        \subfloat[][$n = 73$]{\includegraphics[width=1\linewidth]{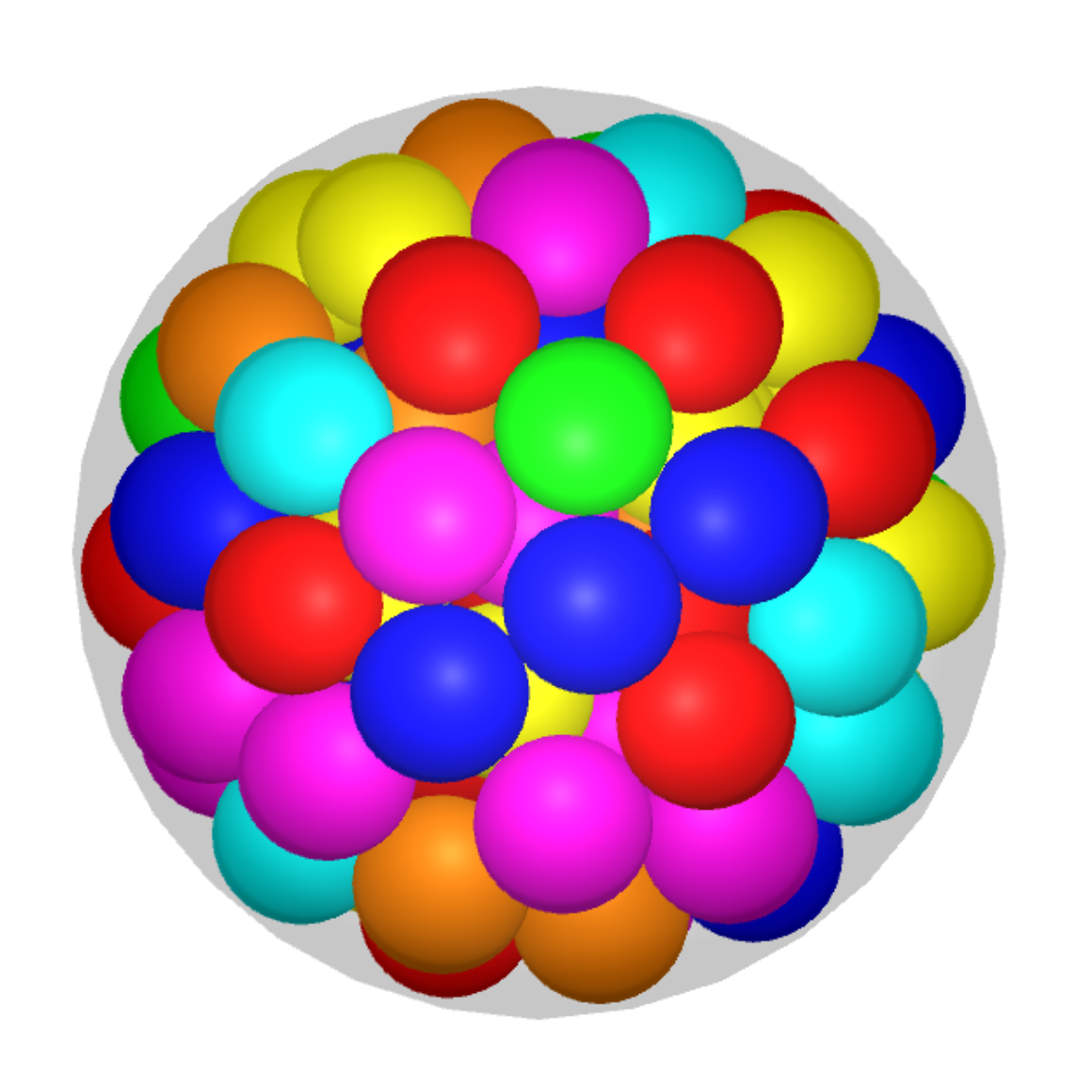}}
    \end{minipage}    
    \begin{minipage}[b]{0.33\linewidth}
        \centering
        \subfloat[][$n = 76$]{\includegraphics[width=1\linewidth]{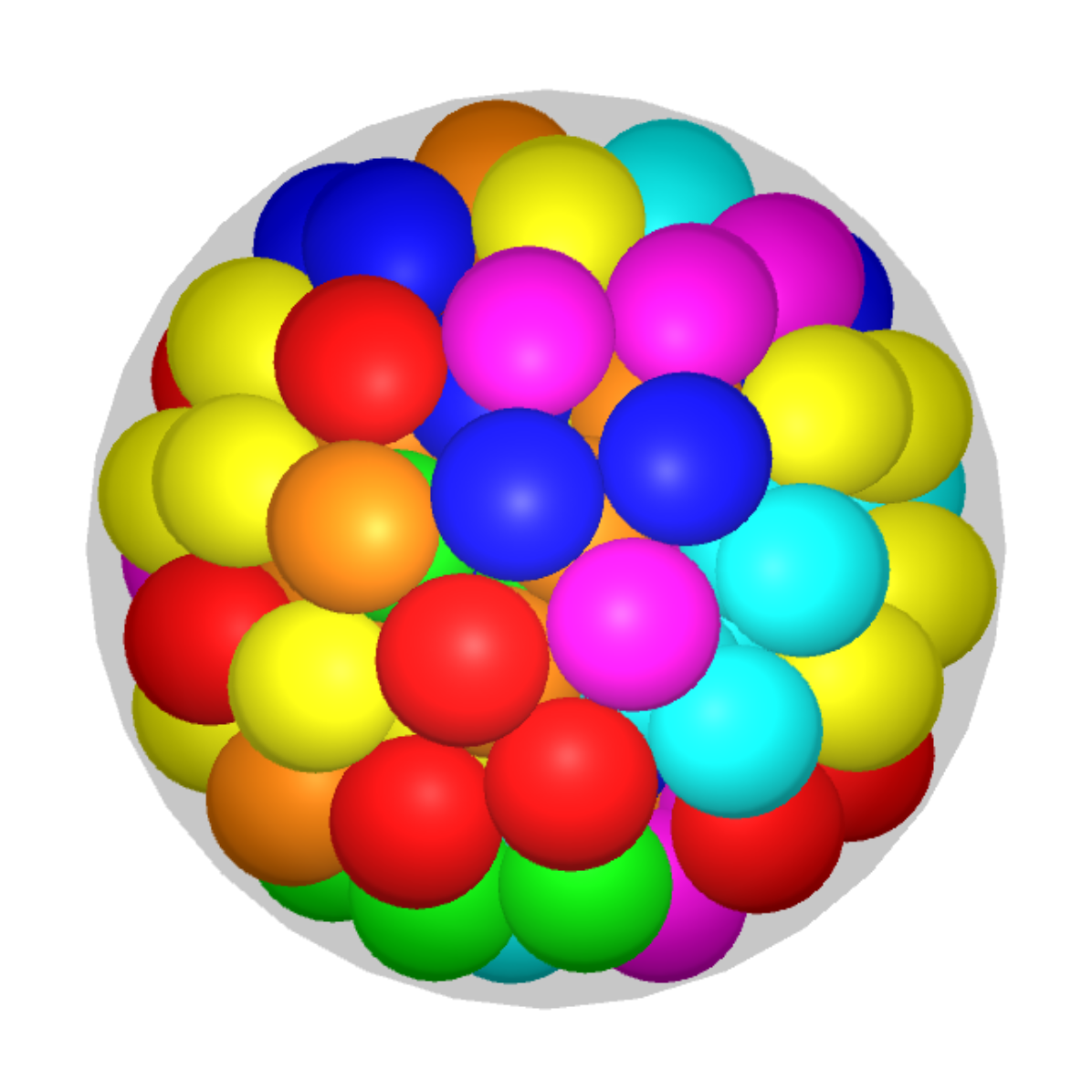}}
    \end{minipage}
    \begin{minipage}[b]{0.33\linewidth}
        \centering
        \subfloat[][$n = 82$]{\includegraphics[width=1\linewidth]{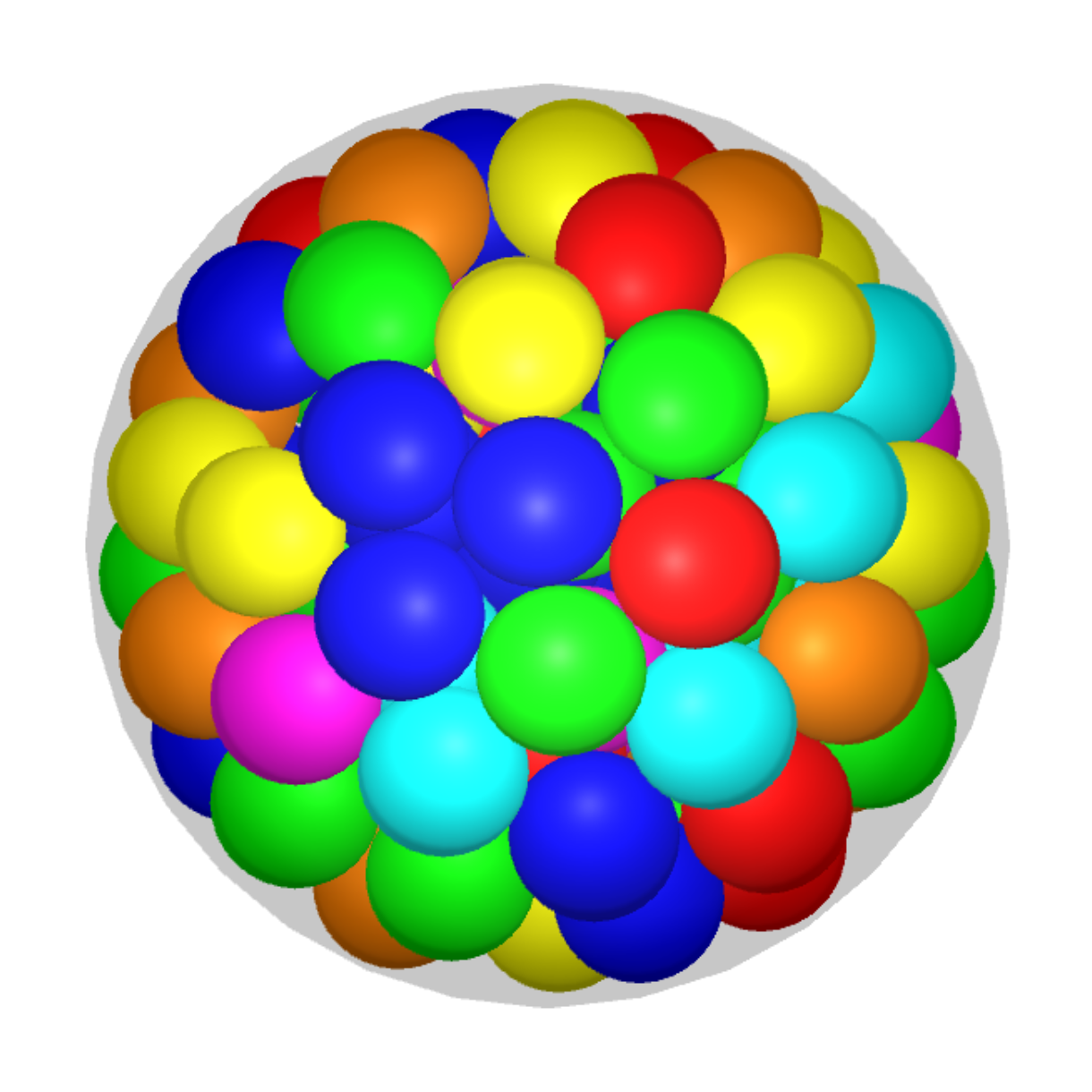}}
    \end{minipage}    
    \begin{minipage}[b]{0.33\linewidth}
        \centering
        \subfloat[][$n = 84$]{\includegraphics[width=1\linewidth]{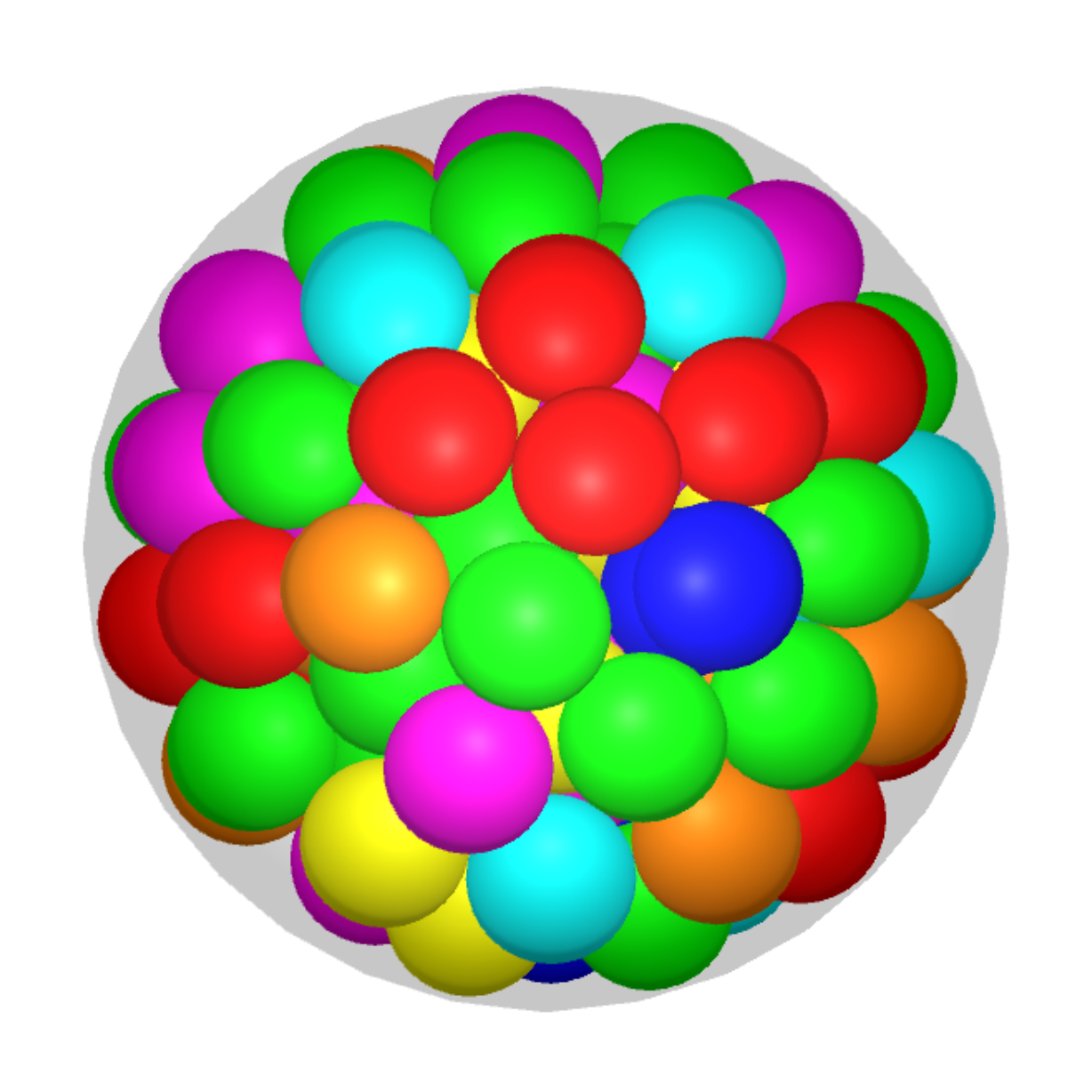}}
    \end{minipage}
    \begin{minipage}[b]{0.33\linewidth}
        \centering
        \subfloat[][$n = 87$]{\includegraphics[width=1\linewidth]{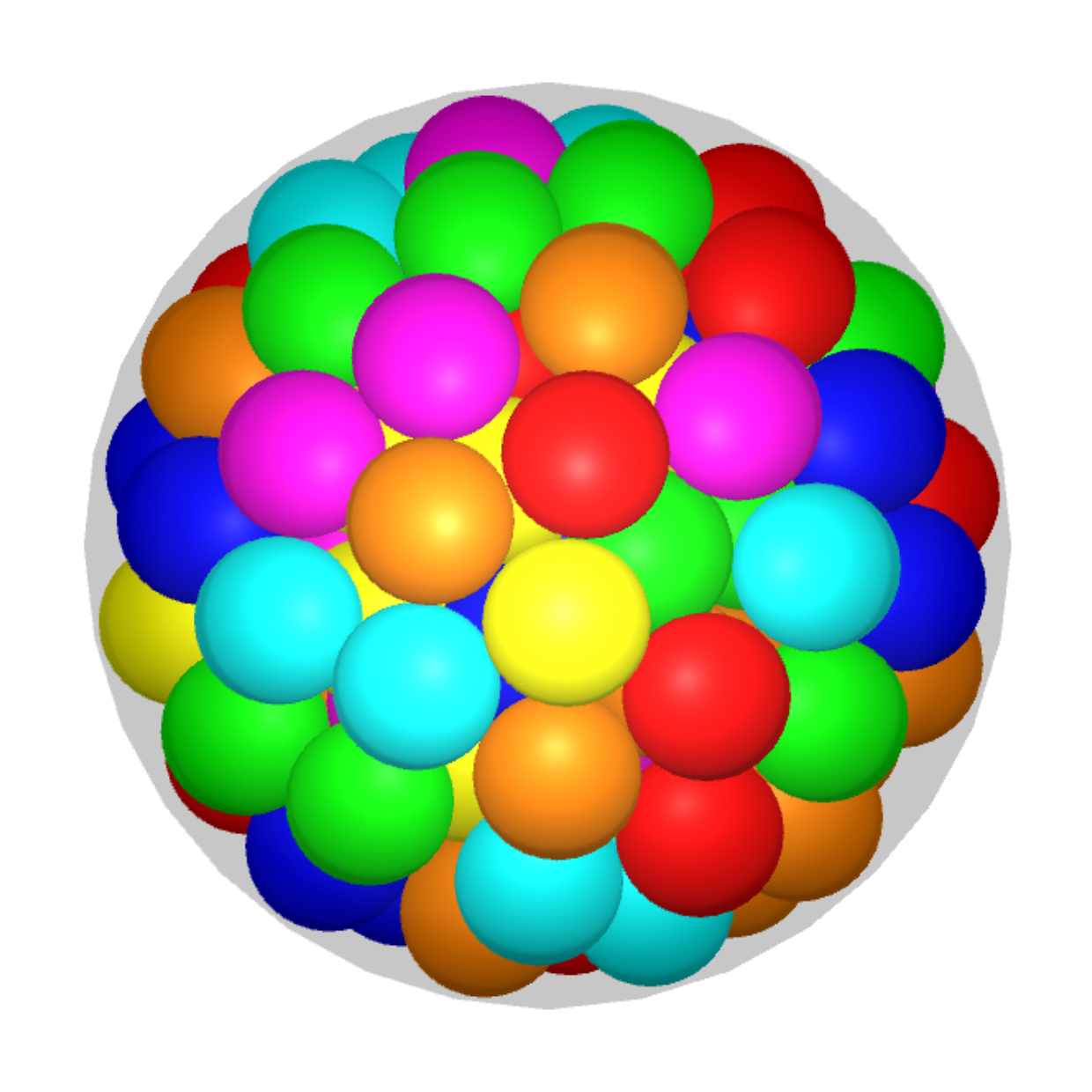}}
    \end{minipage}
    \begin{minipage}[b]{0.33\linewidth}
        \centering
        \subfloat[][$n = 93$]{\includegraphics[width=1\linewidth]{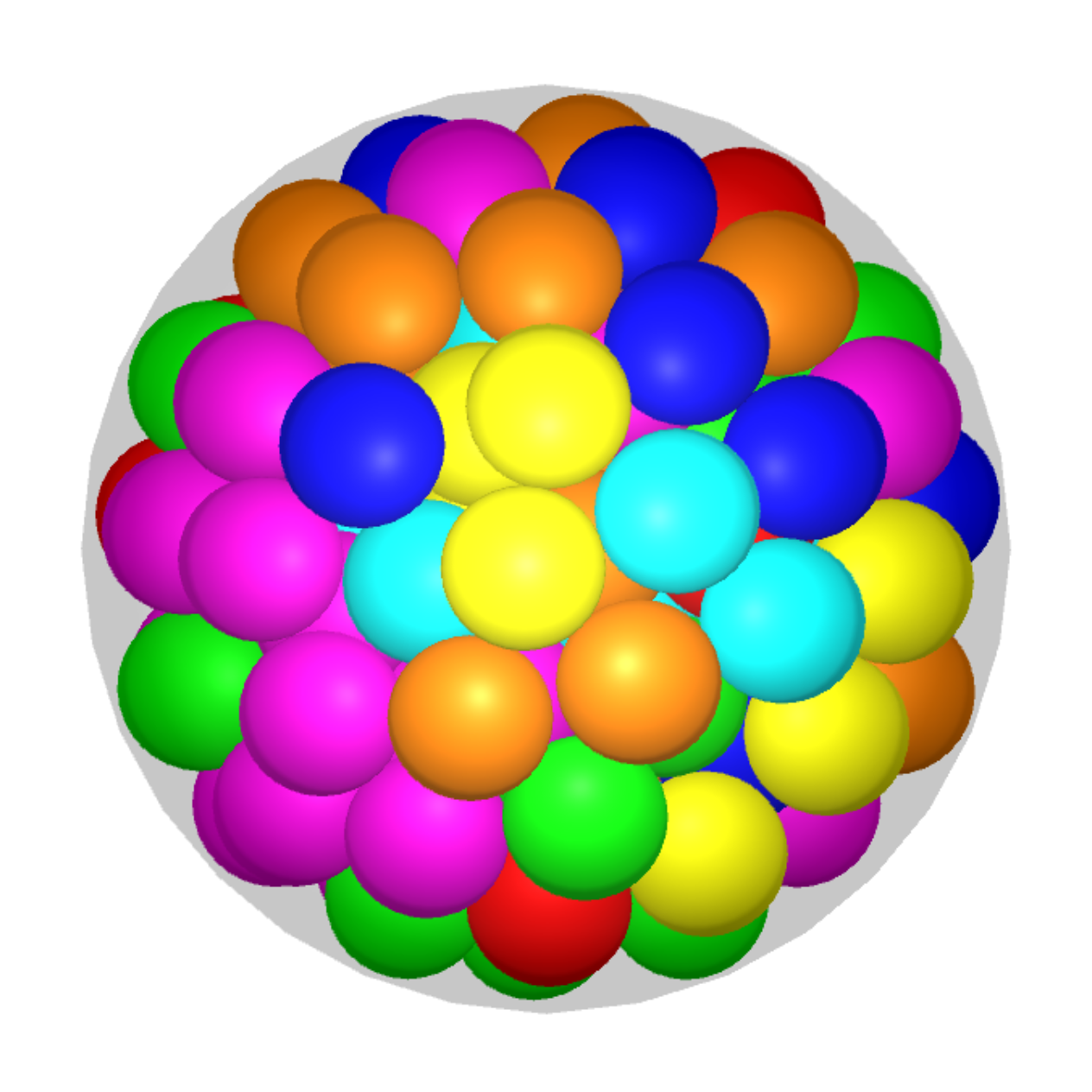}}
    \end{minipage}
    \begin{minipage}[b]{0.33\linewidth}
        \centering
        \subfloat[][$n = 95$]{\includegraphics[width=1\linewidth]{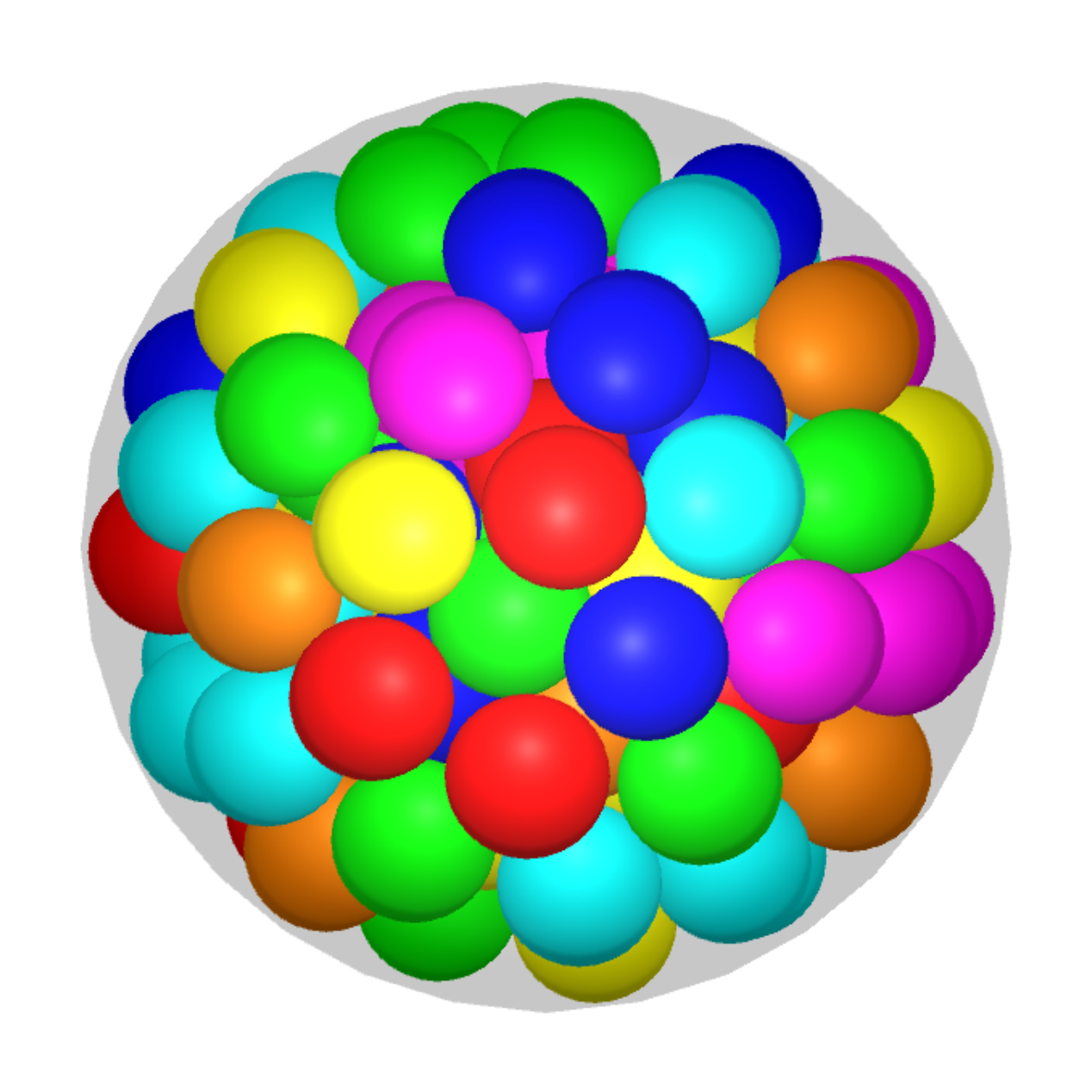}}
    \end{minipage}    
    \begin{minipage}[b]{0.33\linewidth}
        \centering
        \subfloat[][$n = 100$]{\includegraphics[width=1\linewidth]{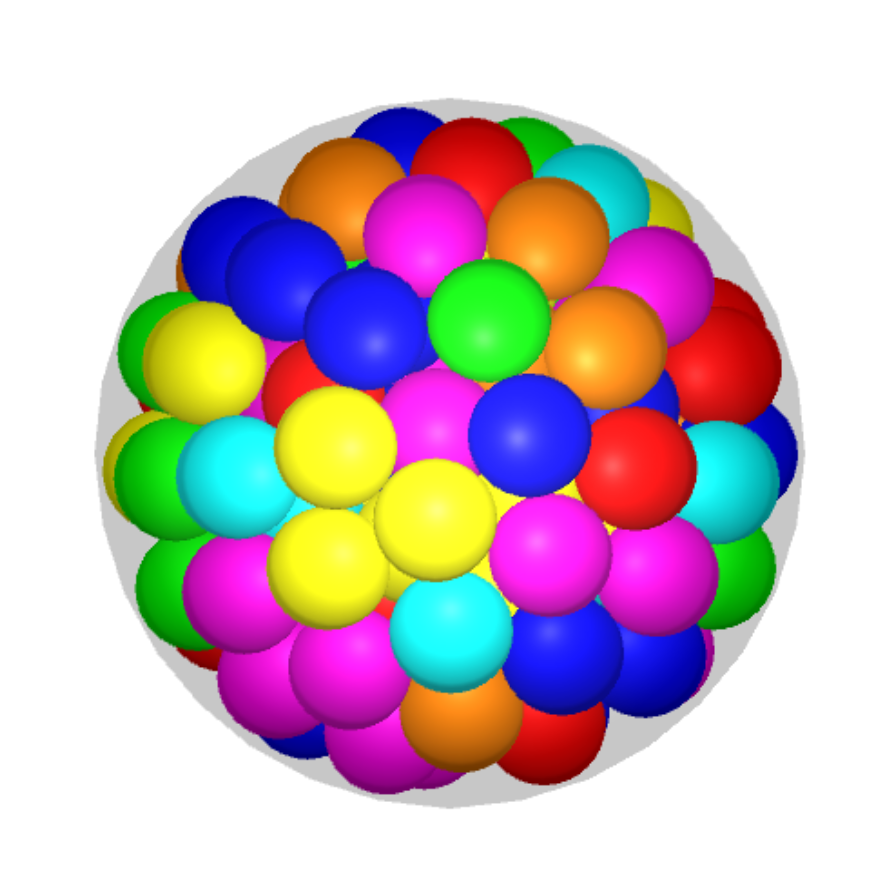}}
    \end{minipage}
    
    \caption{New improved solutions found by our algorithm sampled from the small scale instances $50 \leq n \leq 100$.}
    \label{fig_results_s}
\end{figure}

\begin{figure}[tb]
    \centering
    \begin{minipage}[b]{0.33\linewidth}
        \centering
        \subfloat[][$n = 105$]{\includegraphics[width=1\linewidth]{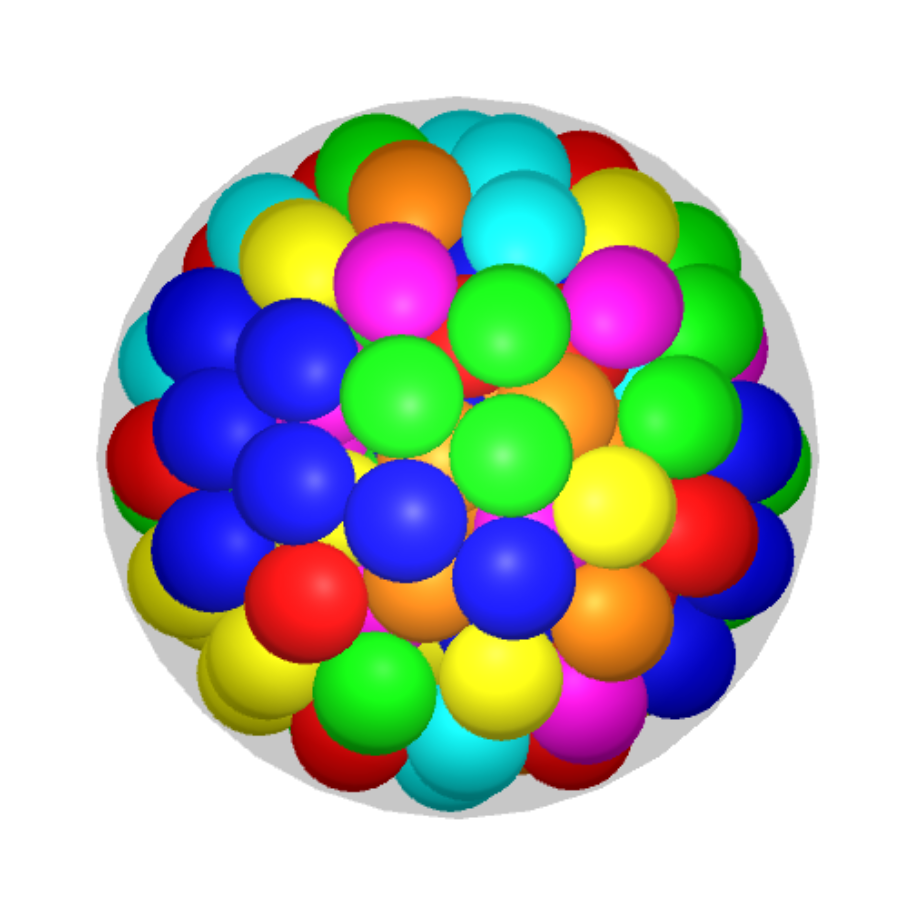}}
    \end{minipage}
    \begin{minipage}[b]{0.33\linewidth}
        \centering
        \subfloat[][$n = 122$]{\includegraphics[width=1\linewidth]{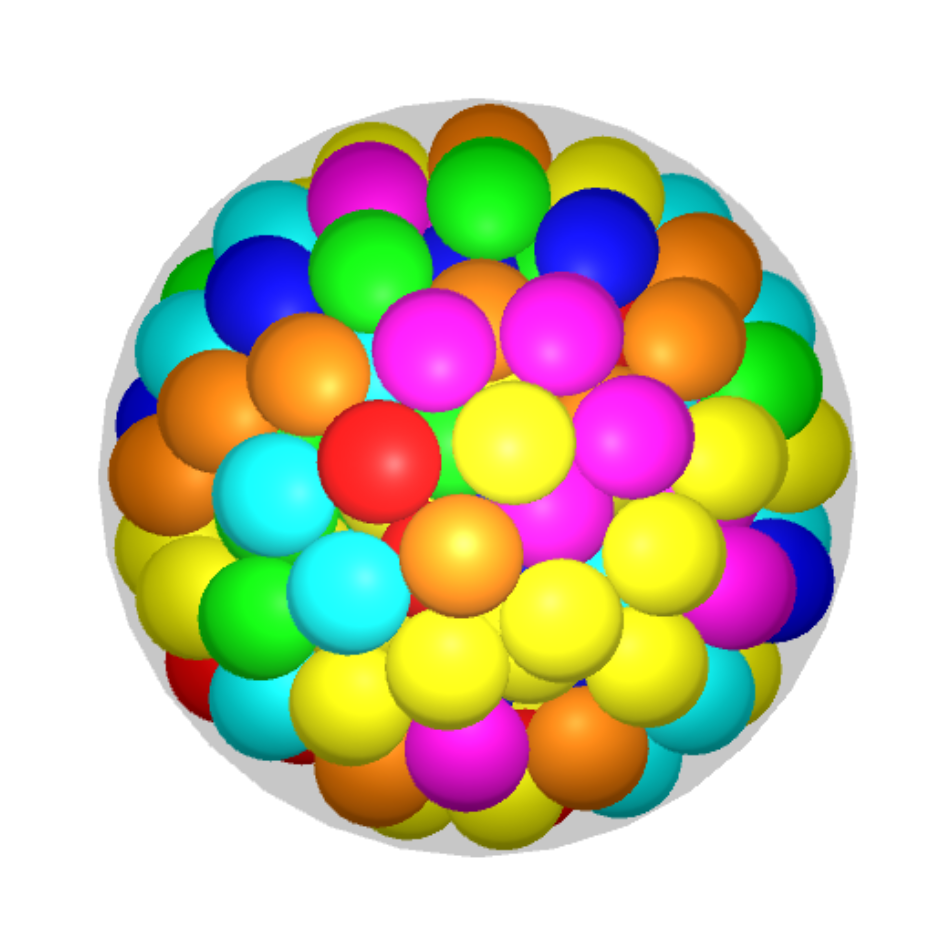}}
    \end{minipage} 
    \begin{minipage}[b]{0.33\linewidth}
        \centering
        \subfloat[][$n = 131$]{\includegraphics[width=1\linewidth]{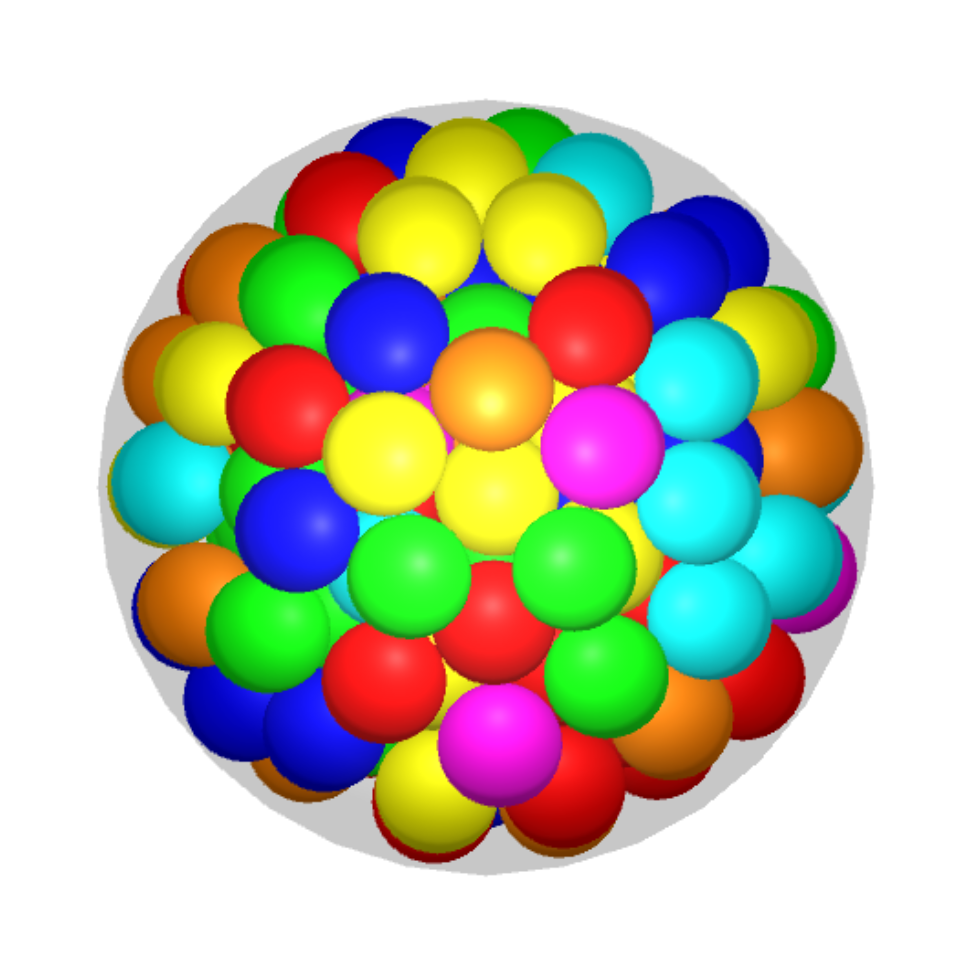}}
    \end{minipage}
    \begin{minipage}[b]{0.33\linewidth}
        \centering
        \subfloat[][$n = 148$]{\includegraphics[width=1\linewidth]{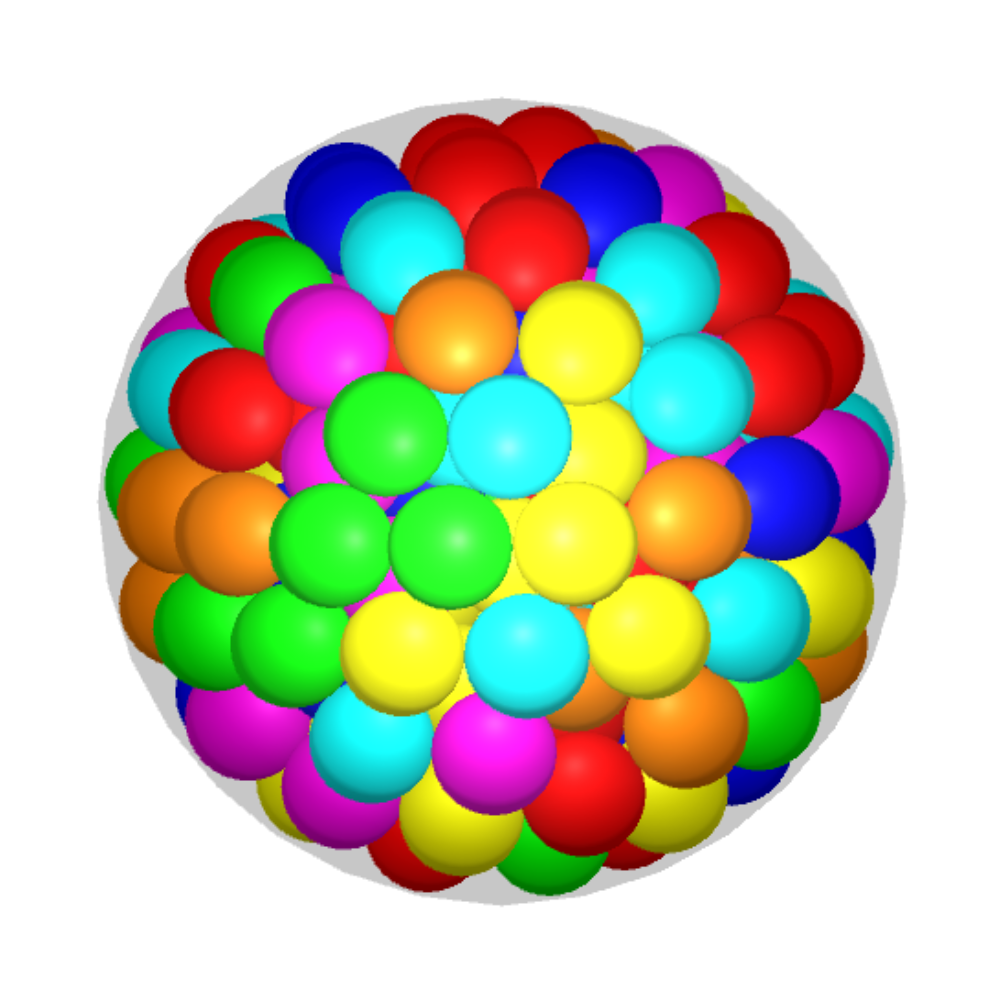}}
    \end{minipage}    
    \begin{minipage}[b]{0.33\linewidth}
        \centering
        \subfloat[][$n = 151$]{\includegraphics[width=1\linewidth]{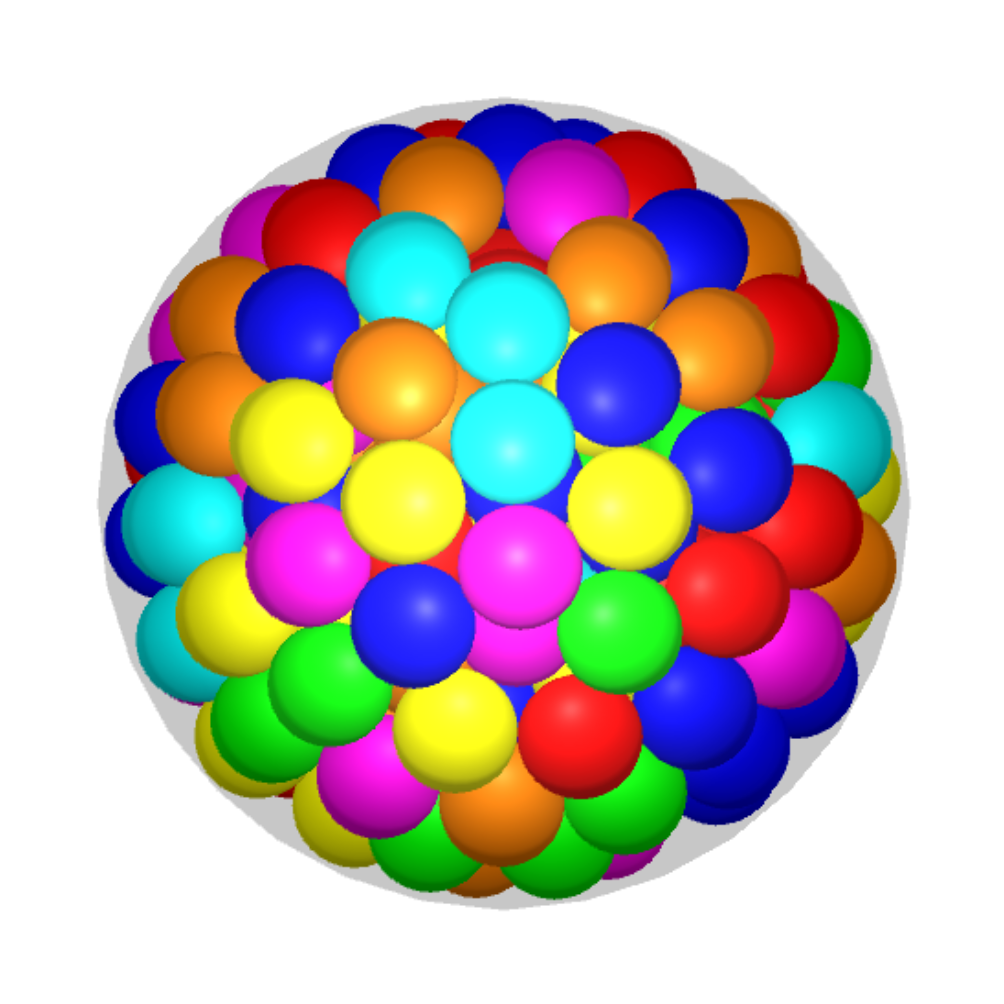}}
    \end{minipage}
    \begin{minipage}[b]{0.33\linewidth}
        \centering
        \subfloat[][$n = 166$]{\includegraphics[width=1\linewidth]{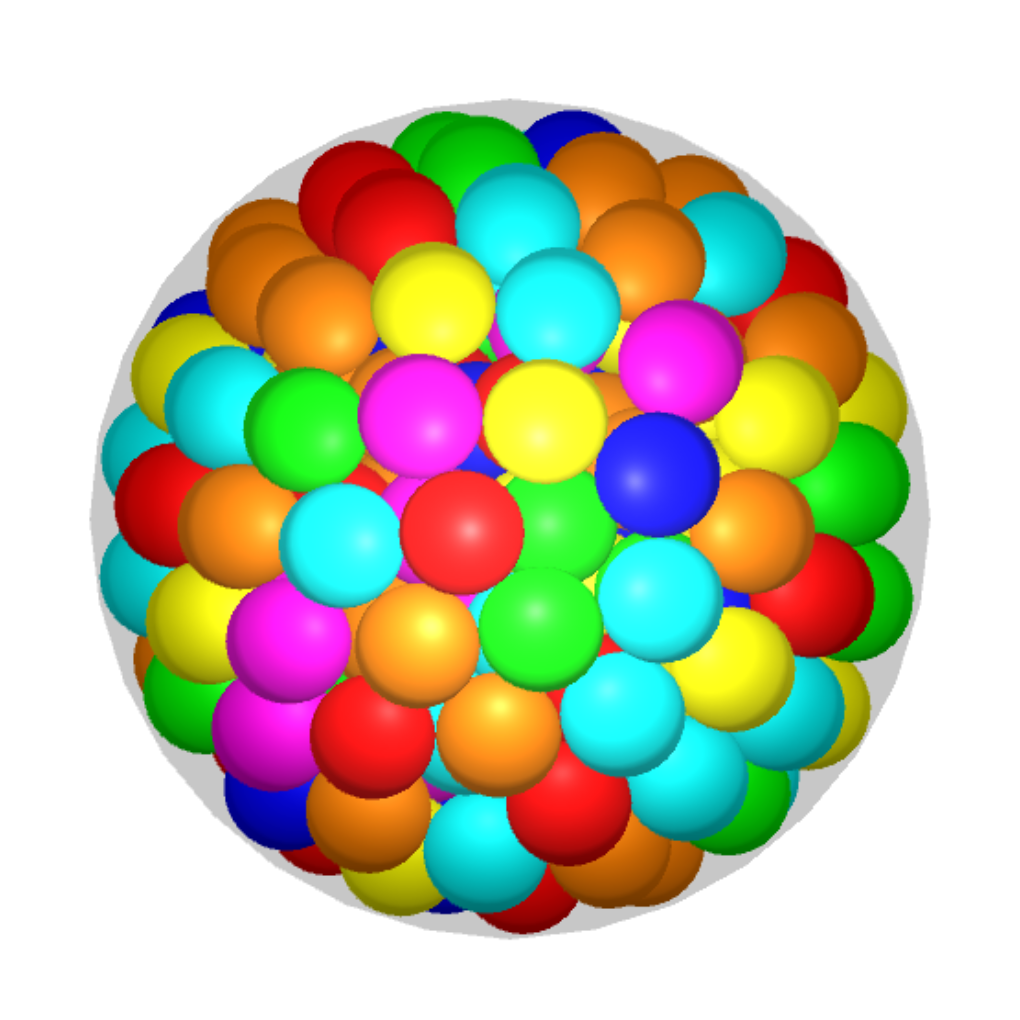}}
    \end{minipage}
    \begin{minipage}[b]{0.33\linewidth}
        \centering
        \subfloat[][$n = 173$]{\includegraphics[width=1\linewidth]{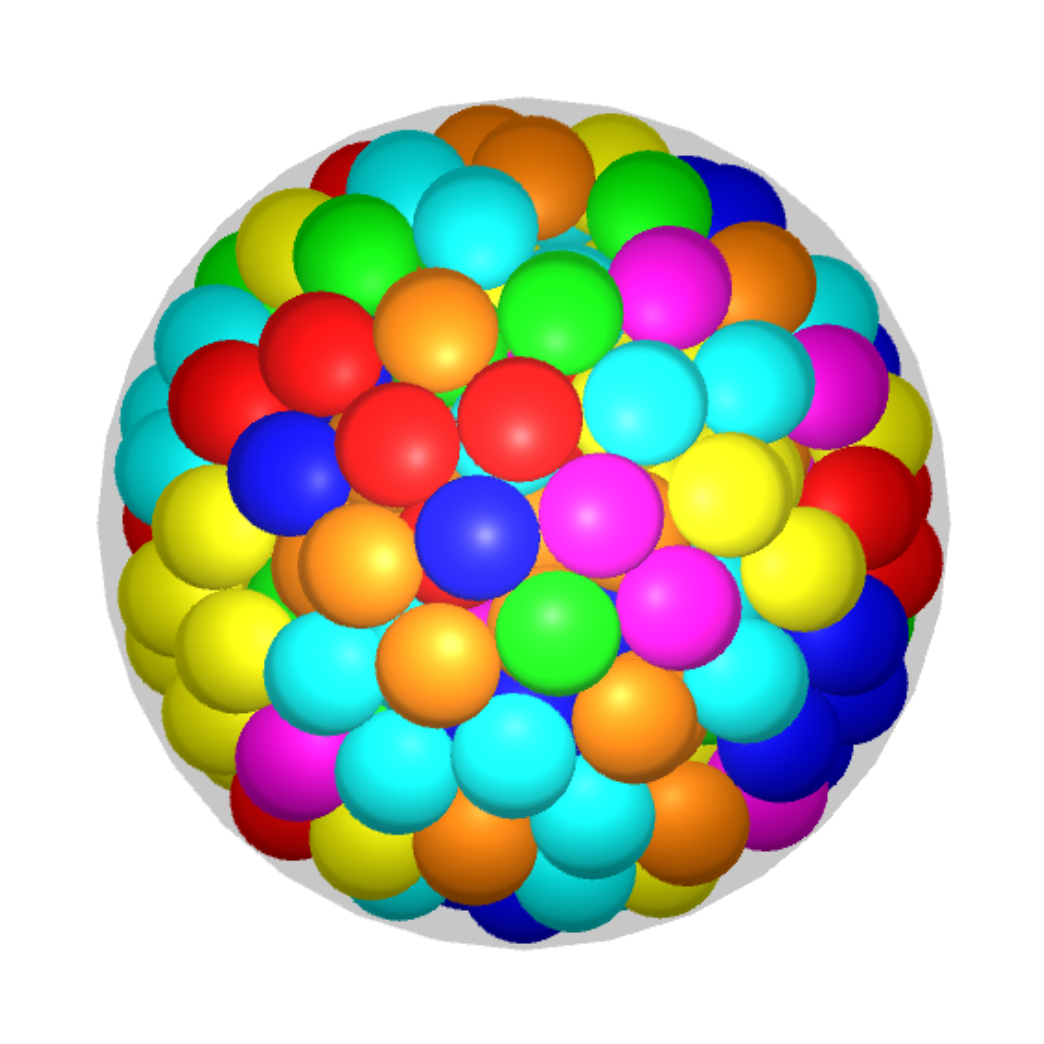}}
    \end{minipage}
    \begin{minipage}[b]{0.33\linewidth}
        \centering
        \subfloat[][$n = 183$]{\includegraphics[width=1\linewidth]{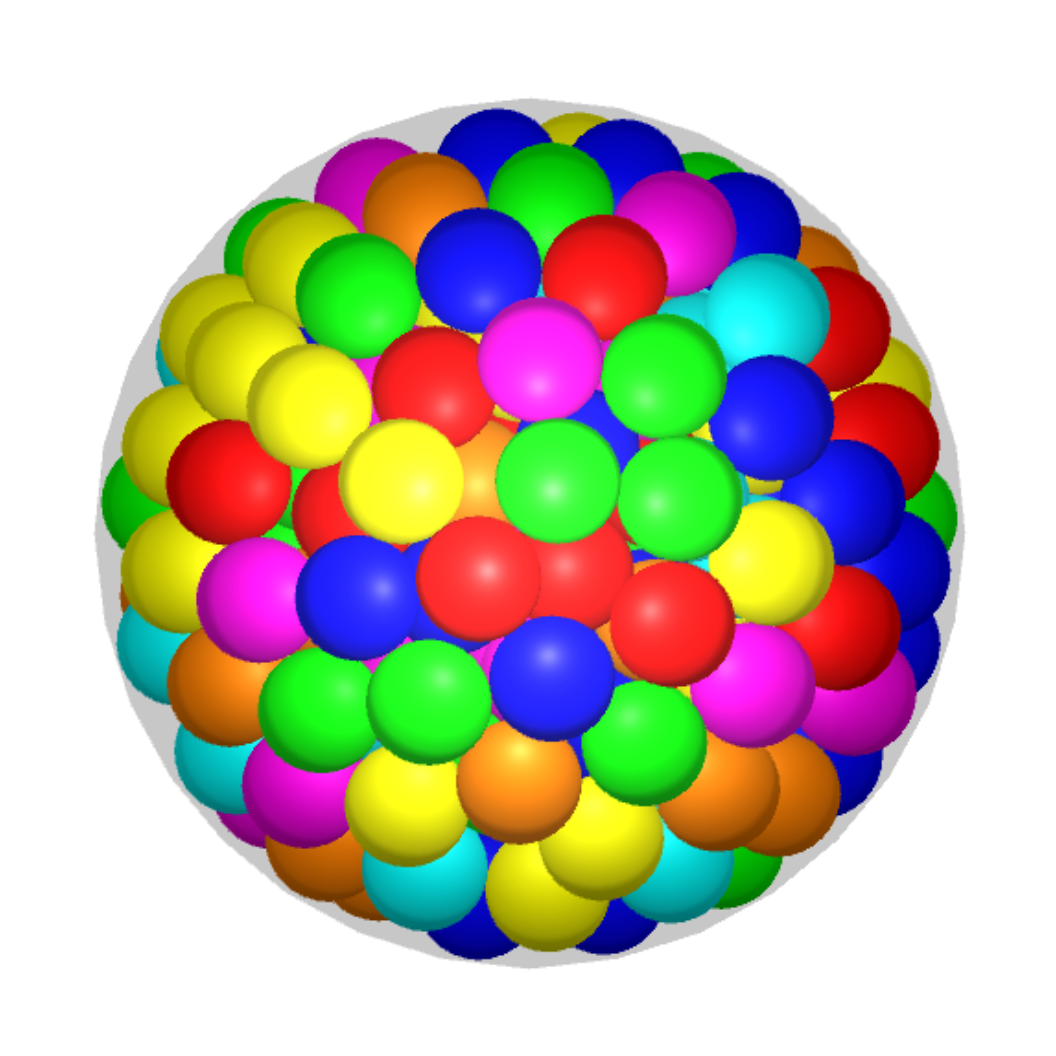}}
    \end{minipage}    
    \begin{minipage}[b]{0.33\linewidth}
        \centering
        \subfloat[][$n = 196$]{\includegraphics[width=1\linewidth]{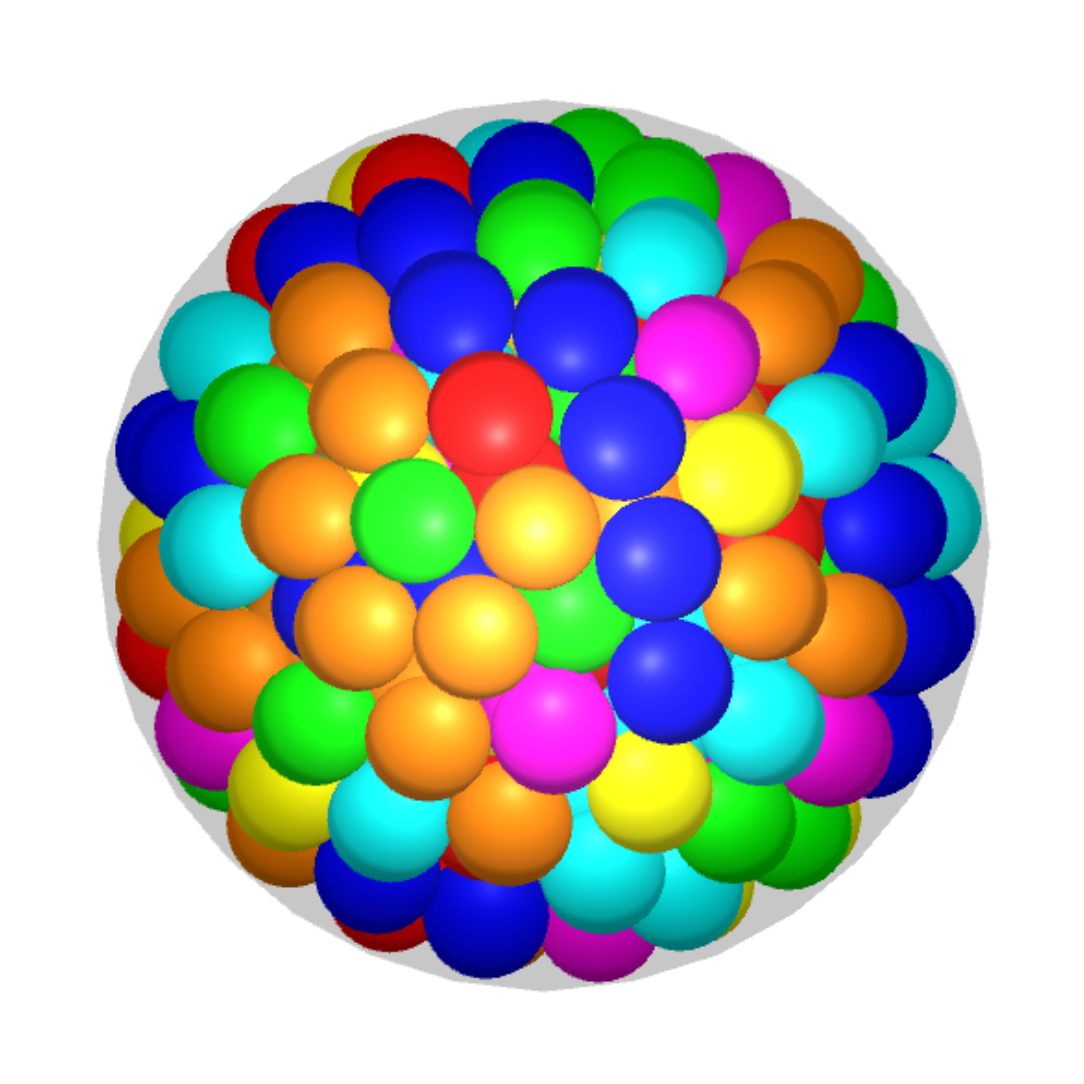}}
    \end{minipage}
    
    \caption{New improved solutions found by our algorithm sampled from the moderate I scale instances $101 \leq n \leq 200$.}
    \label{fig_results_m1}
\end{figure}

\begin{figure}[tb]
    \centering
    \begin{minipage}[b]{0.33\linewidth}
        \centering
        \subfloat[][$n = 205$]{\includegraphics[width=1\linewidth]{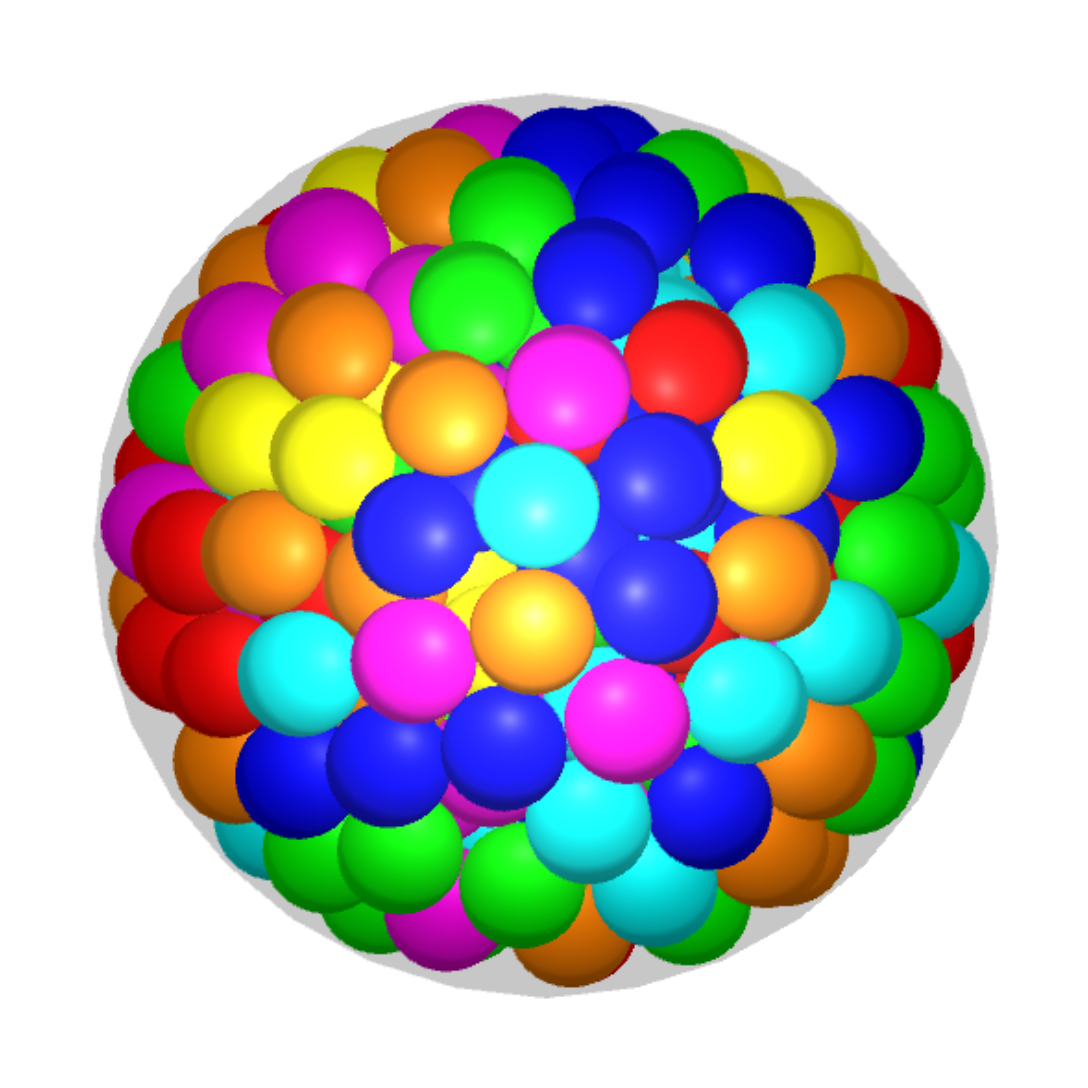}}
    \end{minipage}
    \begin{minipage}[b]{0.33\linewidth}
        \centering
        \subfloat[][$n = 230$]{\includegraphics[width=1\linewidth]{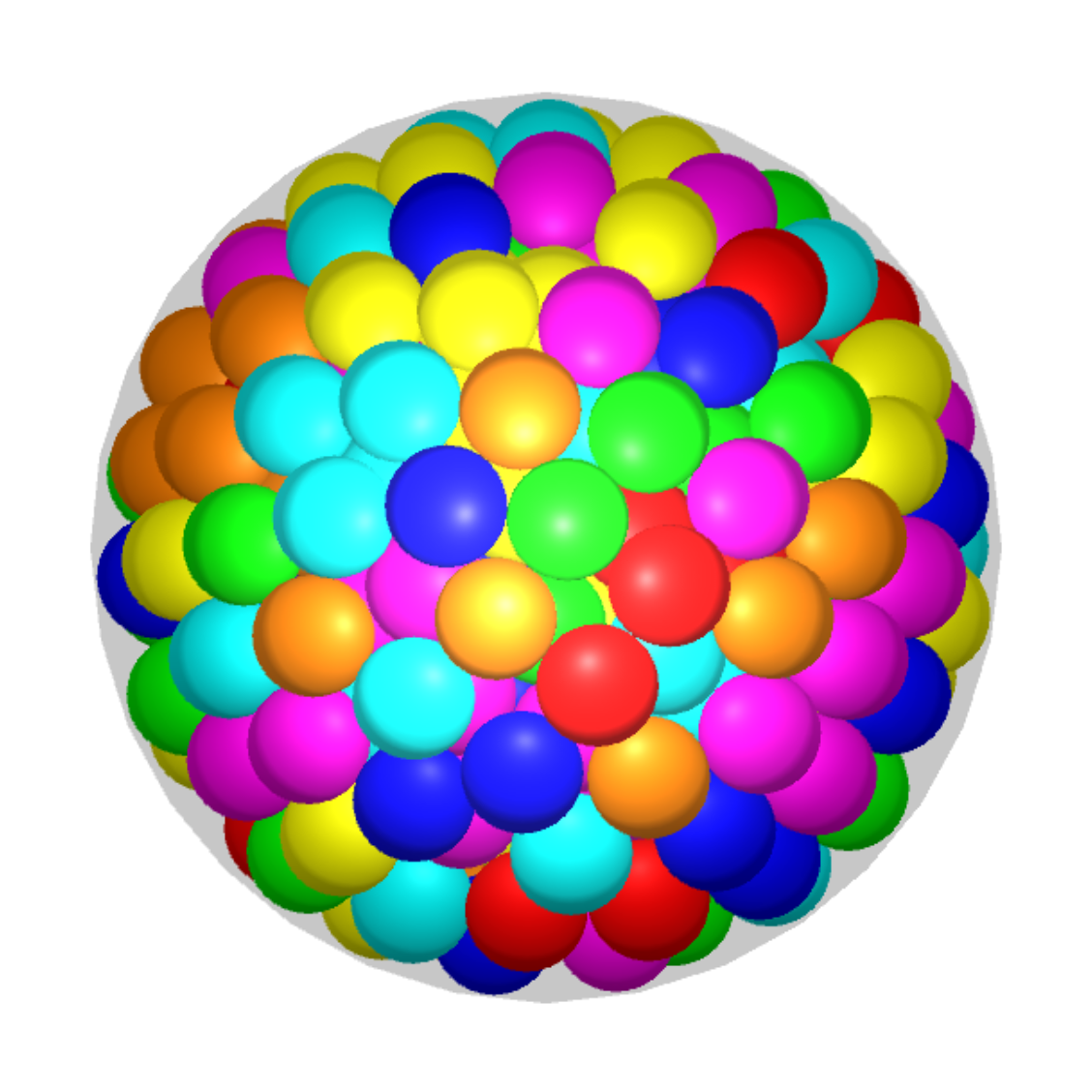}}
    \end{minipage} 
    \begin{minipage}[b]{0.33\linewidth}
        \centering
        \subfloat[][$n = 244$]{\includegraphics[width=1\linewidth]{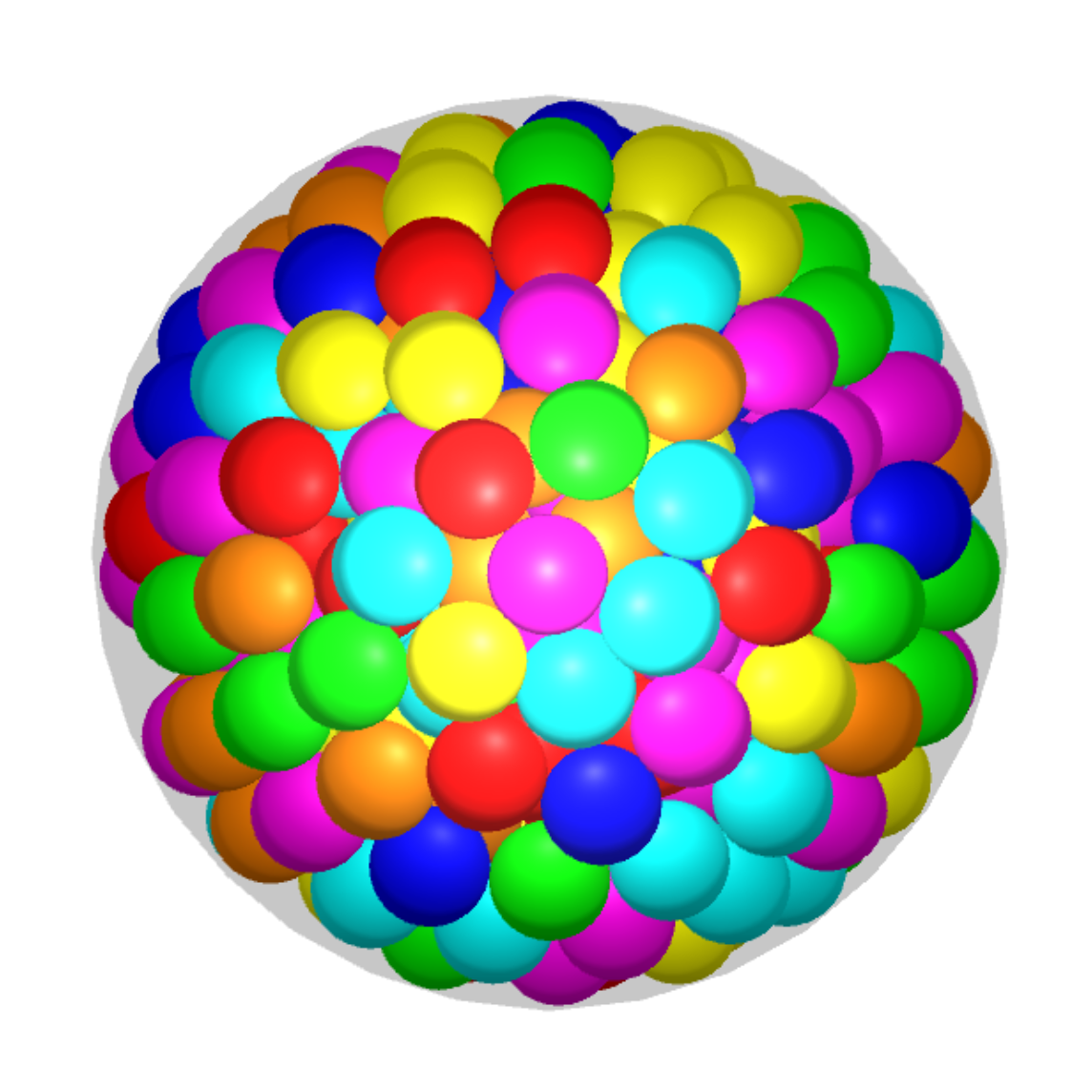}}
    \end{minipage}
    \begin{minipage}[b]{0.33\linewidth}
        \centering
        \subfloat[][$n = 263$]{\includegraphics[width=1\linewidth]{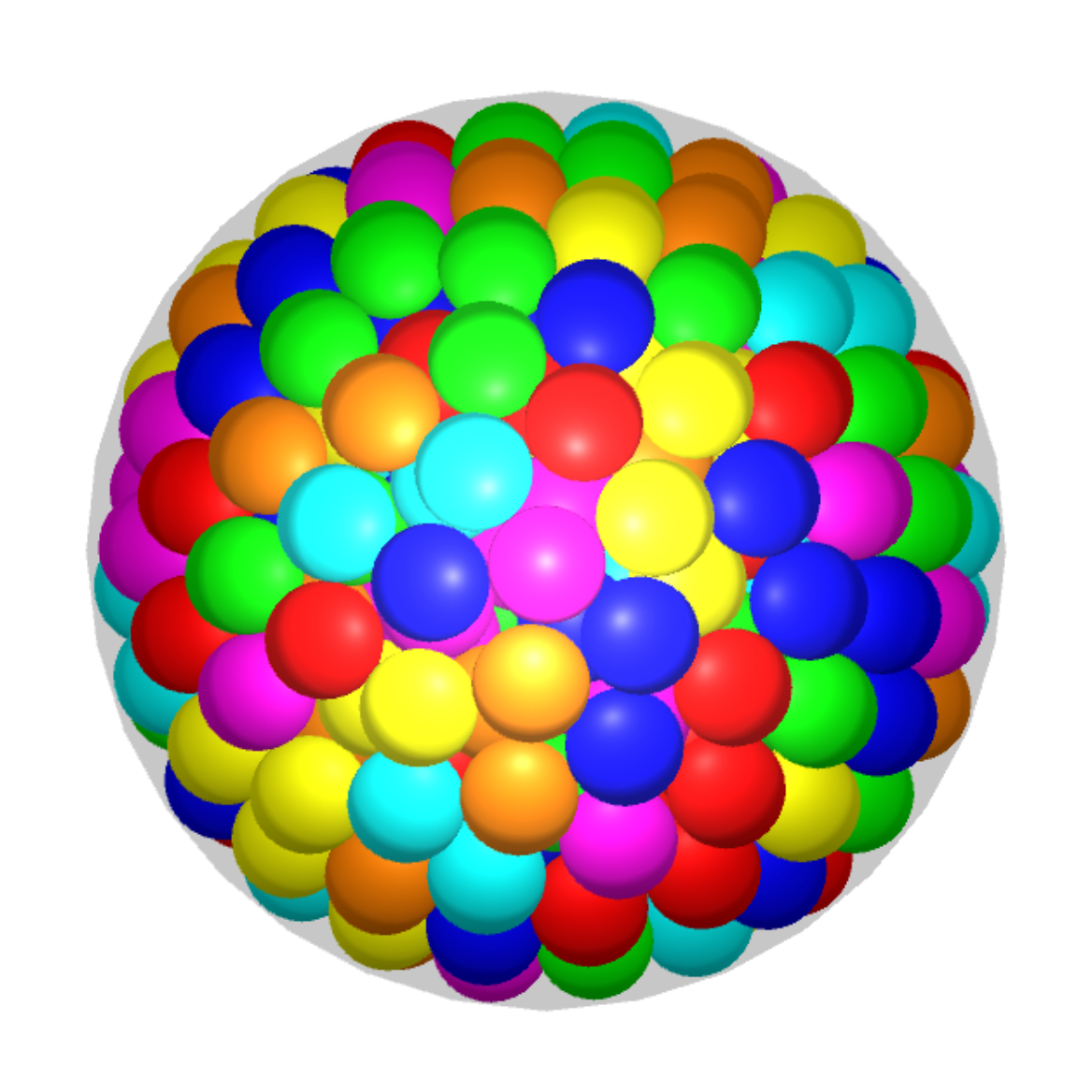}}
    \end{minipage}    
    \begin{minipage}[b]{0.33\linewidth}
        \centering
        \subfloat[][$n = 298$]{\includegraphics[width=1\linewidth]{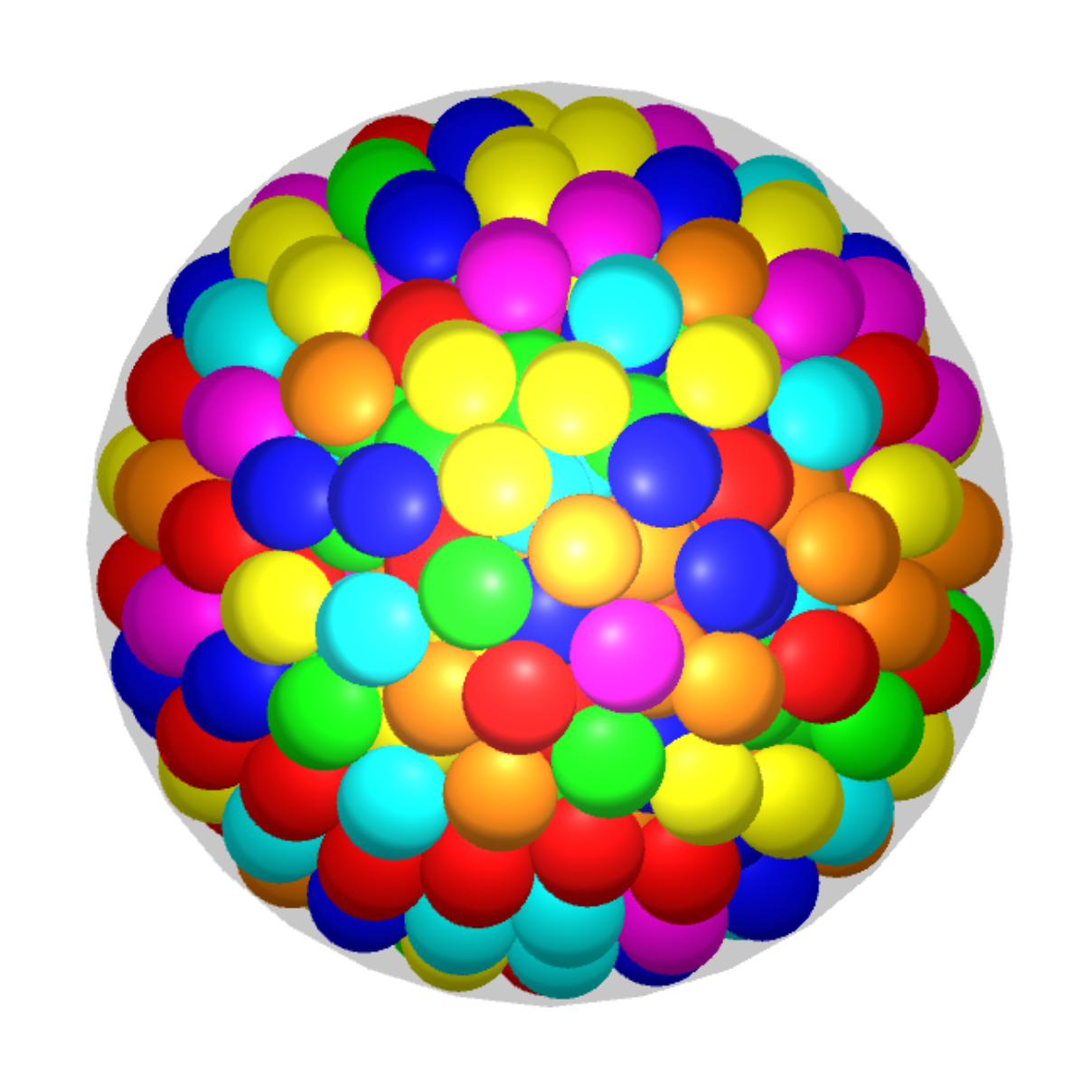}}
    \end{minipage}
    \begin{minipage}[b]{0.33\linewidth}
        \centering
        \subfloat[][$n = 317$]{\includegraphics[width=1\linewidth]{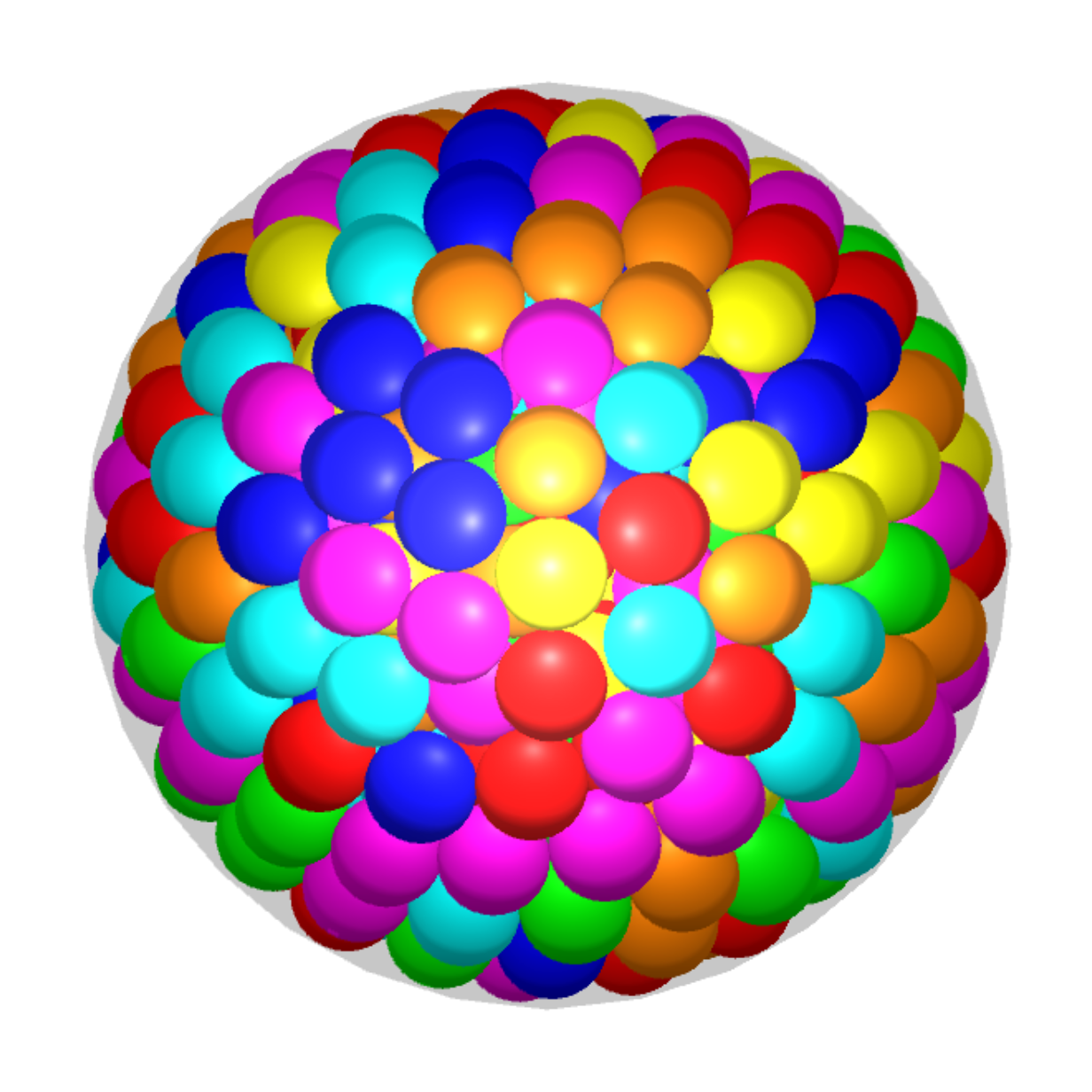}}
    \end{minipage}
    \begin{minipage}[b]{0.33\linewidth}
        \centering
        \subfloat[][$n = 355$]{\includegraphics[width=1\linewidth]{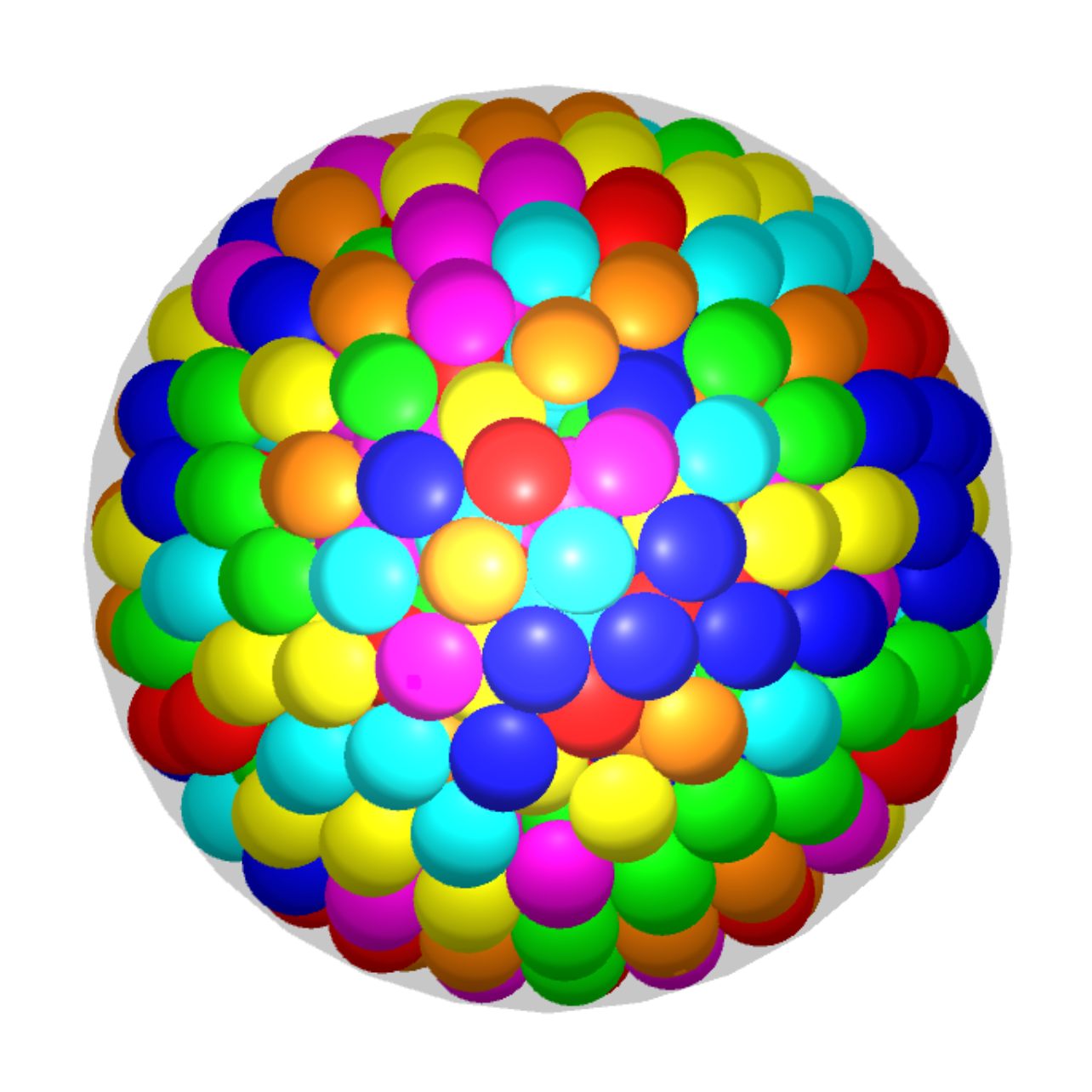}}
    \end{minipage}
    \begin{minipage}[b]{0.33\linewidth}
        \centering
        \subfloat[][$n = 361$]{\includegraphics[width=1\linewidth]{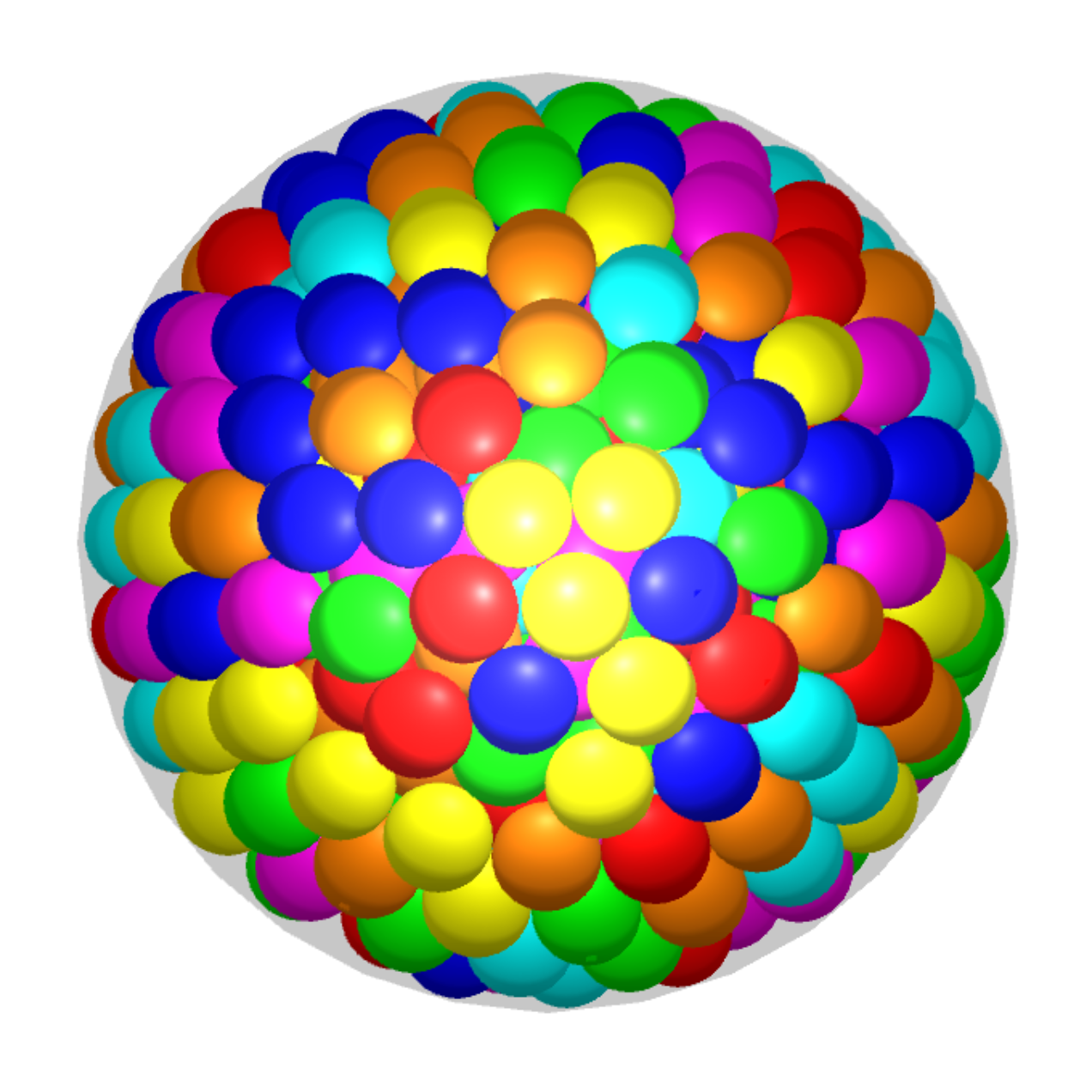}}
    \end{minipage}    
    \begin{minipage}[b]{0.33\linewidth}
        \centering
        \subfloat[][$n = 395$]{\includegraphics[width=1\linewidth]{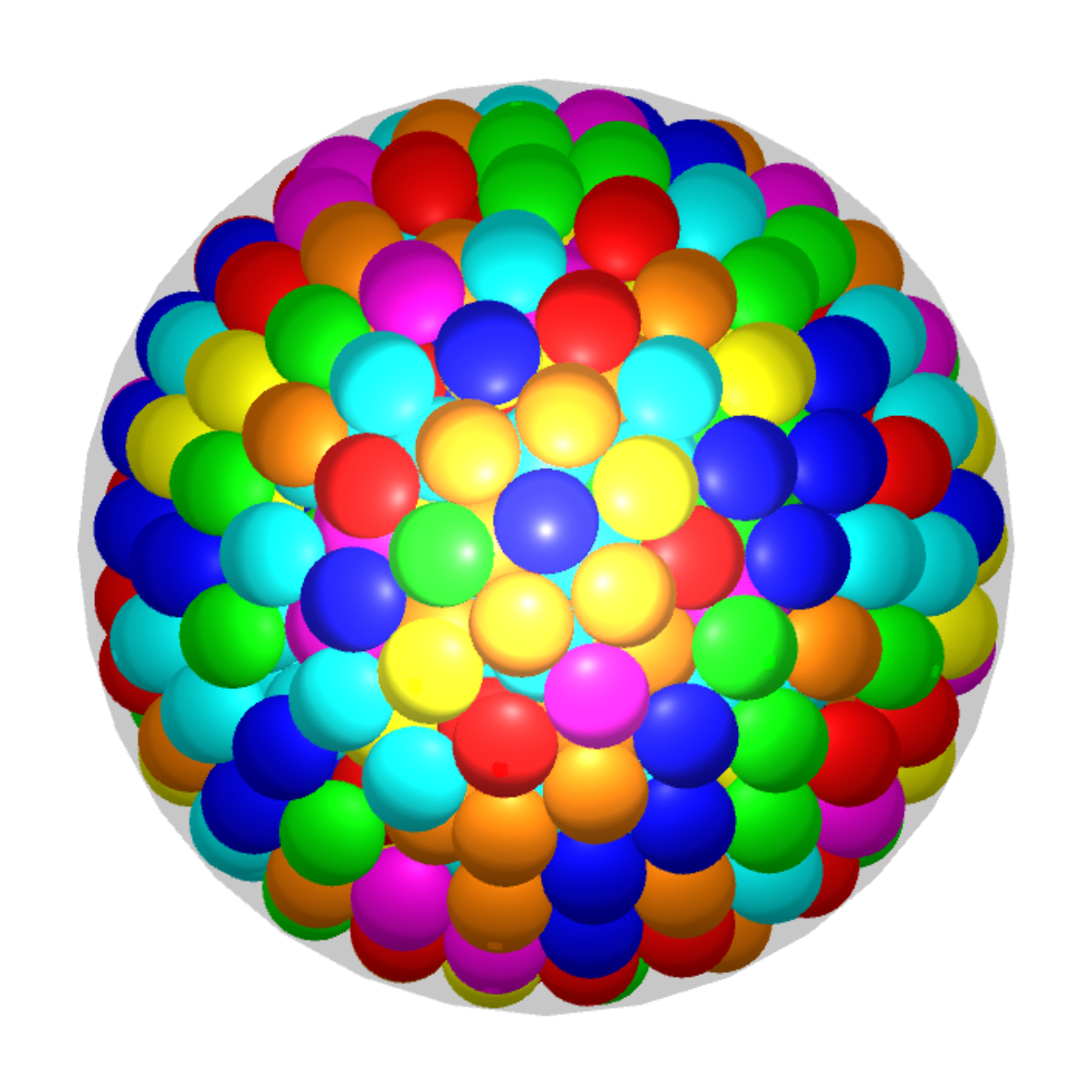}}
    \end{minipage}
    
    \caption{New improved solutions found by our algorithm sampled from the moderate II scale instances $201 \leq n \leq 400$.}
    \label{fig_results_m2}
\end{figure}

Figures~\ref{fig_results_s}-\ref{fig_results_m2} show the new improved results obtained by our algorithm in this work where each figure has 9 representative configurations selected from the corresponding scale instances. 


\subsection{Parameter Study} \label{ssec:05-03PS}

There are three parameters that need to be tuned in this work, including the maximum iteration step $S_{iter}$ of the heuristic SED (see in Algorithm~\ref{alg_sed}), the controlling coefficient $c$ of  the function $J$ (see in Section~\ref{ssec:04-03sed}), and the perturbing parameter $\theta$ of the uniform distribution $U$ (see in Section~\ref{ssec:04-03sed}). We randomly select 2, 4, and 8 instances from the small, moderate I, and moderate II scales, respectively. We tune the three parameters on the 14 selected test instances. The tuning experiments and parameter study are presented as follows.

\begin{table}[tb]
\centering
\caption{Computational results on the average result $R_{avg}$ with the comparison of the maximum iteration step $S_{iter}$ of SED for the 14 selected instances where the best results obtained among the tested parameter values are presented in bold.}
\label{tb_para_iter}

\scalebox{0.87}{
\begin{tabular}{llllllllllll}
\toprule
\diagbox[]{$n$}{$S_{iter}$} & 500                   &  & 600                   &  & 700                   &  & 800                   &  & 900                   &  & 1000                  \\ \midrule
60              & \textbf{4.7749335903} &  & \textbf{4.7749335903} &  & \textbf{4.7749335903} &  & \textbf{4.7749335903} &  & \textbf{4.7749335903} &  & \textbf{4.7749335903} \\
85              & 5.3685586845          &  & 5.3675428274          &  & \textbf{5.3670410532} &  & 5.3670476557          &  & 5.3685604097          &  & 5.3695772339          \\
125             & \textbf{6.0353740583} &  & 6.0365514182          &  & 6.0356903597          &  & 6.0356615862          &  & 6.0365558169          &  & 6.0356733701          \\
140             & \textbf{6.2341210801} &  & 6.2341254599          &  & 6.2342623742          &  & 6.2351625245          &  & 6.2353866960          &  & 6.2353877254          \\
177             & \textbf{6.7386089529} &  & 6.7388866702          &  & 6.7390210504          &  & 6.7391547201          &  & 6.7393323039          &  & 6.7389221430          \\
199             & \textbf{7.0059307131} &  & 7.0079084917          &  & 7.0065604316          &  & 7.0076963008          &  & 7.0072536007          &  & 7.0071925355          \\
205             & 7.0719017567          &  & 7.0714566652          &  & \textbf{7.0702898675} &  & 7.0715287949          &  & 7.0734399411          &  & 7.0725053780          \\
212             & 7.1499807252          &  & 7.1501900954          &  & \textbf{7.1490794335} &  & 7.1492131460          &  & 7.1499729965          &  & 7.1495631426          \\
239             & 7.4278085363          &  & \textbf{7.4241903991} &  & 7.4256091484          &  & 7.4258169196          &  & 7.4278516343          &  & 7.4272433070          \\
257             & 7.5829603512          &  & 7.5830260777          &  & \textbf{7.5823156930} &  & 7.5828835977          &  & 7.5832012629          &  & 7.5844114527          \\
285             & 7.8300728675          &  & 7.8356930128          &  & 7.8311060285          &  & 7.8360062185          &  & \textbf{7.8262750705} &  & 7.8340063708          \\
316             & 8.1005066257          &  & 8.1006066623          &  & 8.0974417541          &  & \textbf{8.0974261420} &  & 8.1008944597          &  & 8.1045266682          \\
364             & 8.5053879884          &  & \textbf{8.5010834702} &  & 8.5015802313          &  & 8.5026067074          &  & 8.5048700031          &  & 8.5047659284          \\
398             & \textbf{8.7635298355} &  & 8.7641287155          &  & 8.7635387623          &  & 8.7636618905          &  & 8.7645112445          &  & 8.7658463846          \\ \midrule
Average         & 7.0421196976          &  & 7.0421659683          &  & \textbf{7.0413192699} &  & 7.0420571282          &  & 7.0423599307          &  & 7.0431825165          \\ \bottomrule
\end{tabular}
}
\end{table}

\textbf{On the maximum iteration step of SED.} 
We perform our algorithm with different maximum iteration step settings of SED for $S_{iter} = 500, 600, ..., 1000$ on the 14 selected instances, and the other parameters are set as default ($c = 7$ and $\theta = 0.8$). Experimental results are summarized in Table~\ref{tb_para_iter}, where the column of $n$ shows the number of packing spheres of the instance, columns 2-7 show the average results $R_{avg}$ of the algorithm over 10 independent runs for each parameter setting $S_{iter}$ on the 14 selected instances. The last row ``Average'' shows the average value of the 14 average results of each column. 

From Table~\ref{tb_para_iter}, we observe that the setting $S_{iter} = 500$ gains the most of best results in terms of $R_{avg}$ among the 6 tested parameter settings for 6 out of the 14 selected instances, followed by the settings $S_{iter} = $ 700, 600, 800, 900 and 1000 gain the number of best results for 5, 3, 2, 2 and 1 out of the 14 selected instances. However, the setting $S_{iter} = 700$ gains the best average value among the 6 tested settings. 
As the setting $S_{iter} = 700$ is only one less than the setting $S_{iter} = 500$ of the best result, but it gains the best average value, so we choose $S_{iter} = 700$ as the default setting.

\begin{table}[tb]
\centering
\caption{Computational results on the average result $R_{avg}$ with the comparison of the controlling coefficient $c$ for the 14 selected instances where the best results obtained among the tested parameter values are presented in bold.}
\label{tb_para_c}

\scalebox{0.9}{
\begin{tabular}{llllllllll}
\toprule
\diagbox[]{$n$}{$c$}   & 5                     &  & 6                     &  & 7                     &  & 8                     &  & 9                     \\ \midrule
60      & \textbf{4.7749335903} &  & \textbf{4.7749335903} &  & \textbf{4.7749335903} &  & \textbf{4.7749335903} &  & \textbf{4.7749335903} \\
85      & 5.3659779921          &  & \textbf{5.3653440298} &  & 5.3670410532          &  & 5.3685624347          &  & 5.3680578744          \\
125     & 6.0359878384          &  & \textbf{6.0352374204} &  & 6.0356903597          &  & 6.0358054004          &  & 6.0358484826          \\
140     & \textbf{6.2341290526} &  & 6.2341662886          &  & 6.2342623742          &  & 6.2353887857          &  & 6.2353827882          \\
177     & 6.7391359616          &  & \textbf{6.7388725013} &  & 6.7390210504          &  & 6.7391934704          &  & 6.7398805469          \\
199     & 7.0070497619          &  & \textbf{7.0064735311} &  & 7.0065604316          &  & 7.0080724710          &  & 7.0069524091          \\
205     & 7.0725672377          &  & 7.0736081776          &  & \textbf{7.0702898675} &  & 7.0725855196          &  & 7.0706347878          \\
212     & 7.1512715786          &  & 7.1502877988          &  & \textbf{7.1490794335} &  & 7.1496716173          &  & 7.1506760487          \\
239     & 7.4292899636          &  & 7.4282991831          &  & \textbf{7.4256091484} &  & 7.4272586794          &  & 7.4262519789          \\
257     & 7.5841140578          &  & \textbf{7.5820822034} &  & 7.5823156930          &  & 7.5823303822          &  & 7.5861163752          \\
285     & 7.8313928737          &  & 7.8345924449          &  & \textbf{7.8311060285} &  & 7.8354999162          &  & 7.8435577572          \\
316     & 8.1038412345          &  & 8.0990948706          &  & \textbf{8.0974417541} &  & 8.0974765499          &  & 8.1045383268          \\
364     & 8.5046370926          &  & 8.5031591953          &  & 8.5015802313          &  & 8.5031584842          &  & \textbf{8.4994721948} \\
398     & 8.7670891784          &  & 8.7658106100          &  & 8.7635387623          &  & 8.7641103181          &  & \textbf{8.7630427590} \\ \midrule
Average & 7.0429583867          &  & 7.0422829889          &  & \textbf{7.0413192699} &  & 7.0424319728          &  & 7.0432389943          \\ \bottomrule
\end{tabular}
}
\end{table}

\textbf{On the controlling coefficient.} The parameter $c$ is a coefficient of the function $J$ (Eq.~\ref{eq-6}) to control the size of perturbing candidate set $C$.
We perform our algorithm with different controlling coefficient settings for $c = 5, 6, ..., 9$ on the 14 selected instances, and the other parameters are set as default ($S_{iter} = 700$ and $\theta = 0.8$). Experimental results are summarized in Table~\ref{tb_para_c}, where the column of $n$ shows the number of packing spheres of the instance, columns 2-6 show the average results $R_{avg}$ of the algorithm over 10 independent runs for each parameter setting $c$ on the 14 selected instances. The last row ``Average'' shows the average value of the 14 average results of each column. 

From Table~\ref{tb_para_c}, we have the following observations. The setting of $c = $ 6 and 7 gains the most of best results in terms of $R_{avg}$ among the 5 tested parameter settings for 6 out of the 14 selected instances, followed by the settings $c = $ 9, 5 and 8 gain the number of 3, 2 and 1 out of the 14 selected instances. And the setting $c = 7$ gains the best average value among the 5 tested settings. Therefore, we choose $c = 7$ as the default setting. 

\begin{table}[tb]
\centering
\caption{Computational results on the average result $R_{avg}$ with the comparison of the perturbing parameter $\theta$ for the 14 selected instances where the best results obtained among the tested parameter values are presented in bold.}
\label{tb_para_theta}

\scalebox{0.9}{
\begin{tabular}{llllllllll}
\toprule
\diagbox[]{$n$}{$\theta$} & 0.6                   &  & 0.7                   &  & 0.8                   &  & 0.9                   &  & 1.0                   \\ \midrule
60        & \textbf{4.7749335903} &  & \textbf{4.7749335903} &  & \textbf{4.7749335903} &  & \textbf{4.7749335903} &  & \textbf{4.7749335903} \\
85        & 5.3699068636          &  & 5.3690764628          &  & \textbf{5.3670410532} &  & 5.3670841519          &  & 5.3674077272          \\
125       & 6.0405877816          &  & 6.0367733617          &  & \textbf{6.0356903597} &  & 6.0373515386          &  & 6.0418923290          \\
140       & 6.2402073454          &  & 6.2345660948          &  & \textbf{6.2342623742} &  & 6.2347088694          &  & 6.2353349643          \\
177       & 6.7479598125          &  & 6.7422245246          &  & \textbf{6.7390210504} &  & 6.7424724758          &  & 6.7516510345          \\
199       & 7.0147025938          &  & 7.0071138997          &  & \textbf{7.0065604316} &  & 7.0198344634          &  & 7.0210409221          \\
205       & 7.0829570570          &  & 7.0774460733          &  & \textbf{7.0702898675} &  & 7.0868646404          &  & 7.0886441012          \\
212       & 7.1555605981          &  & 7.1506092628          &  & \textbf{7.1490794335} &  & 7.1633692354          &  & 7.1643829465          \\
239       & 7.4325759003          &  & 7.4277873712          &  & \textbf{7.4256091484} &  & 7.4412882629          &  & 7.4417633709          \\
257       & 7.6062088590          &  & 7.5961504907          &  & \textbf{7.5823156930} &  & 7.6128361575          &  & 7.6163300946          \\
285       & 7.8562965133          &  & 7.8394809703          &  & \textbf{7.8311060285} &  & 7.8655086824          &  & 7.8715087731          \\
316       & 8.1216621740          &  & 8.1152536425          &  & \textbf{8.0974417541} &  & 8.1343618238          &  & 8.1375496725          \\
364       & 8.5111844242          &  & 8.5026266687          &  & \textbf{8.5015802313} &  & 8.5216024571          &  & 8.5245154455          \\
398       & 8.7606948783          &  & \textbf{8.7551382934} &  & 8.7635387623          &  & 8.7737582321          &  & 8.7743796038          \\ \midrule
Average   & 7.0511027422          &  & 7.0449414791          &  & \textbf{7.0413192699} &  & 7.0554267558          &  & 7.0579524697          \\ \bottomrule
\end{tabular}
}
\end{table}

\textbf{On the perturbing parameter.} Finally, we evaluate the performance on different perturbing parameter settings. The perturbing parameter $\theta$ of the uniform distribution $U$ controls the random shifting value of packing spheres' coordinates in the perturbing operator. 
We perform our algorithm with different perturbing parameter settings for $\theta = 0.6, 0.7, ..., 1.0$ on the 14 selected instances, and the other parameters are set as default ($S_{iter} = 700$ and $c = 7$). Experimental results are summarized in Table~\ref{tb_para_theta}, where the column of $n$ shows the number of packing spheres of the instance, columns 2-6 show the average results $R_{avg}$ of the algorithm over 10 independent runs for each parameter setting $\theta$ on the 14 selected instances, and the last row ``Average'' shows the average value of the 14 average results of each column. 

Table~\ref{tb_para_theta} shows the algorithm with the setting $\theta = 0.8$ gains the best performance in terms of $R_{avg}$ for 13 out of the 14 selected instances, which is obviously better than the other 4 tested parameter settings. And it also gains the best average value among the 5 tested parameter settings. Therefore, we choose $\theta = 0.8$ as the default setting.

\subsection{Analysis of ANM}

ANM is an adaptive maintenance module described in Section~\ref{ssec:04-05anm}. It is adapted to maintain the efficient neighbor structure of packing objects and accelerate the convergence speed of the continuous optimization process. Here we design an experiment and present a comparison to evaluate the performance of our proposed ANM module. 

\begin{figure}[tb]
    \centering
    \begin{minipage}[b]{0.49\linewidth}
        \centering
        \subfloat[Runtime comparison of brute force and ANM]{\includegraphics[width=1\linewidth]{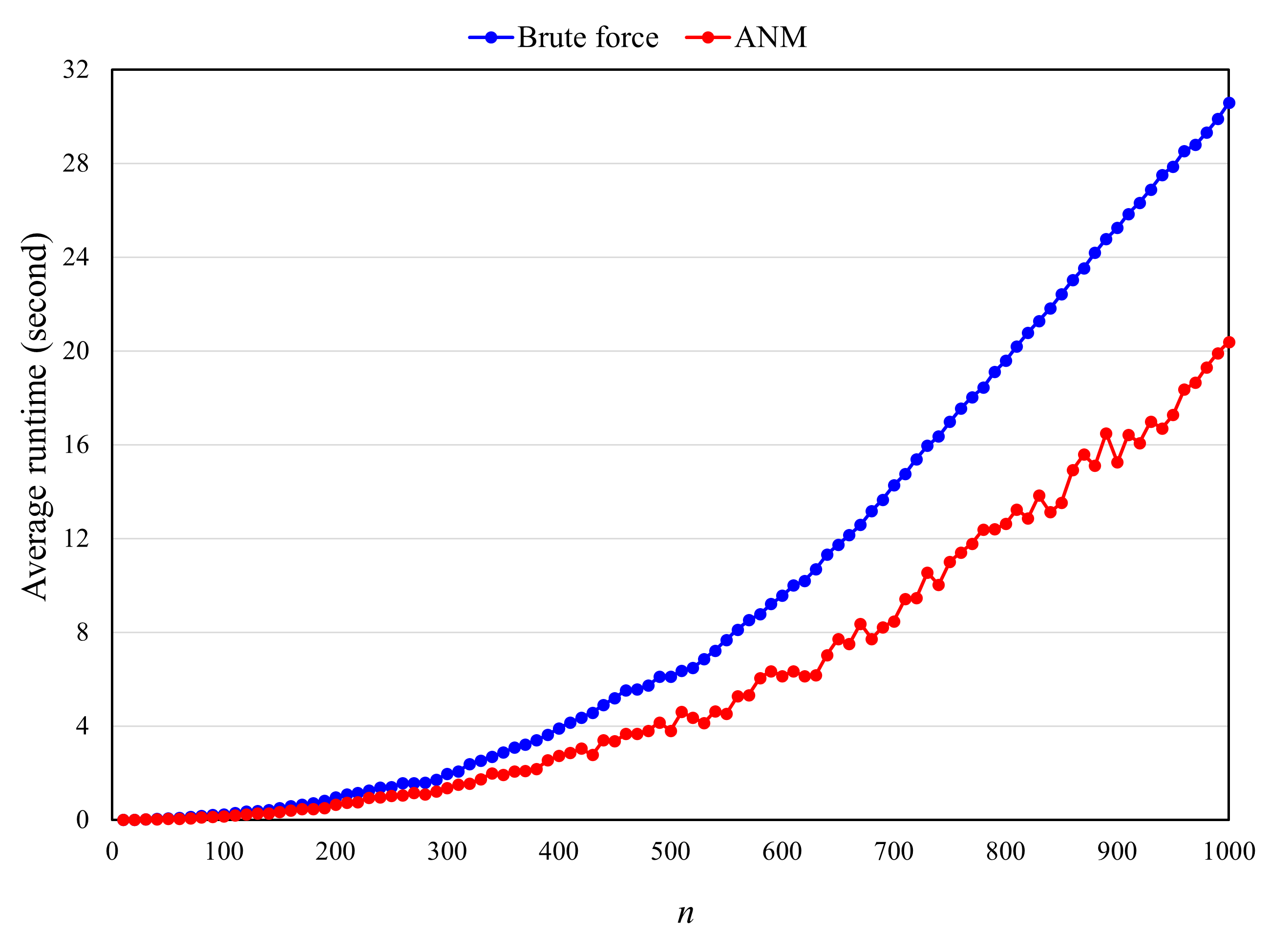} \label{fig_cmp_ANM_runtime}}
    \end{minipage}
    \begin{minipage}[b]{0.49\linewidth}
        \centering
        \subfloat[Deferring ratio of ANM]{\includegraphics[width=1\linewidth]{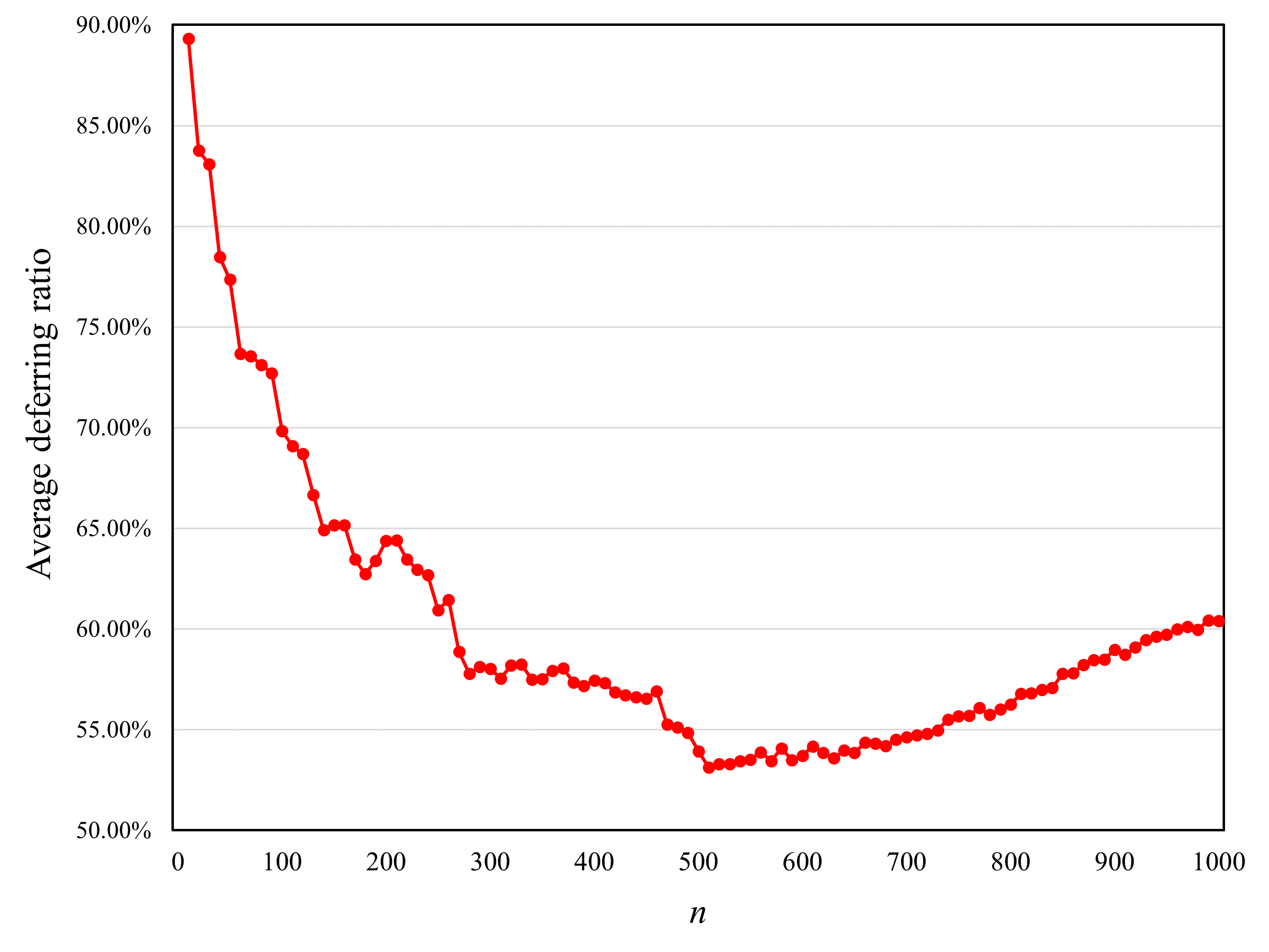} \label{fig_cmp_ANM_ratio}}
    \end{minipage}    

    \caption{Experimental results and comparison of the brute force approach and ANM. The left shows the average runtime plot of the two approaches and the right shows the average deferring ratio plot of ANM.}
    \label{fig_cmp_ANM}
\end{figure}

The ANM module only affects the performance of the continuous optimization process. Therefore, we make a comparison of the brute force approach and ANM on the continuous optimization process of the PESS elastic system. The brute force approach reconstructs the neighbor structure at each iteration. 

Starting from a random initial layout with the radius $R = \sqrt[3]{\frac{n}{0.6}}$, we perform the continuous optimization method (i.e., the L-BFGS algorithm in this work) with the brute force approach and ANM, respectively, until the convergence condition is met. And the experiment is performed for $n = $ 10, 20, 30, ..., 1000 and each setting $n$ performs 1,000 independent runs, the results and comparison are presented in Figure~\ref{fig_cmp_ANM}.

Figure~\ref{fig_cmp_ANM_runtime} shows the runtime comparison of the brute force approach and ANM where the X-axis indicates the number $n$ of packing items and the Y-axis indicates the average runtime consumption over 1,000 independent runs. Figure~\ref{fig_cmp_ANM_ratio} shows the deferring ratio plot of ANM where the X-axis indicates the number $n$ of packing objects and the Y-axis indicates the average deferring ratio over 1,000 independent runs of the ANM module, the deferring ratio is defined as $\mathrm{Ratio(\%)} = 1 - \frac{N_{rec.}}{N_{iter.}}$ where $N_{rec.}$ indicates the number of triggering the neighbor reconstruction of ANM and $N_{iter.}$ indicates the number of the continuous optimization iterations. 

From Figure~\ref{fig_cmp_ANM}, we can observe that adopting the AMN module instead of the brute force approach can significantly speed up the convergence of the continuous optimization process, and the gap between the brute force and AMN increases with increment on the number of packing items. For example, the average runtimes of the brute force approach on the settings $n = $ 200, 400, 600, 800 and 1,000 are 0.96, 3.90, 9.56, 19.58 and 30.58 seconds, and the corresponding average runtimes of ANM are 0.66, 2.73, 6.14, 12.62 and 20.38 seconds where the average runtime ratio of ANM to brute force between 64.23\% and 70.00\%, which means ANM can reduce over 30\% time consumption compared with the brute force approach. 
And the ANM module defers over 50\% unnecessary maintenance during the continuous optimization process for $10 \leq n \leq 1000$. 

In experiments, we observe that the deferring ratio can be increased by tuning the reset value and the multiplying factor of the deferring length $len$ (see in Algorithm~\ref{alg_opti} lines 1,11,13), which can further reduce the time consumption. For example, the reset value can be fixed to 10 (i.e., using $len \leftarrow 10$ instead of $len \leftarrow 1$). In this way, ANM degenerates into the simple method~\citep{he2018efficient}, reconstructing the neighbor every 10 iterations, in the unstable situation. And the multiplying factor can be set to a larger value to enhance the deferring feature (i.e., using $len \leftarrow k \times len$ instead of $len \leftarrow 2 \times len$ where $k > 2$). However, we do not tune these settings and hold the generality and effectiveness of ANM. 

It is worth noting that we employ the efficient scan line approach to construct the neighbor structure, gaining a low time complexity $O(n\sqrt{n})$. The gap between the brute force approach and AMN will dramatically increase if we employ the naive construct approach (i.e., enumerating pairwise objects $O(n^2)$).

\section{Conclusion}
In this work, we present an effective search algorithm to solve the packing equal spheres in a spherical container problem, which is computationally very challenging. Two methods are proposed to achieve this goal. The first method is the Solution space Exploring and Descent heuristic, denoted as SED, that is an efficient local search algorithm for discovering a high-quality solution. The second method is the Adaptive Neighbor object Maintenance method, denoted as ANM, for maintaining the efficient neighbor structure to solve packing problems, which can significantly reduce unnecessary maintenance and speed up the convergence of the continuous optimization process. Our algorithm's excellent performance was demonstrated on the well-known benchmark instances with up to 400 spheres. Specifically, our algorithm improved the best-known result for 274 instances and matched the best-known result for 84 instances out of the 396 benchmark instances. 
Besides, the idea of SED and ANM is general in nature and can be easily extended to other packing and global optimization problems. 


\bibliographystyle{cas-model2-names}

\bibliography{mybibfile}



\end{document}